\documentclass{article}
\usepackage{fullpage}

\usepackage{paralist}
\usepackage{mathpartir}
\usepackage{xspace}
\usepackage{amsfonts}
\usepackage{amsthm}
\usepackage{amssymb}
\usepackage{amsmath}
\usepackage{stmaryrd} 
\usepackage{verbatim}
\usepackage{paralist}
\usepackage{graphicx}
\usepackage{fancyhdr}
\usepackage{placeins} % For \FloatBarrier
\usepackage[usenames,dvipsnames]{color}
\usepackage{soul}
\usepackage{fancyvrb}
\usepackage{enumerate}
\usepackage[all,cmtip]{xy}
\usepackage{times}

\newcommand{\eod}{\end{document}}
\newcommand{\im}[1]{\ensuremath{#1}}

\newcommand{\kw}[1]{\im{\mathtt{#1}}}

\newcommand{\ra}{\im{\rightarrow}}
\newcommand{\red}{\Downarrow}

\renewcommand{\hole}{\im{\square}}

\newcommand{\omitthis}[1]{}

\newenvironment{nop}{}{}

\newenvironment{sdisplaymath}{
\begin{nop}\small\begin{displaymath}}{
\end{displaymath}\end{nop}\ignorespacesafterend}

\newenvironment{smathpar}{
\begin{nop}\small\begin{mathpar}}{
\end{mathpar}\end{nop}\ignorespacesafterend}

\newenvironment{stackAux}[2]{%
\setlength{\arraycolsep}{0pt}
\begin{array}[#1]{#2}}{
\end{array}}

\newtheoremstyle{athm}{\topsep}{\topsep}%
     {\itshape}%         Body font
     {}%         Indent amount (empty = no indent, \parindent = para indent)
     {\bfseries}% Thm head font
     {.}%        Punctuation after thm head
     {.8em}%     Space after thm head (\newline = linebreak)
     {\thmname{#1}\thmnumber{ #2}\thmnote{~\,(#3)}}%         Thm head spec

\theoremstyle{athm}
\newtheorem{theorem}{Theorem}[section]
\newtheorem{lemma}[theorem]{Lemma}

\newtheorem{definition}[theorem]{Definition}
\newtheorem{proposition}[theorem]{Proposition}
\newtheorem{example}[theorem]{Example}
\newtheorem{remark}[theorem]{Remark}

\newenvironment{mathfig}{\begin{sdisplaymath}}{\end{sdisplaymath}}
\newenvironment{syntaxfig}{\begin{mathfig}\begin{array}{@{}l@{\quad}r@{~~}c@{\quad}l}}{\end{array}\end{mathfig}}

\renewcommand{\vec}[1]{\overline{#1}}

\newcommand{\Traces}{\mathcal{T}}
\newcommand{\Queries}{\mathcal{Q}}
\newcommand{\Bool}{\mathbb{B}}

\newcommand{\bindsto}{{\mapsto}}
\newcommand{\pl}{\textsf{TML}\xspace}
\newcommand{\plz}{\text{Core \textsf{TML}}\xspace}

\newcommand{\ts}{\vdash}
\newcommand{\tst}{\ts_\mathtt{T}}

\newcommand{\novec}[1]{\vec{#1}}

\newcommand{\lbl}{\im{\ell}\xspace}

\newcommand{\kwcase}{\kw{case}}
\newcommand{\kwof}{\kw{of}}
\newcommand{\kwinl}{\kw{inl}}
\newcommand{\kwinr}{\kw{inr}}
\newcommand{\kwroll}{\kw{roll}}
\newcommand{\kwunroll}{\kw{unroll}}
\newcommand{\kwfun}{\kw{fun}}

\newcommand{\kwtrue}{\kw{true}}

\newcommand{\kwf}{\oplus}

\newcommand{\kwfst}{\kw{fst}}
\newcommand{\kwsnd}{\kw{snd}}

\newcommand{\kwif}{\kw{if}}
\newcommand{\kwlet}{\kw{let}}
\newcommand{\kwapp}{\kw{app}}
\newcommand{\kwin}{\kw{in}}

\newcommand{\kwthen}{\kw{then}}
\newcommand{\kwelse}{\kw{else}}

\newcommand{\tracemark}{\triangleright}

\newcommand{\trvar}[1]{#1}
\newcommand{\trfst}[1]{\exfst{#1}}
\newcommand{\trsnd}[1]{\exsnd{#1}}

\newcommand{\trc}{c}
\newcommand{\trf}[1]{\exf{#1}}

\newcommand{\trpair}[2]{\expair{#1}{#2}}

\newcommand{\trcasel}[6]{\excase{#5}{#1}{#2}{#3}{#4}\tracemark_{\kwinl}#1.#6}
\newcommand{\trcaser}[6]{\excase{#5}{#1}{#2}{#3}{#4}\tracemark_{\kwinr}#3.#6}
\newcommand{\trcaseml}[4]{\kwcase~#2 \tracemark_{\kwinl} #3.#4}
\newcommand{\trcasemr}[4]{\kwcase~#2  \tracemark_{\kwinr} #3.#4}
\newcommand{\trcaseshortl}[3]{(\kwcase~#1) \tracemark_{\kwinl} #2.#3}
\newcommand{\trcaseshortr}[3]{(\kwcase~#1) \tracemark_{\kwinr} #2.#3}

\newcommand{\trfunk}[1]{\exfunk{#1}}

\newcommand{\trapp}[6]{(#4~#5) \tracemark_{#1(#2).#3} #1(#2).#6}
\newcommand{\trappk}[6]{(#2~#3)\tracemark_{#1} #4(#5).#6}
\newcommand{\trlet}[3]{\exlet{#1}{#2}{#3}}
\newcommand{\tremp}{\hole}
\newcommand{\trroll}[1]{\exroll{#1}}
\newcommand{\trunroll}[1]{\exunroll{#1}}
\newcommand{\trinl}[1]{\exinl{#1}}
\newcommand{\trinr}[1]{\exinr{#1}}

\newcommand{\vclos}[2]{\langle #1,#2 \rangle}
\newcommand{\vc}{c}
\newcommand{\vpair}[2]{\expair{#1}{#2}}
\newcommand{\vany}{{\im{\Diamond}}}
\newcommand{\vhole}{\im{\hole}}
\newcommand{\vinl}[1]{\exinl{#1}}
\newcommand{\vinr}[1]{\exinr{#1}}
\newcommand{\vroll}[1]{\exroll{#1}}

\newcommand{\exvar}[1]{#1}
\newcommand{\exc}{c}
\newcommand{\exf}[1]{\oplus(#1)}
\newcommand{\expair}[2]{(#1,#2)}
\newcommand{\exfst}[1]{\kwfst(#1)}
\newcommand{\exsnd}[1]{\kwsnd(#1)}

\newcommand{\exfun}[3]{\im{\kwfun~\fn{#1}{#2}{#3}}}
\newcommand{\exfunk}[1]{\im{\kwfun~#1}}
\newcommand{\fn}[3]{\im{#1(#2). #3}}
\newcommand{\exapp}[2]{(#1~#2)}
\newcommand{\exlet}[3]{\kwlet~{#2=#1}~\kwin~{#3}}
\newcommand{\exletin}[2]{\kwlet~{#1}~\kwin~{#2}}
\newcommand{\exif}[3]{\kwif~#1~\kwthen~#2~\kwelse~#3}
\newcommand{\excase}[5]{\excasem{#1}{\match{#2}{#3}{#4}{#5}}}
\newcommand{\excasem}[2]{\kwcase~#1~\kwof~#2}
\newcommand{\match}[4]{\{\kwinl(#1).#2 ;  \kwinr(#3).#4\}}
\newcommand{\exinl}[1]{\kwinl(#1)}
\newcommand{\exinr}[1]{\kwinr(#1)}
\newcommand{\exroll}[1]{\kwroll(#1)}
\newcommand{\exunroll}[1]{\kwunroll(#1)}
\newcommand{\tyrec}[2]{\mu{#1}.#2}
\newcommand{\eval}{\red}

\newcommand{\trrun}{\curvearrowright}

\newcommand{\sqgeq}{\sqsupseteq}
\newcommand{\sqleq}{\sqsubseteq}
\newcommand{\eqat}[1]{\im{\eqsim_{#1}}}
\newcommand{\eqxat}[1]{\im{\approx_{#1}}}
\newcommand{\witness}{\mathsf{Witness}}

\newcommand{\Val}{\mathrm{Val}}
\newcommand{\Env}{\mathrm{Env}}

\newcommand{\dom}{\mathrm{dom}}

\newcommand{\where}{\mathsf{W}}
\newcommand{\dep}{\mathsf{D}}
\newcommand{\expr}{\mathsf{E}}
\newcommand{\extract}{\mathsf{F}}

\newcommand{\bwdslice}{\stackrel{\mathrel{\mathsf{disc}}}{\longrightarrow}}

\newcommand{\fwdslice}{\stackrel{\mathrel{\mathsf{obf}}}{\longrightarrow}}

\newcommand{\thmref}[1]{Theorem~\ref{thm:#1}}
\newcommand{\lemref}[1]{Lemma~\ref{lem:#1}}

\newcommand{\figref}[1]{Figure~\ref{fig:#1}}
\newcommand{\appref}[1]{Appendix~\ref{app:#1}}

\newcommand{\av}{\widehat{v}}
\newcommand{\agamma}{\widehat{\gamma}}
\newcommand{\Loc}{\mathrm{Loc}}
\newcommand{\comp}{\circ}

\newcommand{\occ}{\mathit{occ}}
\newcommand{\occnb}{\occ^{\not\bot}}

\newcommand{\REPLAY}{\mathsf{Replay}}
\newcommand{\IN}{\mathsf{IN}}
\newcommand{\OUT}{\mathsf{OUT}}

\newcommand{\Disc}{\mathsf{Disc}}
\newcommand{\Obf}{\mathsf{Obf}}

\newcommand{\path}{\mathsf{path}}

\newcounter{remarkcounter}[section]

\newcommand{\uremark}[1]{}%\myremark{Umut}{U}{#1}}
\newcommand{\rremark}[2]{}%\myremark{Roly}{R}{#1}}

\definecolor{lightGray}{RGB}{200,200,200}
\begin{document}

\title{A Core Calculus for Provenance}

%\author{Umut A. Acar\inst{1} \and Amal Ahmed\inst{2} \and James Cheney\inst{3} \and Roly Perera\inst{1}}
\author{Umut A. Acar \and Amal Ahmed \and James Cheney \and Roly
  Perera}

%\institute{
 %  Max Planck Institute for Software Systems \email{\{umut,rolyp\}@mpi-sws.org} 
%   \and 
%   Indiana University \email{amal@cs.indiana.edu}
%   \and 
%   University of Edinburgh \email{jcheney@inf.ed.ac.uk}
%}

\maketitle

\begin{abstract}
  Provenance is an increasing concern due to the ongoing revolution in
  sharing and processing scientific data on the Web and in other
  computer systems. It is proposed that many computer systems will
  need to become provenance-aware in order to provide satisfactory
  accountability, reproducibility, and trust for scientific or other
  high-value data. To date, there is not a consensus concerning
  appropriate formal models or security properties for provenance. In
  previous work, we introduced a formal framework for provenance
  security and proposed formal definitions of properties called
  disclosure and obfuscation.  

    In this article, we study refined notions of positive and negative
    disclosure and obfuscation in a concrete setting, that of a
    general-purpose programing language.  Previous models of
    provenance have focused on special-purpose languages such as
    workflows and database queries. We consider a higher-order,
    functional language with sums, products, and recursive types and
    functions, and equip it with a tracing semantics in which traces
    themselves can be replayed as computations.  We present an
    annotation-propagation framework that supports many provenance
    views over traces, including standard forms of provenance studied
    previously. We investigate some relationships among provenance
    views and develop some partial solutions to the disclosure and
    obfuscation problems, including correct algorithms for disclosure
    and positive obfuscation based on trace slicing.
\end{abstract}

%%% Local Variables: 
%%% mode: latex
%%% TeX-master: "main"
%%% End: 

% LocalWords:  reproducibility workflows

\section{Introduction}

Provenance, or meta-information about the origin, history, or
derivation of an object, is now recognized as a central challenge in
establishing trust and providing security in computer systems,
particularly on the Web.  Essentially, provenance management involves
instrumenting a system with detailed monitoring or logging of
auditable records that help explain how results depend on inputs or
other (sometimes untrustworthy) sources. The security and privacy
ramifications of provenance must be understood in order to safely meet
the needs of users that desire provenance without introducing new
security vulnerabilities or compromising the confidentiality of other
users.

The lack of adequate provenance information can cause (and has caused)
major problems, which we call \emph{provenance
  failures}~\cite{cheney09onward}.  Essentially, a provenance failure
can arise either from failure to \emph{disclose} some key provenance
information to users, or from failure to \emph{obfuscate} some
sensitive provenance information. As an example of failure to disclose
provenance, in 2008 an undated, years-out-of-date story about United
Airlines' 2002 near-bankruptcy was mistakenly put on Google News' main
page, causing investors to panic about its financial stability, which
in turn led to a significant decrease in its share price over the
course of a few hours~\cite{wsj}.  Another example is the
`Climategate' controversy~\cite{climategate}, in which climate
scientists were embarrassed (and widely criticized by climate change
skeptics) when private emails that suggested poor data analysis
practice were leaked. .  As an example of failure to
obfuscate, in 2003 a Word document about British intelligence prior to
the invasion of Iraq was published with supposedly secret
contributors' identities logged in its change
history~\cite{word-sydney}, revealing the influence of political
advisors on the report.  

Obfuscation is obviously closely related to traditional security
concerns, such as confidentiality and anonymity.  Disclosure is, in
our view, also a security property, linked to the traditional security
concern of availability.  In securing provenance, we seek to disclose
some important provenance information while keeping other aspects of
provenance confidential.  If all we cared about was obfuscation, then
security would be easy to achieve by simply not providing any
provenance.  The tension between the two goals of disclosure and
obfuscation makes the analysis of security for provenance a more
challenging problem.

Provenance has predominantly been studied in the context of scientific
computation and databases. A number of forms of provenance have been
proposed for different computational models, including \emph{why} and
\emph{where} provenance~\cite{buneman01icdt},
\emph{how}-provenance~\cite{green07pods}, and \emph{dependency}
provenance~\cite{cheney11mscs} in databases.  In other settings, a
variety of ad hoc techniques have been proposed, largely based on
instrumenting various systems to record a graph diagramming the
procedure calls or dependencies among data and
processes~\cite{moreau10ftws,bose05cs,DBLP:journals/sigmod/SimmhanPG05}.
However, almost all of this work assumes a cooperative setting in
which users are not intentionally trying to subvert or forge
provenance information.  When the accuracy of information used by day
traders, validity and public acceptance of scientific results, and
independence of intelligence reports from political influence depends
on provenance, there is a great deal at stake, so it is important to
develop foundations for correctness and security of provenance in the
face of attacks.

  Although a wide variety of models of provenance have been studied in
  different settings, there has been relatively little progress on
  developing a general understanding of provenance.  By analogy with
  Abadi, Banerjee, Heintze and Riecke's \emph{core calculus of
    dependency}~\cite{abadi99popl}, which elucidated the common ideas
  underlying different techniques such as information flow security,
  program slicing, and binding-time analysis, this article introduces
  a core calculus for provenance: that is, a calculus that illustrates
  and unifies the key ideas underlying a range of provenance
  techniques, including tracing, annotation-propagation, and
  connections to program slicing.  Our main application of this
  framework is to explore the implications of the general definitions
  of disclosure and obfuscation introduced in our prior work, but we
  hope that our approach will also be useful for studying other
  aspects of provenance.

\paragraph{Prior work on provenance and security.}
Despite its apparent importance, there has been relatively little work
on formal foundations of provenance, and work on provenance security
has only begun to appear over the last five years.

Our previous work~\cite{cheney11mscs} appears to have been the first to
explicitly relate information-flow security to a form of provenance,
called \emph{dependency provenance}.  Provenance has been studied in
language-based security by Cirillo
et~al.~\cite{DBLP:conf/esop/CirilloJPR08}, who developed a form of
authorization logic with notions of provenance for understanding
information flow among concurrently executing objects, and by Swamy
et~al.~\cite{DBLP:conf/sp/SwamyCH08,swamy11icfp}, who developed
mechanisms for dependency-provenance tracking in a dependently-typed
secure programming language called Fable.  Both  projects focus on
specifying and enforcing security policies involving provenance
tracking alongside many other concerns, and not on defining provenance
semantics or extraction techniques.  Work on secure
auditing~\cite{jia08icfp,guts09esorics} and expressive programming
languages for security~\cite{swamy11icfp} is also related, but this work focuses on
explicitly manipulating proofs of authorization or evidence about
protocol or program runs rather than automatically deriving or
securing provenance information in its own right.

There is also some work directly addressing security for
provenance~\cite{chong09tapp,hasan09tos,cheney11csf,davidson11pods,blaustein11pvldb,dey11ssdbm}.
Chong~\cite{chong09tapp} gave (to our knowledge) the first candidate
formal definitions of \emph{data security} and \emph{provenance
  security} using a trace semantics, based in part on earlier,
unpublished work of ours on traces and provenance~\cite{traces-tr}.
Hasan et al.~\cite{hasan09tos} study security techniques for ensuring
the integrity of a document that changes over time along with its
provenance records.  Davidson et al.~\cite{davidson11pods} studied a
notion of privacy for provenance in scientific workflows, focusing on
complexity lower bounds.  In their approach, the definition of privacy
essentially says that for unknown components in a workflow (i.e. a
simple dataflow diagram), an attacker should not be able to learn
functional behavior; for example, should not be able to narrow down
the possible output values for any input to less than a parameter $k$.
Cheney~\cite{cheney11csf} gave an abstract framework for provenance,
proposed definitions of properties called \emph{obfuscation} and
\emph{disclosure}, and discussed algorithms and complexity results for
instances of this framework including finite automata, workflows, and
the semiring model of database provenance~\cite{green07pods}.  Zhang
et al.~\cite{zhang09sdm} develop tamper-detection techniques for
provenance in databases. Blaustein et al.~\cite{blaustein11pvldb}
studied the problem of rewriting provenance graphs to hide information
while still satisfying some plausibility constraints.  Dey et
al.~\cite{dey11ssdbm} studied provenance publishing policies, aimed at
giving users greater control over what information is shown and
hidden.  They developed a system called ProPub equipped with
well-defined publishing and hiding operators, along with constraints
that such policies should satisfy.  (In this respect, Dey et al.'s
publishing operators, and work on ``provenance
views''~\cite{liu11tods} can be seen retroactively as addressing
disclosure requirements subject to additional conciseness constraints.)

More recently, Lyle and Martin~\cite{lyle10tapp} gave a detailed
comparative survey of topics in provenance and in security, pointing
out many parallel developments, and Martin et al.~\cite{martin12tapp}
advocate study of provenance considered as a security control.  Some
other topics in security, such as
non-repudiation~\cite{schneider98csfw}, plausible deniability or
differential privacy~\cite{dwork11cacm}, also appear analogous to our
disclosure and obfuscation properties, and this connection could be
explored.

In this article, we build on prior work on provenance security by
studying the disclosure and obfuscation properties of different forms
of provenance in the context of a higher-order, pure, functional
language.  To illustrate what we mean by provenance, we present
examples of programming with three different forms of provenance in
Transparent ML ($\pl$), a prototype implementation of the ideas of
this article.  

To ease exposition, we present the examples in terms of a
  hypothetical ML-like toplevel loop extended with labeled values,
  first-class traces, a type $(\Gamma,\tau)~\mathtt{trace}$ that
  consists of traces returning type $\tau$ evaluated in label context
  $\Gamma$, and with various functions that extract different forms of
  provenance from traces.  The tracing and extraction features are
  formalized later in the article, and our prototype supports these
  examples, as well as the disclosure and obfuscation slicing
  algorithms presented later in the paper.  The \textsf{Slicer} and
  LambdaCalc tools of Perera et al.~\cite{perera12icfp,perera13phd}
  employ similar ideas, and have been run on larger examples, but
  focus on slicing as a debugging and program understanding technique
  and does not yet support provenance extraction or disclosure and
  obfuscation slicing.  Developing a unified and mature implementation
  supporting all of these ideas is left for future work; Perera et
  al.~\cite{perera12icfp,perera13phd} should be consulted for further
  implementation details.

\subsection{Examples}
\label{sec:intro-example}

\paragraph{Where-provenance.}
Where-provenance~\cite{buneman01icdt,buneman08tods} identifies at most
one source location from which a part of the output was copied.  For
example, consider the following \pl session:
\begin{verbatim}
- f [(1,2), (4,3), (5,6)];
val it = [(5,6), (3,4), (1,2)]
\end{verbatim}
Without access to the source code, one can guess that $f$ is doing
something like
\[\mathit{reverse}~\comp~(\mathit{map} ~(\lambda (x,y). \exif{x<y}{(x,y)}{(y,x)}))\]
However, by providing where-provenance information, the system can
explain whether the numbers in the result were copied from the input
or constructed in some other way:
\begin{verbatim}
- trace (f [(1@L1,2@L2),(4@L3,3@L4),(5@L5,6@L6)]);
it = <trace> : ({L1:int,...}, (int*int) list) trace
- where it;
val it = [(5@L5,6), (3@L4,4), (1@L1,2)] 
\end{verbatim}
This shows that $f$ contrives to copy the first elements of the
returned pairs but construct the second components.

\paragraph{Dependency provenance.}
Dependency provenance~\cite{cheney11mscs} is an approach that tracks a
set of all source locations on which a result depends.  For example, 
if we have:
\begin{verbatim}
- g [(1,2,3), (4,5,6)];
val it = [6,6] : int list
\end{verbatim}
we again cannot tell much about what $g$ does.  By tracing and asking
for dependency provenance, we can see:
\begin{verbatim}
- trace (g [(1@L1,2@L2,3@L3),(4@L4,5@L5,6@L6)]);
val it = <trace> : ({L1:int,...}, int list) trace 
- dependency it;
val it = [6@{L1,L2,L3}, 6@{L1,L2,L3}]
\end{verbatim}
This suggests that $g$ is  computing
both elements of the result from the first triple and
returning the result twice, without examining the rest of the list.
We can confirm this as follows:
\begin{verbatim}
- trace (g ((1@L1,2@L2,3@L3)::[]@L));
val it = <trace> : ({L1:int,...}, int list) trace 
- dependency it;
val it = [6@{L1,L2,L3}]
\end{verbatim} 
The fact that $\mathtt{L}$ does not appear in the output confirms that
$g$ does not look further into the list.
While it appears that $g$ may be computing $6$ from $1,2,3$ by
  adding them together, the exact process by which $g$ computes $6$
  from $1,2,3$ is not explicit in the dependency annotations; they are
  also consistent with the hypothesis that $g$ multiplies $1,2,3$
  together to compute 6 or even that $g$ simply examines $1,2,3$ and
  then returns the constant 6.

\paragraph{Expression provenance.}
A third common form of provenance is an expression graph or tree that
shows how a value was computed by primitive operations. For example,
consider:
\begin{verbatim}
- (h 3, h 4, h 5)
val it = (6,24,120);
\end{verbatim}
We might conjecture that $h$ is actually the factorial function.  By
tracing $h$ and extracting expression provenance, we can confirm this
guess (at least for the given inputs):
\begin{verbatim}
- trace (h (4@L));
val it = <trace> : ({L:int}, int) trace
- expression it;
val it = 24@{L * (L-1) * (L-2) * (L-3)  * 1}
\end{verbatim}
In this case, both where-provenance and dependency provenance would be
uninformative since the result is not copied from, and obviously
depends on, the input.

This kind of provenance is used extensively in \emph{workflow} systems
often used in e-science~\cite{hidders07dils}, where the main program
is a high-level process coordinating a number of external (and often
concurrent) program or RPC calls, for example, image-processing steps
or bulk data transformations, which we could model by adding primitive
image-processing operations and types to our language.  Thus, even
though the above examples use fine-grained primitive operations, this
model is also useful for coarse-grained provenance-tracking.

\begin{figure}[p]
  \begin{center}
     \includegraphics[scale=0.3]{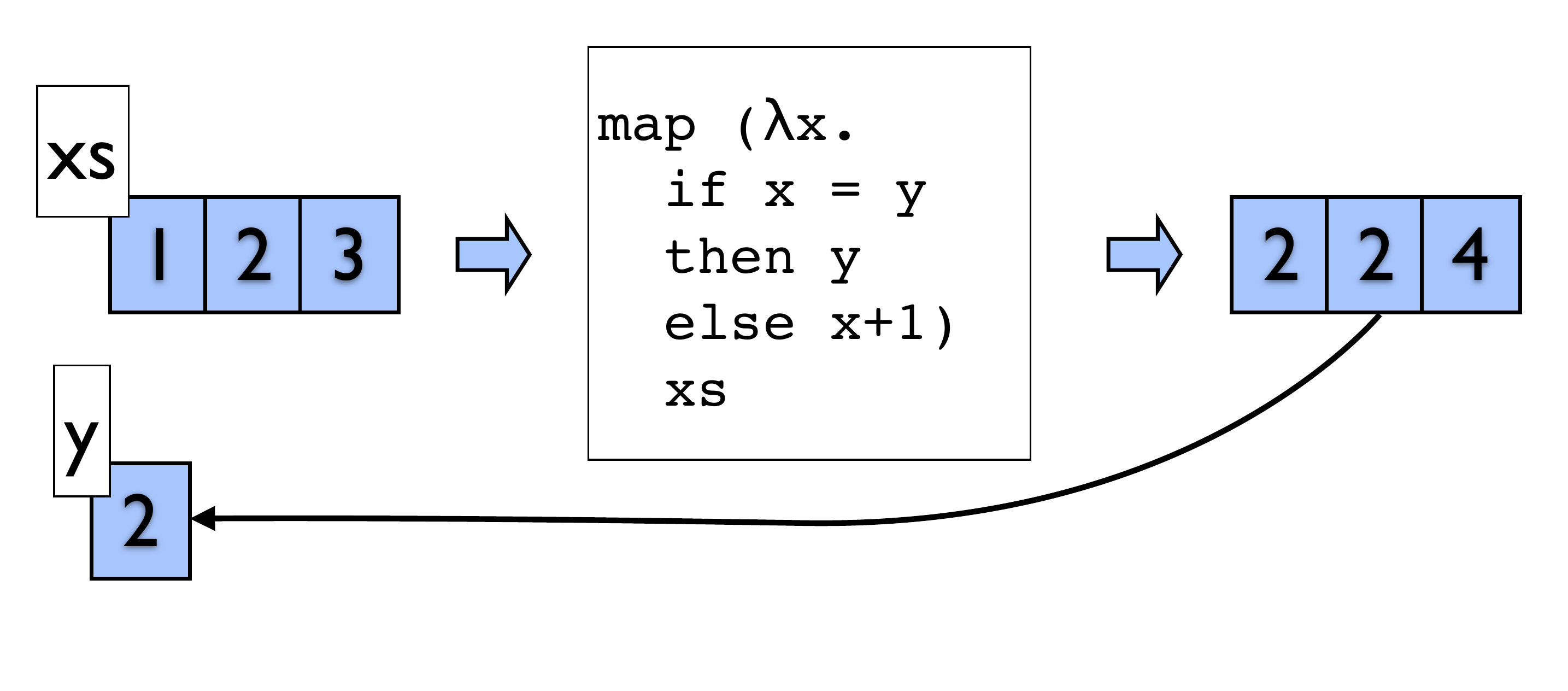}\\
\includegraphics[scale=0.3]{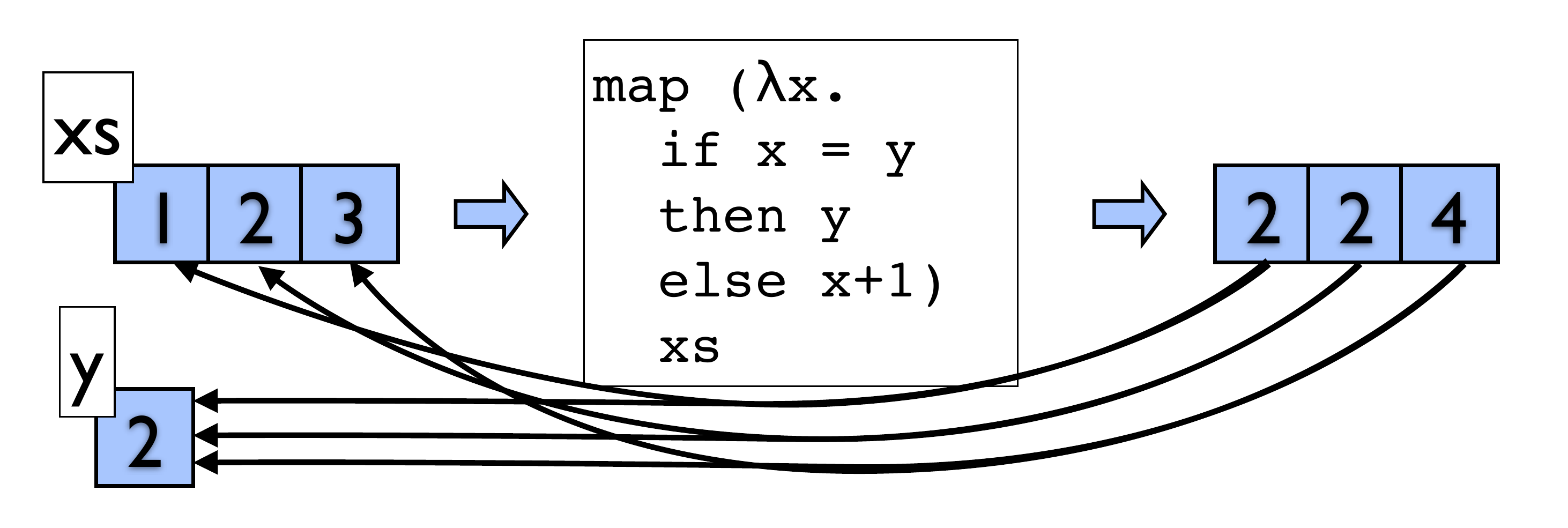}\\
\includegraphics[scale=0.3]{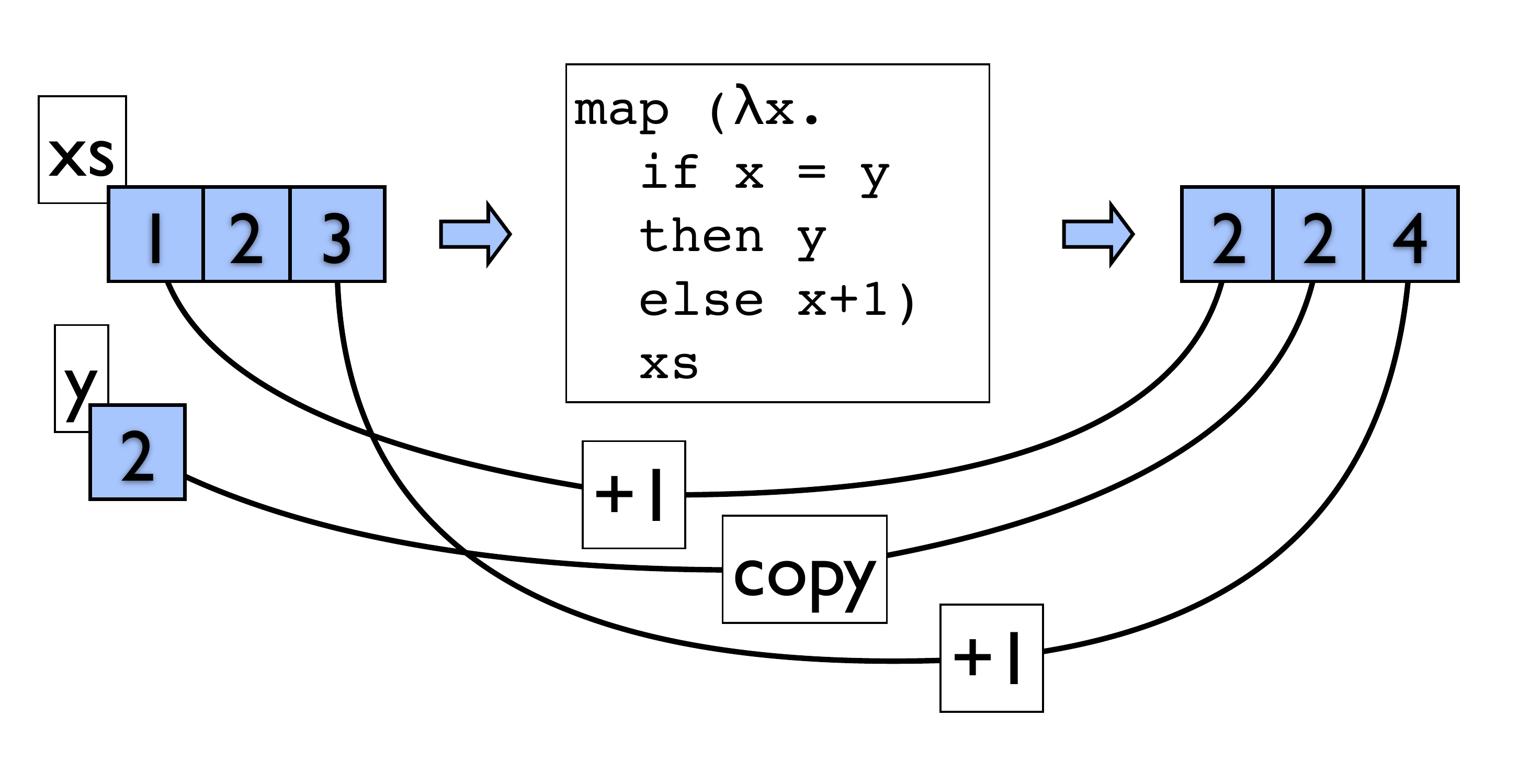}
  \end{center}
  \caption{Illustration of various forms of provenance}
  \label{fig:prov-illustration}
\end{figure}

\paragraph{A running example.}
Figure~\ref{fig:prov-illustration} graphically illustrates these three
forms of provenance on a single example: a simple function mapped over
a list.  This corresponds to the following $\pl$ sessions:
\begin{verbatim}
- val y = 2@L;
- fun f x = if x = y then y else x+1;
- val xs = [1@L1,2@L2,3@L3];
- val t = trace (map f xs);
val t = <trace> : ({L1:int,L2:int,L3:int}, int list) trace
- where t;
val it = [2@{},2@{L},4@{}]
- dependency t;
val it = [2@{L1,L},2@{L2,L},4@{L3,L}];
- expression t;
val it = [2@{L1+1},2@{L},4@{L3+1}];
\end{verbatim}
  Note that this illustrates much of the power of \pl, including
  higher-order, recursive functions and sum, product and recursive
  types. We use this as a running example throughout the paper.

\paragraph{Provenance security.} The three models of provenance above
represent useful forms of provenance that might increase users' trust
or confidence that they understand the results of a program.  However,
if the underlying data, or the structure of the computation, is
sensitive, then making this information available may lead to
inadvertent vulnerabilities, by making it possible for users to infer
sensitive information that they cannot observe directly.  This is a
particular problem if we wish to disclose part of the result of a
program, and provenance that justifies part of the result, while
keeping other parts of the program's execution, input, or output
confidential.

As a simple example, consider a program $\exif{x\neq
  1}{(y,y)}{(z,w)}$.
  Suppose we wish to disclose some information about the computation
  to an untrusted recipient Alice (not necessarily a malicious
  attacker), that none of $x,y$ are visible to Alice, that $z,w$ are
  visible to Alice, and only the value of $x$ is confidential.  If $z$
  and $w$ happen to both equal a common value, say $42$, then the
  result is $(42,42)$.  It is safe to disclose this result to Alice,
  because (without inspecting $x$ or $y$) she cannot be certain
  whether $x = 1$, since there are two scenarios consistent with the
  result: $x=1$ and $y = 42$ or $x\neq 1$ and $y$ arbitrary.
However, any of the above forms of
provenance make it possible to distinguish which branch was taken
because the two different copies of $42$ in $z$ and $w$ will have
different provenance.  Thus, if the provenance information is released
then a principal can infer that the second branch was taken, and
hence, $x = 1$.  In technical terms, we cannot \emph{disclose} any of
the above forms of provenance for the result while \emph{obfuscating}
the fact that $x=1$.

To study these problems systematically, we introduce a single, general
model of provenance that can be instantiated in different ways to
obtain the other models mentioned above (among many others).
Essentially, our approach is to record a detailed trace of evaluation,
whose structure corresponds closely to that of a large-step
operational semantics derivation.  Other forms of provenance can then
be extracted by traversing the trace, and the trace itself can be
viewed as a form of provenance.  Of course, naively recording such a
detailed trace may be prohibitively expensive, and in this article we
are not advocating that such traces be explicitly
constructed in practical systems, only that they are useful as a
formalism for understanding different forms of provenance and their
security properties.

\subsection{Summary}

\paragraph{Contributions.} In this article, we build on, and refine, the
provenance security framework previously introduced by
Cheney~\cite{cheney11csf}.  We introduce a core language with
replayable execution traces for a call-by-value, higher-order
functional language, and make the following technical contributions:
\begin{itemize}
\item Refined definitions of obfuscation and disclosure (Sec. 2).
\item A core calculus defining traced execution for a pure functional
  programming language (Sec. 3).
\item A generic provenance extraction framework that includes several
  previously-studied forms of provenance as instances (Sec. 4).
\item An analysis of disclosure and obfuscation guarantees provided by
  different forms of provenance, including techniques based on slicing
  execution traces (Sec. 5).
\end{itemize}
This article is a revised and expanded version of a conference
paper~\cite{acar12post}.  Compared with the conference paper, this
article includes detailed proofs, a more complete discussion of
related work (including work published recently that was not covered
in the conference paper), and additional examples and discussion of
technical points.  In addition, we encountered a problem with proving
correctness of the disclosure slicing algorithm proposed
in~\cite{acar12post}; specifically, Lemma 2 in the conference
version had a subtle problem, which we avoid through a reformulation
of the disclosure slicing algorithm.

  This article is also closely related to work on using traces for
  program slicing~\cite{perera12icfp} published in ICFP 2012.  The two
  papers present different aspects of a single research project; the
  trace model and some aspects of the slicing algorithms are closely
  related.  However, the two papers make distinct contributions, and
  the system in the ICFP paper incorporates simplifications that are
  appropriate pragmatic choices for its application area (program
  slicing) but not appropriate for security analysis.  Because of
  these differences we have chosen not to attempt to develop a unified
  presentation or implementation, to ensure that focus in this article
  remains on provenance security.  We summarize the key differences
  below.
  \begin{itemize}
  \item In the ICFP paper, traces and slicing are defined in terms of
    an ad hoc semantics over partial values, and justified by a Galois
    connection between them.  Here, we instead define slices for
    disclosure and obfuscation in terms of a standard operational
    semantics.  One important consequence of these different choices
    is that unique minimal disclosure slices do not exist, whereas
    unique minimal backwards slices do exist in the ICFP paper.
\item The slicing algorithms in the ICFP paper differ from those given
  here in certain technical details: specifically, we use value
  patterns and the $\eqat{p}$ equivalence relation instead of partial values, and we use
  $\vany$-patterns to support slicing of primitive operations instead
  of tagging primitive operations with the values of their inputs in traces.
\item The ICFP paper did not present provenance extraction or explore
 the connection to provenance security that is the focus of this paper.
\item The ICFP paper presents work on program slicing, differential
  slicing, and implementation techniques, topics that are beyond the
  scope of this article.
  \end{itemize}

\paragraph{Outline.} Section~\ref{sec:background} briefly
recapitulates the framework introduced by Cheney~\cite{cheney11csf}
and refines some definitions.  We present the (standard) syntax and
tracing semantics of \pl in Section~\ref{sec:language}.  In
Section~\ref{sec:extraction} we introduce a framework for querying and
extracting provenance views from traces, including the three models
discussed above.  Section~\ref{sec:analysis} presents our main results
about disclosure, obfuscation, and trace
slicing. Section~\ref{sec:related} presents related work and
Section~\ref{sec:concl} concludes.

%%% Local Variables: 
%%% mode: latex
%%% TeX-master: "main"
%%% End: 

% LocalWords:  workflow Executability OPM sublanguages subtraces impl concl hoc

% LocalWords:  MapReduce satisfiability subvalues subderivations tuple subtrees
% LocalWords:  unevaluate subtree subvalue Unevaluation executability hingest
% LocalWords:  ahd online curation metadata postprocessed multisets dataset RPC
% LocalWords:  nondeterminstic nondeterministic filesystem workflows reify et
% LocalWords:  recomputation postprocess subproblem dependences auditable al
% LocalWords:  Cirillo Swamy Chong automata semiring replayable Climategate Dey
% LocalWords:  advisors Hasan dataflow Zhang Blaustein ProPub
% LocalWords:  deniability

\section{Background}\label{sec:background}

We recapitulate the main components of the provenance security
framework of Cheney~\cite{cheney11csf}.  The framework assumes a given
set of \emph{abstract traces} $\Traces$, together with a collection $\Queries$
of possible \emph{trace queries} $Q : \Traces \to \Bool$,
where $\Bool = \{0,1\}$ is the set of Boolean truth values.  These
represent properties of traces that the system designer may want to
protect or that legitimate users or attackers of the system may want
to learn.  In the previous paper, we considered refinements to take
into account the knowledge of the principals about the possible system
behaviors.  In this article, we consider a single principal and assume
that all traces $\Traces$ are considered possible,
for simplicity.

Fix a set $\Omega$ of the possible \emph{provenance views}, and a
function $P : \Traces \to \Omega$ mapping each trace to a provenance
view of the trace.  We may write $(\Omega,P:\Traces \to \Omega)$ or
just $(\Omega,P)$ for a provenance view.  Also, we typically write $q
: \Omega \to \Bool$ for a provenance query, that is, a query on a
provenance view.

Given this framework, we proposed the following definitions:
\begin{definition}
  [Disclosure]
  A query $Q : \Traces \to \Bool$ is \emph{disclosed} by a provenance view $(\Omega,P)$ if for every $t,t'\in \Traces$, if $P(t) = P(t')$ then $Q(t) = Q(t')$.  
\end{definition}
In other words, disclosure means that there can be no traces $t,t'$
that have the same provenance view but where one satisfies the query
and the other does not.

\begin{definition}
  [Obfuscation]
  A query $Q : \Traces \to \Bool$ is \emph{obfuscated} by a provenance view $(\Omega,P)$ if for every $t$ in $\Traces$, there exists $t' \in \Traces$ such that $P(t) = P(t')$ and $Q(t) \neq  Q(t')$.
\end{definition}
Thus, obfuscation is not exactly the opposite of disclosure; instead,
it means that for every trace there is another trace with the same
provenance view but different $Q$-value.  This means that a principal
that has access to the provenance view but not the trace cannot be
certain whether or not $Q$ is satisfied by the underlying
trace.

In the previous paper, we gave several examples of instances of this
framework.  Here, for illustration, we just review one such instance,
given by regular languages and finite automata.
\begin{example}[Strings as traces]
  % The set of traces $\Traces_M$ of an automaton $M =
  % (\Sigma,Q,q_0,\delta,F)$ is the set $Q (\Sigma Q)^*$ of alternating
  % sequences of states and alphabet letters.  The queries are simply
  % regular subsets of $\Traces_M$.  The provenance views are given by
  % finite-state transducers.  
  Consider the regular sublanguage of $\{a,b\}^+$
    consisting of nonempty
  strings; these can be viewed as traces of an
  automaton or other sequential process.  Some views of the traces of
  the automaton include a transducer $T_1$ replacing each symbol with
  $a$, a transducer $T_2$ that deletes all of the $a$s, and a
  transducer $T_3$ that deletes alternating symbols.  A query over
  these traces can test whether the number of $b$s is even; this is
  obfuscated by $T_1$, disclosed by $T_2$, and neither
  fully obfuscated nor fully disclosed by
  $T_3$.
\end{example}
When finite automata are used for queries and transducers for
provenance views, we showed that disclosure is decidable for all
queries and views and that obfuscation is decidable for all queries
and views whose range is finite.  It is unknown whether obfuscation is
decidable in the general case.

The definitions above turn out to be too strong for our purposes; in
this paper we will also consider some weaker versions of disclosure
and obfuscation.
  \begin{definition}
    A query $Q : \Traces \to \Bool $ is \emph{positively disclosed} by
    provenance view $(\Omega,P)$ via query $q : \Omega \to \Bool$ if
    for every $t$, if $q(P(t))= 1$ then $Q(t) = 1$.

    A query $Q : \Traces \to \Bool $ is \emph{negatively disclosed} by
    provenance view $(\Omega,P)$ via query $q : \Omega \to \Bool$ if
    for every $t$, if $q(P(t)) = 0$ then $Q(t) = 0$.
  \end{definition}

In other words, positive disclosure means that there is a query $q$
on the provenance that safely overapproximates $Q$ on the underlying
trace.  If $q(P(t))$ holds then we know $Q(t)$ holds but otherwise we
may not learn anything about $t$.  Dually, negative disclosure means
that if $q(P(t))$ is false then we know $Q(t)$ is also false, but
otherwise learn nothing.

\begin{example}
  Suppose $\Traces = \Omega = \{a,b\}^*$.  Define query $Q(t)$ to be true if
  and only if
  $t$ is not of the form $u\cdot abab\cdot v$ for strings $u,v$, and
  let $P(a_1a_2\cdots a_n) = a_2a_4\cdots a_{\lfloor \frac{n}{2}\rfloor}$, the
  function that deletes alternate letters of its argument.  Finally,
  let $q(t)$ be a query on $\Omega$ that is true if and only if $t$
  has no substrings of the form $aa$ or $bb$.  Then $Q$ is positively
  disclosed by $P$ via $q$, for if $P(t)$ has no $aa$ or $bb$
  substring, then $t$ can have no $abab$ substring.  However, $Q$ is
  not negatively disclosed by $P$ (for any $q'$), because, for
  example, $P(abab) = bb = P(bbbb)$.
\end{example}

  \begin{definition}
    A query $Q : \Traces \to \Bool$ is \emph{positively obfuscated} by
    $(\Omega,P)$ if for every $t$ satisfying $Q(t) = 1$
    there exists a trace $t'$ such that $Q(t') = 0$ and
    $P(t) = P(t')$.

    A query $Q : \Traces \to \Bool$ is \emph{negatively obfuscated} by
    $(\Omega,P)$ if for every $t$ satisfying $Q(t) = 0$ there exists a
    trace $t'$ such that $Q(t') = 1$ and $P(t) = P(t')$.
  \end{definition}

In other words, positive obfuscation means that the provenance never
reveals that $Q$ holds of the trace, but it may reveal that $Q$ fails.
This weaker notion is useful for asserting that sensitive data is
protected: if the sensitive data is not present in the trace then it
is harmless to reveal this, but if the sensitive data is present then
the provenance should hide enough information to make its presence
uncertain.  Dually, negative obfuscation means that the
  provenance view does not reveal $\neg Q$.

  \begin{example}
    Again suppose $\Traces =\Omega= \{a,b\}^*$.  Define $Q(t)$ to be
    true if and only if the number of $a$ symbols in $t$ is odd and
    false otherwise.  Define $P(t)$ to be $t$ with all $a$s replaced
    by $b$s.  (In other words, $P(t) = b^{|t|}$, a string of $b$s of
    the same length as $t$.)  $P$ positively obfuscates $Q$ because if
    $Q(t)$ holds, then $t$ is a nonempty string with an odd number of
    $a$s, and we can form $t'$ such that $P(t) = P(t')$ by replacing
    one of the $a$s of $t$ with a $b$.  However, $P$ does not
    negatively obfuscate $Q$ because $P(\epsilon) = \epsilon$ and
    there is no other string $t'$ with $Q(t') = true$ and $P(t') =
    \epsilon$.
  \end{example}

\begin{proposition}\label{prop:pos-neg}
  If $P$ both positively discloses and negatively discloses $Q$ via
  $q$, then $P$ discloses $Q$. Similarly, if $P$ both positively and
  negatively obfuscates $Q$ then $P$ obfuscates $Q$.
\end{proposition}
\begin{proof}
  For the first part, suppose $P$ discloses $Q$ positively and
  negatively via $q$.  Let $t,t' \in \Traces$ be given where $P(t) =
  P(t')$.  If $q(P(t))$ holds then $q(P(t'))$ holds so $Q(t) = 1 =
  Q(t')$.  If $q(P(t)) = 0$ then $q(P(t')) = 0$ and so $Q(t) = 0 =
  Q(t')$.  

  The argument for obfuscation is straightforward.
\end{proof}

  \begin{remark}
    It may seem surprising that positive and negative disclosure
    specify a provenance query $q$ while full disclosure does not
    specify such a parameter.  If full disclosure holds, then there is
    no need to mention the provenance query $q$ that answers trace
    queries over $\Omega$, since it is the characteristic function of
    $\{P(t) \mid Q(t) = 1\}$.  However, if we leave out (or
    existentially quantify over) the provenance query $q$ in positive
    or negative disclosure, then both definitions become trivial,
    since positive disclosure always holds for $q(x) = 0$ and negative
    disclosure always holds for $q(x) = 1$. Moreover, we want to be
    able to decompose proving full disclosure into proving positive
    and negative disclosure, but the argument given above requires
    that the positive and negative disclosure hold with respect to the
    same $q$.
  \end{remark}

We now proceed to instantiate the framework with traces generated by a
richer language, with corresponding notions of trace query and
provenance view.

%%% Local Variables: 
%%% mode: latex
%%% TeX-master: "main"
%%% End: 

% LocalWords:  overapproximates Dually automata

\begin{figure}[t]
\begin{syntaxfig}
\mbox{Types}
&
\tau
& 
::=
& 
b
\mid
% \kwbool
% \mid
\tau_1 \times\tau_2
\mid
\tau_1 + \tau_2
\mid
\tau_1 \ra \tau_2
\mid 
\tyrec{\alpha}{\tau}
\mid 
\alpha
\\[1mm]
\mbox{Type contexts}
&
\Gamma 
&
::= &
[x_1:\tau_1 , \ldots , x_n:\tau_n]
\\[1mm]

\mbox{Code pointers}
  &
  \kappa
  & 
  ::= 
  &
  \fn{f}{x}{e}
  \\[1mm]

\mbox{Matches}
  &
  m
  & 
  ::= 
  &
  \match{x_1}{e_1}{x_2}{e_2}
  \\[1mm]

 \mbox{Values}
  &
  v
  & 
  ::=
  & 
  \vc
  \mid
  \vpair{v_1}{v_2}
  \mid 
  \vinl{v}
  \mid 
  \vinr{v}
  \mid
  \vclos{\kappa}{\gamma}
  \mid 
  \vroll{v}
\\[1mm]
\mbox{Environments}
&
\gamma 
& 
::= 
& 
[x_1 \mapsto v_1,\ldots,x_n \mapsto v_n]
 \\[1mm]
  \mbox{Expressions}
  & 
  e
  & 
  ::=
  & 
   \exc 
  \mid
  x
  \mid  
\exf{\novec{e}}
 \mid
  \exletin{x=e_1}{e_2} 
\\[1mm]
&&\mid
&  \expair{e_1}{e_2}
  \mid
  \exfst{e}
  \mid
  \exsnd{e}
\\[1mm]
&&  \mid
&  \exinl{e}
  \mid
  \exinr{e}
\mid
  \excasem{e}{m}
 \\[1mm]
&&\mid
&
\exfunk{\kappa} 
  \mid
  \exapp{e}{e'}
\\[1mm]
&&  \mid& 
  \exroll{e}
  \mid
  \exunroll{e}
  \\[2mm]

\mbox{Traces}
  & 
  T
  & 
  ::=
  & 
  \exc \mid x \mid \exf{\novec{T}} \mid \exletin{x=T_1}{T_2}
    \\[1mm]
    &&\mid & \expair{T_1}{T_2} \mid \exfst{T} \mid \exsnd{T}
\\[1mm]
&&\mid&  \exinl{T} \mid \exinr{T} 
\mid \trcaseml{m}{T}{x}{T_1}
\mid \trcasemr{m}{T}{x}{T_2}
\\[1mm]
&&\mid& \exfunk{\kappa}
\mid
\trappk{\kappa,\Gamma}{T_1}{T_2}{f}{x}{T} 
    \\[1mm]
    && \mid& \exroll{T} \mid \exunroll{T} 
\end{syntaxfig}
%\hrule
\caption{Abstract syntax of $\plz$.}
\label{fig:syn}
\end{figure}

%%% Local Variables: 
%%% mode: latex
%%% TeX-master: "main"
%%% End: 

\section{Core Language}
\label{sec:pl}
\label{sec:language}

We will develop a core language for provenance based on a standard,
typed, call-by-value, pure language, called Transparent ML, or \pl.
For the purpose of this article, we focus on terminating runs of pure
computations.  We only consider terminating runs since otherwise there is
no trace or end result to analyze; the question of how to deal with
provenance in nonterminating or effectful programs is interesting, but
left for future work.  Allowing for effects or moving to a small-step
semantics each seem likely to complicate the trace semantics (and subsequent
analysis) considerably.

The syntax of \pl types, expressions, and other syntactic
classes is shown in Figure~\ref{fig:syn}.  The syntax of expressions
and values is standard, following common textbook treatments of
languages with binary pairs, binary sums, recursive types, and
recursive functions~\cite{pierce02types}; constructs such as boolean
conditionals, records,
datatypes, or mutually recursive functions can be added without
difficulty following the same pattern.  We parameterize the syntax and
semantics over primitive operations $\oplus$ that take inputs of base
type only; for example, equality on integers, arithmetic and boolean
operations. In $\fn{f}{x}{e}$, both $f$ and
$x$ are variable names; $f$ is the name of the recursively defined
function while $x$ is the name of the argument.  Both $f$ and $x$ are
bound in $e$ in an expression of the form $\fn{f}{x}{e}$; generally,
we adhere to the convention that in an expression of the form $x.e$,
variable $x$ is bound in $e$.

We abbreviate functional terms of the form $\fn{f}{x}{e}$ using the
letter $\kappa$, when convenient; similarly, we often abbreviate the
expression $\{\vinl{x_1}.e_1;\vinr{x_2}.e_2\}$ as $m$.  We sometimes
refer to $\kappa$ or $m$ as a \emph{code pointer} or \emph{match
  pointer} respectively; in a fixed program, there are a fixed finite
number of such terms and so we can share them instead of explicitly
copying them when used in traces.

  The syntax of traces is also defined in Figure~\ref{fig:syn}.
  Trace expressions $T$ have many syntactic forms in common with
  expressions; they differ primarily in the case and application trace
  forms, which include additional information showing how an
  application or case expression was evaluated.  Traces can be viewed
  as witnessing terms for the operational derivation of an expression,
  and so their meaning is explained below along with that of
  the operational semantics rules.  We will refer to trace expressions
  $T$ as \pl-traces when necessary to distinguish them from abstract
  traces $\Traces$ introduced in the previous section.

\subsection{Dynamic Semantics}
\label{sec:pl::dynamic}
\label{sec:pl::dynamic-semantics}

\begin{figure}[tb]
\fbox{$\gamma,e \red v,T$}
\vspace{-4mm}
\begin{smathpar}
     \inferrule*
    {\strut}{\gamma, \exc \red \exc,\trc}
\and
   \inferrule*
    {
      \strut
    }
    {
      \gamma, x \red \gamma(x), \trvar{x}
    }
    \and    \inferrule*
    {
     \gamma,  \novec{e} \red \novec{v},\novec{T}
    }
    {
      \gamma, \kwf(\novec{e}) \red \hat{\oplus}(\novec{v}), \trf{\novec{T}}
    }
\and
    \inferrule*
    {
      \gamma, e_1 \red v_1, T_1
      \\
      \gamma[x\mapsto v_1], e_2 \red v_2, T_2
   }
    {
      \gamma,\exletin{x  = e_1}{e_2} \red v_2, \trlet{T_1}{x}{T_2}
    }
\\
    \inferrule*
    {
      \gamma,  e_1 \red v_1, T_1
      \\
      \gamma,  e_2 \red v_2, T_2
    }
    {
     \gamma,  \expair{e_1}{e_2} \red \expair{v_1}{v_2}, \trpair{T_1}{T_2}
    }
    \and
    \inferrule*
    {
\gamma,  e \red \expair{v_1}{v_2}, T
}
    {
\gamma, \exfst{e} \red v_1, \trfst{T}
}
\and
    \inferrule*
    {
      \gamma,  e \red \expair{v_1}{v_2}, T
    }
    {
      \gamma, \exsnd{e} \red v_2, \trsnd{T}
    }
\\
    \inferrule*
    {
      \gamma , e \red v, T
    }
    {
      \gamma , \exinl{e} \red \vinl{v}, \trinl{T} 
    }
\and \inferrule*
    {
     (\exinl{x_1}.e_1 \in m)\\
      \gamma,e \red \vinl{v}, T
      \\
      \gamma[x_1\mapsto v], e_1 \red v_1, T_1
   }
    {
      \strut
      \gamma,\excasem{e}{m} \red v_1, \trcaseml{m}{T}{x_1}{T_1}
    }
\and
    \inferrule*
    {
      \gamma , e \red v, T
    }
    {
      \gamma , \exinr{e} \red \vinr{v}, \trinr{T} 
    }\and
    \inferrule*
    {
      (\exinr{x_2}.e_2 \in m)\\
      \gamma,e \red \vinr{v}, T
      \\
      \gamma[x_2\mapsto v], e_2 \red v_2, T_2
    }
    {
      \strut
      \gamma,\excasem{e}{m} \red v_2, \trcasemr{m}{T}{x_2}{T_2}
    }
\\
    \inferrule*
    {
      \gamma , e \red v, T
    }
    {
      \gamma , \exroll{e} \red \vroll{v}, \trroll{T} 
    }
    \and
    \inferrule*
    {
      \gamma , e \red \vroll{v}, T
    }
    {
      \gamma , \exunroll{e} \red v, \trunroll{T} 
    }
\\
\inferrule*
    {
      \strut      
    }
    {
     \gamma,  \exfunk{\kappa} \red \vclos{\kappa}{\gamma},\trfunk{\kappa}
    }
\and
\inferrule*
    {
      \gamma,e_1 \red \vclos{\kappa}{\gamma'},T_1
      \\
      (\kappa = \fn{f}{x}{e})
      \\
      \gamma,e_2 \red v_2,T_2
      \\
      \gamma'[f \mapsto \vclos{\kappa}{\gamma'},x \mapsto v_2], e \red v, T
   }
    {
      \gamma,\exapp{e_1}{e_2} \red v, \trappk{\kappa}{T_1}{T_2}{f}{x}{T}
    }
 \end{smathpar}
%\hrule
\caption{Dynamic semantics of $\plz$: rules for expression evaluation.}
\label{fig:dynamic}
\end{figure}
%%% Local Variables: 
%%% mode: latex
%%% TeX-master: "main"
%%% End: 

We augment a standard large-step operational semantics for \pl by
adding a parameter $T$, which records a trace of the evaluation of the
expression.
The judgment $\gamma,e \eval v,T$, defined in
Figure~\ref{fig:dynamic}, says that in environment $\gamma$,
expression $e$ evaluates to value $v$ with trace $T$.  

  If we ignore the trace parameter in this judgment, then the rules
  are essentially the standard ones for a call-by-value, pure
  functional language with pairs, sums, and recursive types and
  functions~\cite{pierce02types}.  In particular, pairs are
  constructed by pairing and can be taken apart using the $\kwfst$ and
  $\kwsnd$ operations.  Values of sum type are constructed using the left and right
  injection operations $\kwinl,\kwinr$ and can be analyzed using the
  case expression, which examines a value of type $\tau_1 + \tau_2$
  and calls the appropriate branch with the injected value bound to
  a variable.  Values of recursive type $\mu \alpha.\tau$ are constructed using
  $\kwroll$ and destructed using $\kwunroll$; these operations
  indicate the explicit isomorphisms in the \emph{isorecursive} treatment of
  recursive types.  Finally, functions are (as usual) constructed
  using the function expression $\exfun{f}{x}{e}$ and applied using
  function application $e_1~e_2$.

  Now if we consider the trace parameter, note that each rule has its
  own trace form, which builds the trace up from sub-traces obtained
  by the hypotheses of the rule.
Traces can contain bound variables, reflecting the binding structure
of the original expression.  To illustrate, for let expressions,
traces are similar to expressions:
\[
\trlet{T_1}{x}{T_2}
\]
where we bind the variable in $T_2$.  

The case and application evaluation traces record additional
information about control flow.  In either case, the first argument is
evaluated to determine what expression to evaluate to obtain the final
result.
For case expressions, traces are of the
form:
\[\trcaseml{m}{T}{x_1}{T_1} \quad \trcasemr{m}{T}{x_2}{T_2}\]
where we record the
trace of the case scrutinee $T$ and the taken branch ($T_1$ or
$T_2$), and we re-bind the variable ($x_1$ or $x_2$) in the trace of the
taken branch.  The subscript indicates which branch was taken.
Similarly, for an application expression:
\[
\trappk{\kappa}{T_1}{T_2}{f}{x}{T}
\]
we record the traces of the function subexpression $T_1$, the argument
subexpression $T_2$, and the trace $T$ of the evaluation of the body
of the function.  The subscript $\kappa = \vclos{\fn{f}{x}{e}}{\Gamma}$ is a code pointer
indicating the function and the typing environment of the call.  The
$\Gamma$ annotation is needed only to typecheck traces, so we usually
elide it.  Again, since the body trace can mention the function and
argument names as free variables, we re-bind these variables.

We want to emphasize at this point that we do not necessarily expect
that implementations routinely construct fully detailed traces along
the above lines.  Rather, the trace semantics is proposed here as a
candidate for the most detailed form of provenance we will consider.
Recording and compressing or filtering relevant information from
traces in an efficient way is beyond the scope of this paper.
However, some preliminary experiments in this direction have been
performed in a recent paper on slicing for higher-order functional
programs, based on a similar trace
model~\cite{perera12icfp}. 

\begin{example}
Consider the factorial program expressed in $\plz$:
\begin{verbatim}
let f = fun f(x). if x = 0 then 1 else x*(f(x-1))
in f 4
\end{verbatim}
The trace of this program has the form
\begin{verbatim}
let f = fun f(x). e
in f 4    |> f(x).(e |>_else x * (
    f(x-1) |> f(x).(e |>_else x * (
     f(x-1) |> f(x).(e |>_else x * (
      f(x-1) |> f(x).(e |>_else x * (
       f(x-1) |> f(x).(e |>_then 1)))))))))
\end{verbatim}
where \verb|e = if x = 0 then 1 else x*(f(x-1))|.   The trace reflects
that $f$ calls itself four additional times when evaluating $f~4$ and
the $\kw{else}$-branch is taken four times, and finally the $\kw{then}$-branch
is taken.  Here, we use subscripts $\kw{then}$ and $\kw{else}$ to indicate the
branch taken instead of $\kw{inl}$ and $\kw{inr}$.  
\end{example}

  \begin{remark}
    The syntax of traces and expressions, and their corresponding
    evaluation rules, exhibit some redundancy.  The syntax and semantics of
    expressions and traces could be fused so that both 
    fall out as subsystems of one joint syntax / semantics.  We adopt
    an explicit treatment for clarity, despite the resulting redundancy.

    The operational semantics rules in Figures~\ref{fig:dynamic} and
    \ref{fig:replay} illustrate a recipe that appears straightforward
    to follow in order to extend the system to a more realistic (pure)
    language; it is less clear how to extend the trace semantics to
    handle effects, nontermination, or other features.  It may be
    interesting to try to capture the recipe as a formal construction
    over operational semantic specifications.
  \end{remark}

\paragraph{Trace Replay.}

\begin{figure}[tb!]
\fbox{$\gamma,T \trrun v$}
\vspace{-1mm}
\begin{smathpar}\
\and
\inferrule*
{\strut}
{\gamma, \trc \trrun \exc%, \trc
}\and
\inferrule*
{
  \strut
}
{
  \gamma, \trvar{x} \trrun \gamma(x)%, \trvar{x}
}
\and
\inferrule*
{
  \gamma, \novec{T} \trrun \novec{v}%, \novec{T'}
}
{
  \gamma, \trf{\novec{T}}
  \trrun 
  \hat{\oplus}(\novec{v})%, \trf{\novec{T'}}
}\and
\inferrule*
{
  \gamma,T_1 \trrun v_1%,T_1'
\\
\gamma[x\mapsto v_1],T_2 \trrun v_2%,T_2'
}
{
\gamma,\trlet{T_1}{x}{T_2} \trrun v_2%,\trlet{T_1}{x}{T_2}
}
\\
\inferrule*
{
 \gamma, T_1 \trrun v_{1}%, T_1'
\\
 \gamma, T_2 \trrun v_{2}%, T_2'
\\
}
{
  \gamma,  \trpair{T_1}{T_2}
  \trrun
  \vpair{v_{1}}{v_{2}}%, \trpair{T_1'}{T_2'}
}
\and
\inferrule*
{
\gamma, T \trrun \vpair{v_{1}}{v_{2}}%, T'
}
{
\gamma , \trfst{T} 
\trrun
v_{1}% , \trfst{T'}
}
\and
\inferrule*
{
\gamma, T \trrun \vpair{v_{1}}{v_{2}}%, T'
}
{
\gamma, \trsnd{T}
\trrun
v_2%, \trsnd{T'}
}
\\
\inferrule*
{
  \gamma, T \trrun v%, T'
}
{
\gamma, \trinl{T} \trrun \vinl{v}%, \trinl{T'}
}
\and
\inferrule*
{
\gamma, T \trrun \vinl{v}%, T'
\\
\gamma[x_1 \mapsto v], T_1 \trrun v_1%, T_1'
}
{
\gamma, \trcaseml{m}{T}{x_1}{T_1}
\trrun
v_1%, \trcaseml{m}{T'}{x_1}{T_1'}
}
\\
\inferrule*
{
  \gamma, T \trrun v%, T'
}
{
\gamma, \trinr{T} \trrun \vinr{v}%, \trinr{T'}
}
\and 
\inferrule*
{
\gamma, T \trrun \vinr{v}%, T'
\\
\gamma[x_2 \mapsto v], T_2 \trrun v_2%, T_2'
}
{
\gamma, \trcasemr{m}{T}{x_2}{T_2}
\trrun
v_2%, \trcasemr{m}{T'}{x_2}{T_2'}
}
\\
\inferrule*
{
  \gamma, T \trrun v%, T'
}
{
\gamma, \trroll{T} \trrun \vroll{v}%, \trroll{T'}
}
\and
\inferrule*
{
  \gamma, T \trrun \vroll{v}%, T'
}
{
\gamma, \trunroll{T} \trrun {v}%, \trunroll{T'}
}
\\
\inferrule*
{
  \strut
}
{
  \gamma,\trfunk{\kappa}
  \trrun 
  \vclos{\kappa}{\gamma}%, \trfunk{\kappa}
}
\and
\inferrule*
{
\gamma,T_1 \trrun \vclos{\kappa}{\gamma'}%,T_1'
\\
\gamma,T_2 \trrun v_2%,T_2'
\\
\gamma'[f\mapsto\vclos{\kappa}{\gamma'}, x\mapsto v_2],T \trrun v%,T'
}
{
\gamma,\trappk{\kappa}{T_1}{T_2}{f}{x}{T} \trrun v%,\trappk{\kappa}{T_1'}{T_2'}{f}{x}{T'}
}
% \and
%  \inferrule*
% {
% \gamma, T \trrun \vinl{v}%, T'
% \\
% \gamma[x_1 \mapsto v], T_1 \trrun v_1%, T_1'
% }
% {
% \gamma, \trcaseml{m}{T}{x_1}{T_1}
% \trrun
% v_1%, \trcaseml{m}{T'}{x_1}{T_1'}
% }
% \and 
% \inferrule*
% {
% \gamma, T \trrun \vinr{v}%, T'
% \\
% \gamma[x_2 \mapsto v], T_2 \trrun v_2%, T_2'
% }
% {
% \gamma, \trcasemr{m}{T}{x_2}{T_2}
% \trrun
% v_2%, \trcasemr{m}{T'}{x_2}{T_2'}
% }
% \and
% \inferrule*
% {
% \gamma,T_1 \trrun \vclos{\kappa}{\gamma'}%,T_1'
% \\
% \gamma,T_2 \trrun v_2%,T_2'
% \\
% \gamma'[f\mapsto\vclos{\kappa}{\gamma'}, x\mapsto v_2],T \trrun v%,T'
% }
% {
% \gamma,\trappk{\kappa}{T_1}{T_2}{f}{x}{T} \trrun v%,\trappk{\kappa}{T_1'}{T_2'}{f}{x}{T'}
% }
% \and
% \inferrule*[Right=($\dagger$1)]
% {
% \gamma, T \eval \vinr{v}, T'
% \\
% \kwinr(x_2).e_2 \in m
% \\
% \gamma[x_2 \mapsto v], e_2 \red v_2, T_2'
% }
% {
% \gamma, \trcaseml{m}{T}{x_1}{T_1}
% \eval
% v_2, \trcasemr{m}{T'}{x_2}{T_2'}
% }
% \and
% %
% \inferrule*[Right=($\dagger$2)]
% {
% \gamma, T \eval \vinl{v}, T'
% \\
% \kwinl(x_1).e_1 \in m
% \\
% \gamma[x_1 \mapsto v], e_1 \red v_1, T_1'
% }
% {
% \gamma, \trcasemr{m}{T}{x_2}{T_2}
% \eval
% v_1, \trcaseml{m}{T'}{x_1}{T_1'}
% }
% \and
% \inferrule*[Right=($\dagger$3)]
% {
% T_1 \eval \vclos{\kappa'}{\gamma'},T_1'
% \\
% (\kappa' = \exfun{f}{x}{e'} \neq \kappa)
% \\
% T_2 \eval v_2,T_2'
% \\
% \gamma'[f\mapsto \vclos{\kappa'}{\gamma'}, x\mapsto v_2] , e' \red v', T_3'
% }
% {
% \gamma, \trappk{\kappa}{T_1}{T_2}{f}{x}{T_3} \eval v', \trappk{\kappa'}{T_1'}{T_2'}{f}{x}{T_3'}
% }
% \and
% \inferrule*
% {\strut}
% {\gamma, \tremp \eval \vhole, \tremp}
\end{smathpar}
%\hrule
\caption{Dynamic semantics of $\plz$: rules for trace replay.}
\label{fig:replay}
\end{figure}
%%% Local Variables: 
%%% mode: latex
%%% TeX-master: "main"
%%% End: 

%

We equip traces with a semantics that relates them to expressions.  We
write $\gamma,T \trrun v$ for the \emph{replay} relation that
reruns a trace on an environment (possibly different from the one
originally used to construct $T$).  Figure~\ref{fig:replay} shows the rules for
replaying traces. The rules for most trace forms are
the same as the standard rules for evaluating the corresponding
expression forms.   Essentially, these rules
require that the same control flow branches are taken as in the
original run.  If the input environment is different enough that the
same branches cannot be taken, then replay fails.

\begin{remark}
  This behavior should be contrasted with traces used in
  self-adjusting computation~\cite{acar06toplas,Acar09}.  Such traces
  must always recompute the updated result; however, they typically
  operate at a coarser granularity by tracking reads and writes to
  memory locations.  Moreover, the traces are essentially graphs built
  in memory using references and closures, so it is not
  straightforward to traverse such traces to obtain fine-grained
  information about what happened at run-time, as we shall do in
  Section~\ref{sec:extraction}.
\end{remark}

Like evaluation, replay is deterministic, in that if a trace can be
replayed on an environment then the resulting value is unique:
  \begin{theorem}\label{thm:replay-det}
    If $\gamma,T \trrun v_1$ and $\gamma,T \trrun v_2$ then
    $v_1 = v_2$.
  \end{theorem}
  \begin{proof}
    Proof is by (straightforward) structural induction on the first
    derivation and inversion on the second.
  \end{proof}

\subsection{Basic Properties of Traces}
\label{sec:traces::properties}

In this section, we identify key properties of traces, including type
safety, and the consistency and fidelity properties that
characterize how traces record the evaluation of an expression.

\paragraph{Determinacy and Type Safety.}
We employ a standard type system for expressions.
\figref{types} shows the (standard) typing rules for
  expressions and \figref{value-types} shows the rules for values and
  environments.  Type checking requires a variable context $\Gamma$
that maps variables to types.  We write $\Gamma \ts e : \tau$ to
indicate that $e$ has type $\tau$ in context $\Gamma$.  Similarly, we
write $\Gamma \tst T : \tau$ to indicate that $T$ is a well-formed
trace of type $\tau$ in context $\Gamma$.  \figref{trace-types} shows
the typing rules for traces.  The unusual rules are those for case and
application traces, whose form differs from the corresponding
expression forms.  One important point is that in the rule for
application traces, the trace of the function body needs to typecheck
in the same context $\Gamma'$ as the body of $e$.  This is why we
allow the annotation $\Gamma'$ indicating the environment of the
called function in application traces.
 
  As noted above, expressions (and hence also traces) can be
  well-formed at more than one type, but this does not matter since we
  are not concerned with typechecking algorithms here.

    \begin{figure}[tb]
\fbox{$\Gamma \ts e:\tau$}
\begin{smathpar}
\inferrule { \exc : \tau \in \Sigma} {\Gamma \ts \exc: \tau
    } 
    \and 
 \inferrule
    {
      x: \tau \in \Gamma
    }
    {\Gamma \ts \exvar{x}: \tau}
\and 
\inferrule {
      \Gamma \ts \novec{e}: \novec{\tau}  \\
      \kwf : \novec{\tau} \to \tau' \in \Sigma } {\Gamma \ts
      \exf{\novec{e}} : \tau'} 
\and
     \inferrule
     {
      \Gamma \ts e_1 : \tau_1 \\
      \Gamma, x:\tau_1 \ts e_2 : \tau_2 \\
    }
    {\Gamma \ts \exlet{e_1}{x}{e_2}: \tau_2}
\\
 \inferrule {
      \Gamma \ts e_1 : \tau_1 \\
      \Gamma \ts e_2 : \tau_2 } {\Gamma \ts \expair{e_1}{e_2} :
      \tau_1 \times \tau_2} 
\and 
\inferrule {\Gamma \ts e: \tau_1
      \times \tau_2} {\Gamma \ts \exfst{e} : \tau_1} 
\and 
\inferrule
    {\Gamma \ts e: \tau_1 \times \tau_2} {\Gamma \ts \exsnd{e} :
      \tau_2} 
\\ 
\inferrule {\Gamma \ts e : \tau_1} {\Gamma \ts
      \exinl{e} : \tau_1 + \tau_2} 
\and
\inferrule {\Gamma \ts e :
      \tau_2} {\Gamma \ts \exinr{e} : \tau_1 + \tau_2}
\and
   \inferrule
    {
      \Gamma \ts e : \tau_1+\tau_2 \\
     \Gamma,x_1:\tau_1 \ts e_1 : \tau \\
      \Gamma,x_2:\tau_2 \ts e_2 : \tau 
    }
    {\Gamma \ts \excase{x_1}{e_1}{x_2}{e_2}{e}: \tau}
    \\
   \inferrule{\Gamma,f:\tau_1\to\tau_2,x:\tau_1 \ts e : \tau_2}
    {
      \Gamma \ts \exfunk{\fn{f}{x}{e}} : \tau_1 \to \tau_2
    }
\and
    \inferrule
    {
      \Gamma \ts e_1 : \tau_1 \to \tau_2 \\
     \Gamma \ts e_2 : \tau_1 
    }
    {\Gamma\ts \exapp{e_1}{e_2}: \tau_2}
  \\
    \inferrule
    {\Gamma \ts e : \tyrec{\alpha}{\tau}}
    {\Gamma \ts \exunroll{e} :  \tau[\tyrec{\alpha}{\tau}/\alpha]}
    \and    
    \inferrule
    {\Gamma \ts e : \tau[\tyrec{\alpha}{\tau}/\alpha]}
    {\Gamma \ts \exroll{e} :  \tyrec{\alpha}{\tau}}
\end{smathpar}
  \caption{Well-typed expressions of \pl.}
\label{fig:types}
\fbox{$ \ts v:\tau$}
\begin{smathpar}
\inferrule { 
\exc : \tau \in \Sigma
}{\ts \trc: \tau
}
\and
 \inferrule {
       \ts v_1 : \tau_1 \\
       \ts v_2 : \tau_2 } {
\ts \vpair{v_1}{v_2} :  \tau_1 \times \tau_2
} 
\and
\and 
\inferrule { \ts v : \tau_1} { \ts
      \vinl{v} : \tau_1 + \tau_2} 
\and
\inferrule { \ts v :
      \tau_2} {\ts \vinr{v} : \tau_1 + \tau_2}
\and
\inferrule{
\ts \gamma : \Gamma\\
\Gamma,f:\tau_1\to\tau_2,x:\tau_1 \ts e:\tau_2
}{
\ts \vclos{\vclos{\fn{f}{x}{e}}{\Gamma}}{\gamma} : \tau_1 \to \tau_2
}
\and
    \inferrule
    { \ts v : \tau[\tyrec{\alpha}{\tau}/\alpha]}
    {\ts \vroll{v} :  \tyrec{\alpha}{\tau}}
\and 
  \inferrule{
    \ts v_1 : \tau_1\\ 
    \cdots \\ 
    \ts v_n : \tau_n}{
    \ts [x_1\mapsto v_1,\ldots,x_n\mapsto v_n] :
    [x_1:\tau_1,\ldots,x_n:\tau_n]
  }
\end{smathpar}
\caption{Value and environment typing.}
\label{fig:value-types}
\end{figure}

%%% Local Variables: 
%%% mode: latex
%%% TeX-master: "paper"
%%% End: 

\begin{figure}[tb]
\fbox{$\Gamma \tst T:\tau$}
\begin{smathpar}
\inferrule { \exc : \tau \in \Sigma} {\Gamma \tst \trc: \tau
    } 
    \and 
 \inferrule
    {
      x: \tau \in \Gamma
    }
    {\Gamma \tst \trvar{x}: \tau}
\and 
\inferrule {
      \Gamma \tst \novec{T}: \novec{\tau}  \\
      \kwf : \novec{\tau} \to \tau' \in \Sigma } {\Gamma \tst
      \trf{\novec{T}} : \tau'} 
\and
     \inferrule
     {
      \Gamma \tst T_1 : \tau_1 \\
      \Gamma, x:\tau_1 \tst T_2 : \tau_2 \\
    }
    {\Gamma \tst \trlet{T_1}{x}{T_2}: \tau_2}
\\
 \inferrule {
      \Gamma \tst T_1 : \tau_1 \\
      \Gamma \tst T_2 : \tau_2 } {\Gamma \tst \trpair{T_1}{T_2} :
      \tau_1 \times \tau_2} 
\and 
\inferrule {\Gamma \tst T: \tau_1
      \times \tau_2} {\Gamma \tst \trfst{T} : \tau_1} 
\and 
\inferrule
    {\Gamma \tst T: \tau_1 \times \tau_2} {\Gamma \tst \trsnd{T} :
      \tau_2} 
\\ 
\inferrule {\Gamma \tst T : \tau_1} {\Gamma \tst
      \trinl{T} : \tau_1 + \tau_2} 
\and
   \inferrule
    {
      \Gamma \tst T : \tau_1+\tau_2 \\
     \Gamma,x_1:\tau_1 \ts e_1 : \tau \\
      \Gamma,x_2:\tau_2 \ts e_2 : \tau \\
      \Gamma,x_1:\tau_1 \tst T_1 : \tau \\
    }
    {\Gamma \tst \trcasel{x_1}{e_1}{x_2}{e_2}{T}{T_1} : \tau}
    \\
\inferrule {\Gamma \tst T :
      \tau_2} {\Gamma \tst \trinr{T} : \tau_1 + \tau_2}
\and
\inferrule
    {
      \Gamma \tst T : \tau_1+\tau_2 \\
     \Gamma,x_1:\tau_1 \ts e_1 : \tau \\
      \Gamma,x_2:\tau_2 \ts e_2 : \tau \\
      \Gamma,x_2:\tau_2 \tst T_2: \tau \\
   }
    {\Gamma \tst \trcaser{x_1}{e_1}{x_2}{e_2}{T}{T_2} : \tau}
    \\
   \inferrule{\Gamma,f:\tau_1\to\tau_2,x:\tau_1 \ts e : \tau_2}
    {
      \Gamma \tst \trfunk{\fn{f}{x}{e}} : \tau_1 \to \tau_2
    }
\and
    \inferrule
    {
      \Gamma \tst T_1 : \tau_1 \to \tau_2 \\
      \Gamma',f{:}\tau_1\to\tau_2,x{:}\tau_1\ts e: \tau_2\\
     \Gamma \tst T_2 : \tau_1 \\
      \Gamma',f{:}\tau_1\to\tau_2,x{:}\tau_1 \tst T : \tau_2
    }
    {\Gamma\tst \trappk{\fn{f}{x}{e},\Gamma'}{T_1}{T_2}{f}{x}{T}: \tau_2}
  \\
    \inferrule
    {\Gamma \tst T : \tyrec{\alpha}{\tau}}
    {\Gamma \tst \trunroll{T} :  \tau[\tyrec{\alpha}{\tau}/\alpha]}
    \and    
    \inferrule
    {\Gamma \tst T : \tau[\tyrec{\alpha}{\tau}/\alpha]}
    {\Gamma \tst \trroll{T} :  \tyrec{\alpha}{\tau}}
\end{smathpar}
  \caption{ Well-typed traces of \pl.}
\label{fig:trace-types}
\end{figure}

%%% Local Variables: 
%%% mode: latex
%%% TeX-master: "paper"
%%% End: 

  \begin{theorem}\label{thm:red-safe}
    If $\Gamma \ts e : \tau$ and $\ts \gamma : \Gamma$ and
    $\gamma,e\eval v,T$ then $\ts v : \tau$ and $\Gamma \tst T :
    \tau$.
  \end{theorem}
  \begin{proof}
    Proof is by induction on the structure of the evaluation
    derivation, using inversion on the typing derivation.  The only
    nonstandard cases are for the well-formedness of the trace, but
    these cases are straightforward.
  \end{proof}
  Replay is also type-safe in the obvious sense:
  \begin{theorem}\label{thm:replay-safe}
    If $\Gamma \tst T : \tau$ and $\ts \gamma : \Gamma$ and
    $\gamma,T \trrun v$ then $\ts v : \tau$.
  \end{theorem}
  \begin{proof}
    Proof is by induction on the structure of the replay derivation;
    most cases are similar to analogous cases for \thmref{red-safe}.
 \end{proof}

\paragraph{Consistency and Fidelity.}
We say that a trace $T$ is \emph{consistent} with an environment
$\gamma$ if there exists $v$ such that $\gamma,T \trrun v$.  A trace
can easily be inconsistent with an environment, either because it is untyped nonsense and can
never run (e.g. $\exfst{42}$), or, more interestingly, because
replaying leads to situations that disagree with the control flow of
the trace (e.g. while replaying $\trcaseml{m}{T}{x_1}{T_1}$, the
replay of $T$ yields $\vinr{v}$).

Evaluation produces consistent traces, and replaying a trace on the
same input yields the same value:

\begin{theorem}[Consistency]\label{thm:red-consistent}
  If $\gamma, e \red v,T$ then $\gamma,T \trrun v$.
\end{theorem}
\begin{proof}
  Proof is by (straightforward) induction on the
  structure of derivations.  
\end{proof}
The converse does not hold: a trace can be consistent without ever
being produced by running a program.  In particular, consistency does
not check that the traces corresponding to bodies of function calls
match the code pointers recorded in the trace.  It is possible to
refine the definition of replay so that the function bodies are
checked against the traces, providing a stronger notion of
consistency.  However, this would complicate the replay semantics. In
the rest of this article we usually consider traces obtained by
running the tracing semantics.  When this is the case, the derivation
of $\gamma,e \eval v,T$ is itself a witness to this, so there is no
need to introduce an additional judgment that captures this invariant.

Furthermore, the trace produced by evaluation is faithful to the
original expression, in the sense that whenever the trace can be
successfully replayed on a different input, the result (and its trace)
is the same as what we would obtain by rerunning $e$ from scratch, and
the resulting trace is the same as well.  We call this property
\emph{fidelity}.
\begin{theorem}[Fidelity]\label{thm:fidelity-trace}
  If $\gamma,e \eval v,T$ and $\gamma',T \trrun v'$ then $\gamma',e
  \eval v',T$.
\end{theorem}
\begin{proof}
  Straightforward proof by induction on the structure of derivations.
  The interesting cases are for case and application expressions; in
  each case, the induction hypothesis ensures that the intermediate
  sum or function value encountered when recomputing $e$ in $\gamma'$
  matches that in the original derivation, so that the subtraces
  contingent on this value can be reused.
\end{proof}
Intuitively, fidelity corresponds to a repeatability or
reproducibility property: it does not just guarantee that we get the
same results when the trace is replayed on the same input, it also
guarantees that the trace tells us what would happen if we rerun on
inputs that are similar enough to the original input that replay can
succeed.  Thus, traces correspond to a form of \emph{explanation},
analogous to forms of explanation explored in causal models and
workflow
provenance~\cite{cheney10dcm,halpern05bjps-1,halpern05bjps-2}.  While
we do not make more of this connection here, fidelity is also related
to the correctness properties for various forms of slicing, including
disclosure slicing (as discussed in Section~\ref{sec:analysis}).

  \begin{remark}
    As noted at the beginning of the section, we made two simplifying
    assumptions: we consider traces only for terminating runs, and we
    exclude side-effects from the language.  These assumptions are
    reasonable for many application areas of provenance (for example,
    in scientific computation and databases), but it is naturally of
    interest to consider extending our approaches to trace
    nonterminating or effectful computations.  These raise potential
    complications: for example, adapting the trace semantics to a
    small-step semantics seems nontrivial, and it is not as clear what
    the appropriate correctness properties are for traces involving
    effects (including nondeterminism or allocation).  These are
    interesting areas for exploration in future work.
  \end{remark}

%%% Local Variables: 
%%% mode: latex
%%% TeX-master: "main"
%%% End: 

% LocalWords:  boolean metavariables isorecursive Booleans datatypes datatype
% LocalWords:  unicity typechecking subtraces subexpression subexpressions iso
% LocalWords:  typecheck coinductive Coinduction intensional unfoldings dep
% LocalWords:  nontermination scrutinee determinacy

\section{Provenance Views and Extraction}
\label{sec:extraction}

In this section we consider different kinds of views and queries over
provenance traces.  To be specific, we consider a consistent triple
$(\gamma,T,v)$ where $\gamma,T \trrun v$ to be the ``traces'' in the
sense of the provenance security framework.  Then queries over these
triples correspond to sets of triples (generally definable using some
compact syntax), and views correspond to functions from triples to
some other data.  We first consider a general class of views definable
using \emph{annotation propagation}, by giving a generic framework for
extracting other kinds of provenance from execution traces.  These
forms of provenance induce provenance views in a natural way if we
allow for initial annotations that uniquely identify each part of a
value by a path.  

\subsection{Annotations, Paths, and Provenance Extraction}

Many previous approaches to provenance can be viewed as performing a
form of {\em annotation propagation}.  The idea is to decorate the
input with annotations (often, initially, unique identifiers) and
propagate the annotations through the evaluation.  For example, in
where-provenance, annotations are optional tags that can be thought of
as pointers showing where output data was copied from in the
source~\cite{buneman01icdt,buneman08tods}.  Other techniques, such as
why-, how-, and dependency provenance, can also be defined in terms of
annotation
propagation~\cite{green07pods,foster08pods,buneman08pods,cheney09ftdb}.
We gave similar definitions of different forms of provenance using a
common framework for XQuery~\cite{cheney09planx}; some of the properties proved are
generalizations of properties shown there or in~\cite{buneman08tods}.

Based on this observation, we define a provenance extraction framework
in which values are decorated with annotations and extraction
functions take traces and return annotated values that can be
interpreted as useful provenance information.  We first define
annotated values and give a generic annotation-propagation operation.
We apply this framework
to specify several concrete annotation schemes and extraction
functions.

\paragraph{Annotations.}
Let $A$ be an arbitrary set of \emph{annotations} $a$, which we
usually assume includes a blank annotation $\bot$ and a countably
infinite set of identifiers $\lbl \in \Loc$, called \emph{locations}.  We define \emph{$A$-annotated values} $\av$ (or just
\emph{annotated values}, when $A$ is clear) using the following
grammar:
\begin{eqnarray*}
  \av &::=& w^a   \\
\agamma &::=& [x_1 \mapsto
  \av_1,\ldots,x_n \mapsto \av_n] \\
  w &::=& \vc \mid \vpair{\av_1}{\av_2} \mid
  \vinl{\av} \mid \vinr{\av} \mid  \vclos{\kappa}{\agamma}\mid\vroll{\av} 
\end{eqnarray*}
We write
$\agamma$ for \emph{annotated environments} mapping variables to
annotated values.  We define an erasure function $|\av|$ that maps
each annotated value to an ordinary value by erasing the annotations.
Similarly, $|\agamma|$ is the ordinary environment obtained by erasing
the annotations from the values of $\agamma$.
  This function is defined mutually recursively on annotated values
  and environments as shown in Figure~\ref{fig:erasure}.   We also introduce a notation
for the set of annotated values occurring in a value in
Figure~\ref{fig:occ}.
Moreover, we write 
\[\occnb(\av) = \{w^a\in \occ(\av) \mid a \neq \bot\}\]
for the set of annotated values with annotation $a \neq \bot$.

\begin{figure}[tb]
\small
  \begin{eqnarray*}
    |[x_1\mapsto v_1,\ldots,x_n\mapsto v_n]| &=& [x_1\mapsto
    |v_1|,\ldots,x_n \mapsto |v_n|]\\
    |c^a| &=& c\\
    |\vpair{\av_1}{\av_2}^a| &=& (|\av_1|,|\av_2|)\\
    |\vinl{\av}^a| &=& \vinl{|\av|}\\
    |\vinr{\av}^a| &=& \vinr{|\av|}\\
    |\vclos{\kappa}{\agamma}^a| &=& \vclos{\kappa}{|\agamma|}\\
    |\vroll{\av}^a| &=& \vroll{|\av|}
  \end{eqnarray*}
\caption{Erasure operation.}\label{fig:erasure}
 \begin{eqnarray*}
    \occ(c^a) &=& \{c^a\}\\
    \occ((\av_1,\av_2)^a) &= &\{(\av_1,\av_2)^a\} \cup \occ(\av_1) \cup \occ(\av_2)\\
    \occ((\vinl{\av})^a) &=&\{(\vinl{\av})^a\} \cup \occ(\av)\\
    \occ((\vinr{\av})^a) &=& \{(\vinr{\av})^a\} \cup \occ(\av)\\
    \occ((\vroll{\av})^a) &= &\{(\vroll{\av})^a\} \cup \occ(\av)\\
    \occ(\vclos{\kappa}{\agamma}^a) &=& \{\vclos{\kappa}{\agamma}^a\} \cup
    \occ(\agamma)\\
    \occ(\agamma) &=& \bigcup_{x \in \dom(\agamma)} \occ(\agamma(x))
\end{eqnarray*}
\caption{Occurrences of annotated values.}\label{fig:occ}
\end{figure}

\paragraph{Paths as annotations.}  For annotations to be useful when the
full input is unavailable, we consider annotations where the locations
$\lbl$ are \emph{paths} that uniquely address parts of the input
environment.  Paths $\pi$ have syntax:
\[
\pi ::= \epsilon \mid x.\pi \mid 1.\pi \mid 2.\pi
\]
and we consider path concatenation $\pi.\pi'$ to be associative with unit
$\epsilon$, so that we may write $\pi.i$ to construct a pattern ending
in $i$.  Paths address parts of values or environments; we write
$v[\pi]$ or $\gamma[\pi]$ for the part of $v$ or $\gamma$ addressed by
$\pi$, defined in Figure~\ref{fig:path-lookup}.
\begin{figure}[tb]
\small
  \begin{eqnarray*}
    v[\epsilon] &=& v\\
    (v_1,v_2)[1.\pi] &=& v_1[\pi]\\
    (v_1,v_2)[2.\pi] &=& v_2[\pi]\\
    (\vinl{v})[1.\pi] &=& v[\pi]\\
    (\vinr{v})[1.\pi] &=& v[\pi]\\
    (\vroll{v})[1.\pi] &=& v[\pi]\\
    (\vclos{\kappa}{\gamma})[1.\pi] &=& \gamma[\pi]\\
    \gamma[x.\pi] &=& \gamma(x)[\pi]
  \end{eqnarray*}
    \caption{Path lookup operation.}
\label{fig:path-lookup}
\end{figure}

We write $\path(\gamma)$ for the environment $\gamma$ with each
component annotated with the path to that component.  More generally,
we define $\path_\pi(\gamma$) and $\path_\pi(v)$ as shown in
Figure~\ref{fig:path-annotation}.  Then $\path(v) = \path_\epsilon(v)$
and $\path(\gamma) = \path_\epsilon(\gamma)$.  For example,
  $\path([x\mapsto(1,2),y\mapsto\vinl{4}]) =
  [x\mapsto(1^{x.1},2^{x.2})^x,y\mapsto\vinl{4^{y.1}}^y]$.  

\begin{figure}[tb]
\small
  \begin{eqnarray*}
    \path_\pi(\gamma) &=& [x_1 \mapsto
      \path_{\pi.x_1}(\gamma(x_1)),\ldots,x_n \mapsto
      \path_{\pi.x_n}(\gamma(x_n))]\\
    \path_\pi(c) &=& c^\pi \\
    \path_\pi(\vpair{v_1}{v_2}) &=& \vpair{\path_{\pi.1}(v_1)}{\path_{\pi.2}(v_2)}^\pi\\
    \path_\pi(\vinl{v}) &=& \vinl{\path_{\pi.1}(v)}^\pi\\
    \path_\pi(\vinr{v}) &=& \vinr{\path_{\pi.1}(v)}^\pi\\
    \path_\pi(\vclos{\kappa}{\gamma}) &=& \vclos{\kappa}{\path_{\pi.1}(\gamma)}^\pi
  \end{eqnarray*}
  \caption{Path annotation operation.}
\label{fig:path-annotation}
\end{figure}

\paragraph{Extraction framework.}
We will define a family of provenance extraction functions
$\extract(T,\agamma)$ that take a trace $T$ and an environment
$\agamma$ and return an annotated value.  Each such $\extract$ can be
specified by giving the following annotation-propagation functions:
\begin{eqnarray*}
  \extract_\trc,\extract_\kappa &:& A\\
 \extract_1,\extract_2,\extract_L,\extract_R,\extract_{\kwapp},\extract_{\kwunroll}
 &:& A \times  A \to A\\
 \extract_\kwf &:& A^n \to A \qquad (\text{where $\kwf$ is $n$-ary})
\end{eqnarray*}
Each function shows how the annotations involved in the corresponding
computational step propagate to the result.  For example,
$\extract_1(a,b)$ gives the annotation on the result of a
$\kwfst$-projection, where $a$ is the annotation on the pair and $b$
is the annotation of the first element.  
\figref{generic-from-trace} shows how to propagate annotations through
a trace given basic annotation-propagation functions.

\begin{figure}[tb]
  \[\small\begin{array}{rclll}
%    \extract(\tremp,\agamma) &=& \vhole\\[1pt]
   \extract(\trvar{x},\agamma) &=& \agamma(x)\\[1pt]
    \extract (\trlet{T_1}{x}{T_2},\agamma) &=& \extract(T_2,\agamma[x\bindsto \extract(T_1,\agamma)])\\[1pt]
    \extract (\trc,\agamma) &=& c^{\extract_c}\\[1pt]
    \extract (\trf{T_1,\ldots,T_n},\agamma) &=&
    (\hat{\oplus}(c_1,\ldots,c_n))^{\extract_\kwf(a_1,\ldots,a_n)}
    &\text{where} &c_i^{a_i} = \extract(T_i,\agamma)\\[1pt]
   \extract (\trpair{T_1}{T_2},\agamma) &=& (\extract(T_1,\agamma),\extract(T_2,\agamma))^\bot\\[1pt]
    \extract (\trfst{T},\agamma) &=& v_1^{\extract_1(a,b)}
&\text{where}& (v_1^b,\av_2)^a = \extract(T,\agamma)\\[1pt]
   \extract (\trsnd{T},\agamma) &=&  v_2^{\extract_2(a,b)}
&\text{where}& (\av_1,v_2^b)^a = \extract(T,\agamma)\\[1pt]
   \extract (\trinl{T},\agamma) &=&\vinl{\extract(T,\agamma)}^\bot\\[1pt]
  \extract (\trinr{T},\agamma) &=& \vinr{\extract(T,\agamma)}^\bot\\
    [1pt]
   \extract (\trcaseshortl{T}{x}{T_1},\agamma) &=& v^{\extract_{L}(a,b)}
     &\text{where}&\vinl{\av}^a = \extract(T,\agamma)\\
&&&\text{and}& v^b = \extract(T_1,\agamma[y\bindsto \av])
 \\[1pt]
\extract (\trcaseshortr{T}{y}{T_2},\agamma) &=& v^{\extract_{R}(a,b)}
&\text{where}&\vinr{\av}^a = \extract(T,\agamma) \\
&&&\text{and}& v^b = \extract(T_2,\agamma[y\bindsto \av]) \\[1pt]
   \extract (\trfunk{\kappa},\agamma) &=& \vclos{\kappa}{\agamma}^{\extract_\kappa}\\[1pt]
    \extract (\trappk{\kappa}{T_1}{T_2}{f}{x}{T},\agamma) &=&
    v^{\extract_{\kwapp}(a,b)}
   &\text{where}&\vclos{\kappa}{\agamma'}^a =
    \extract(T_1,\agamma)\\
&&&\text{and}& \av_2 =\extract(T_2,\agamma) \\
&&&\text{and}& v^b = \extract(T,\agamma'[f\bindsto
    \vclos{\kappa}{\agamma'}^a, x \bindsto \av_2]))
 \\[1pt]
 \extract (\trroll{T},\agamma) &=& \vroll{\extract(T,\agamma)}^\bot\\[1pt]
    \extract (\trunroll{T},\agamma) &=&  v^{\extract_\kwunroll(a,b)}
    &\text{where}& \vroll{v^b}^a = \extract(T,\agamma) 
\end{array}
\]
%\hrule
 \caption{Generic extraction.}
 \label{fig:generic-from-trace}
\end{figure}

\begin{remark}
    The extraction framework hard-wires the behavior of certain
    operations such as $\mathtt{let}$, $\vinl{}$,
    $\vinr{}$,$\vroll{}$, and pairing, using $\bot$ to handle all of
    them. On the other hand, even though these constructors are
    hard-wired so that the top-level annotation is always $\bot$, this
    does not imply that the first arguments supplied to the
    corresponding extraction functions $\extract_1,\extract_2$,
    etc. are always $\bot$; see
    Example~\ref{ex:first-argument-not-always-bot} for an illustration
    of this point.

    It would also be possible to extend the framework to allow greater
    customization; however, this functionality is not needed by any of
    the forms of provenance in this article.  We believe that the
    framework presented in this paper is general enough to be of use
    beyond the three provenance models we considered, but we do not
    know how one could prove that it is general enough for all
    purposes --- or how one could prove that any alternative framework
    is general enough for all purposes.  It is also possible that
    there are natural forms of provenance that do not fit (a
    reasonable generalization of) the framework.
\end{remark}

\begin{theorem}
  Every generic provenance extraction function is compatible with
  replay: that is, for any $\agamma,T,v$, if $|\agamma|,T \trrun v$
  then $|\extract(T,\agamma) | = v$.
\end{theorem}
\begin{proof}
  Straightforward induction on replay derivations.
\end{proof}

  \begin{remark}
    Consider the trivial annotation structure $\mathsf{Triv}$ with
    underlying annotation set $\{\bot\}$.  In this setting, the
    erasure function $|-|$ is bijective; its inverse just decorates
    each part of a value with $\bot$.  Consider also a trivial
    instance of the generic provenance framework for $\{\bot\}$ such
    that $\mathsf{Triv}_\kappa = \bot$ and $\mathsf{Triv}_c = \bot$
    and all of the annotation-propagation functions are constant
    functions returning $\bot$, that is, $\mathsf{Triv}_1(x,y) =
    \bot$, etc.  This instance of the provenance framework is
    essentially the same as the trace replay semantics defined by the
    judgment $\gamma,T \trrun v$.  Thus, the generic extraction
    framework can be viewed as a denotational presentation of the
    replay semantics of traces, generalized to allow for annotated
    values.
  \end{remark}

\paragraph{Where-provenance.}
Where-provenance can be defined via an annotation-propagating
semantics where annotations are either labels $\lbl$ or the blank
annotation $\bot$. 
Intuitively, for where-provenance, an explicit label $\lbl$
annotating a part of the input indicates that that part ``comes from''
a part of the input with the same label; an annotation $\bot$
provides no information about where the output part ``comes from''
in the input (if anywhere).
We define the where-provenance semantics
$\where(T,\agamma)$ using the following annotation-propagation
functions:
\begin{eqnarray*}
  \where_c, \where_\kappa &=& \bot\\
  \where_1,\where_2,\where_L,\where_R,\where_{\kwapp},\where_{\kwunroll}
  &=& \lambda (x,y). y\\
  \where_\kwf&=& \lambda (a_1,\ldots,a_n). \bot
\end{eqnarray*}
Essentially, these functions preserve the annotations of data that are
copied, and annotate computed or constructed data with $\bot$.  This
semantics is similar to that in Buneman et al.~\cite{buneman08tods}
and previous treatments of where-provenance in databases, adapted to
\pl.
 \figref{where-from-trace} shows the generic semantics specialized to
  where-provenance.  Note that for a function like ``factorial'', the
where-provenance of the output is always $\bot$.  

\begin{figure}[tb]
  \[\small\begin{array}{rclll}
%  \where(\trvar{x},\agamma) &=& \agamma(x)\\[1pt]
 %   \where (\trlet{T_1}{x}{T_2},\agamma) &=& \where(T_2,\agamma[x\bindsto \where(T_1,\agamma)])\\[1pt]
    \where (\trc,\agamma) &=&\exc^\bot\\[1pt]
    \where (\trf{T_1,\ldots,T_n},\agamma) &=&
    (\hat{\oplus}(c_1,\ldots,c_n))^\bot
    &\text{where}&c_i^{a_i} = \where(T_i,\agamma)\\[1pt]
    % \where (\trpair{T_1}{T_2},\agamma) &=& (\where(T_1,\agamma),\where(T_2,\agamma))^\bot\\[1pt]
    \where (\trfst{T},\agamma) &=& \av_1
    & \text{where}& (\av_1,\av_2)^a = \where(T,\agamma)\\[1pt]
  \where (\trsnd{T},\agamma) &=&  \av_2 
   &\text{where}& (\av_1,\av_2)^a = \where(T,\agamma)\\[1pt]
   % \where (\trinl{T},\agamma) &=& \vinl{\where(T,\agamma)}^\bot\\[1pt]
   %  \where (\trinr{T},\agamma) &=&  \vinr{\where(T,\agamma)}^\bot\\
   %   [1pt]
   \where (\trcaseshortl{T}{x}{T_1},\agamma) &=& \where
    (T_1,\agamma[x\bindsto \av])
    &\text{where}&\vinl{\av}^a = \where(T,\agamma) \\[1pt]
 \where (\trcaseshortr{T}{y}{T_2},\agamma) &=& \where
     (T_2,\agamma[y\bindsto \av])
     &\text{where}&\vinr{\av}^a = \where(T,\agamma) \\[1pt]
   \where (\trfunk{\kappa},\agamma) &=& \vclos{\kappa}{\agamma}^\bot\\[1pt]
    \where (\trappk{\kappa}{T_1}{T_2}{f}{x}{T},\agamma) &=&  \where(T,\agamma'[f\bindsto
    \vclos{\kappa}{\agamma'}^a, x \bindsto \av]) 
    &\text{where}&\vclos{\kappa}{\agamma'}^a =
    \where(T_1,\agamma)\\
&& & \text{and} & \av =\where(T_2,\agamma) 
 \\[1pt]
 % \where (\trroll{T},\agamma) &=& \vroll{\where(T,\agamma)}^\bot\\[1pt]
    \where (\trunroll{T},\agamma) &=&  \av 
    &\text{where}& \vroll{\av}^a = \where(T,\agamma)
\end{array}\]
%\hrule
 \caption{Where-provenance extraction (selected cases).}
 \label{fig:where-from-trace}
\end{figure}

%%% Local Variables: 
%%% mode: latex
%%% TeX-master: "main"
%%% End: 

\begin{example}
  Recall the example program from Section~\ref{sec:intro-example}:
\begin{verbatim}
let y = 2@L in 
let f x = if x = y then y else x+1 in 
map f [1@L1,2@L2,3@L3]
\end{verbatim}
The result of the where-provenance extraction semantics applied to
this program is  $[2,2^L,4]$, as shown graphically in
Figure~\ref{fig:prov-illustration}, showing that the second result element is
copied from $y$ and giving no information about the other two.
\end{example}

To state the key property of where-provenance, we use the function
$\occ$ introduced earlier in this section.  The key property of
where-provenance is that if an annotated value $w^a$
appears in $\where(T,\agamma)$ with annotation $a \neq \bot$, then
$w^a$ is an exact copy (including any nested annotations) of a part of
$\agamma$.
\begin{theorem}\label{thm:where}
  Suppose $|\agamma|,e \red v,T$. Then $\occnb(\where(T,\agamma))
  \subseteq \occnb(\agamma)$.
\end{theorem}
 \begin{proof}
    See \appref{where-proof}.
  \end{proof}
    \begin{remark}
      Buneman et al.~\cite{buneman08tods} consider a where-provenance
      semantics for database query and update languages (with nested
      collection types and pairs, but no recursion or datatypes),
      which we adapt here to a conventional functional language (with
      recursion and datatypes, but no collection types).  The basic
      idea, propagating annotations from the input to output when data
      are copied, is the same.  They did not propose a tracing model
      of their calculus, but instead defined where-provenance via a
      syntactic translation that inserts annotation propagation code.
      The correctness property we discuss here corresponds to their
      \emph{copying} property~\cite[Prop. 5.5]{buneman08tods}. They
      studied additional query normalization and semantic
      expressiveness properties that we do not address here.
    \end{remark}

\paragraph{Expression provenance.}
To model expression provenance, we consider \emph{expression
  annotations} $t$ consisting of labels $\lbl$, blanks $\bot$,
constants $\trc$, or primitive function applications $\trf{t_1,\ldots,t_n}$.
\[t ::= \lbl \mid \trc \mid \trf{t_1,\ldots,t_n} \mid \bot\]
Intuitively, a label $\lbl$ indicates that a part of the output is
copied from a part of the input with the same label; a constant $\trc$
indicates an output part that is built by evaluating constant in the
program; a term $\trf{t_1,\ldots,t_n}$ indicates a part of the output
that is computed by evaluating $\kwf$ on values obtained from
$t_1,\ldots,t_n$, and $\bot$ provides no information about how a part
of the output was computed from the input.
 We define expression-provenance extraction $\expr(T,\agamma)$ in
  much the same way as $\where$, with the following differences:
  \[
  \expr_c = c \qquad \expr_\kwf(t_1,\ldots,t_n) =\trf{t_1,\ldots,t_n}
  \]
 \figref{expr-from-trace} shows the generic semantics specialized to
expression-provenance.  
\begin{figure}[tb]
\[ \small \begin{array}{rclll}
 % \expr(\trvar{x},\agamma) &=& \agamma(x)\\[1pt]
 %  \expr (\trlet{T_1}{x}{T_2},\agamma) &=& \expr(T_2,\agamma[x\mapsto\expr(T_1,\agamma)])\\
  \expr (\trc,\agamma) &=& \vc^\trc\\[1pt]
  \expr (\trf{T_1,\ldots,T_n},\agamma) &=&
 (\hat{\oplus}(c_1,\ldots,c_n))^{\trf{t_1,\ldots,t_n}}
  &\text{where}&c_i^{t_i} = \expr(T_i,\agamma)
   \\   [1pt]
    % \expr (\trpair{T_1}{T_2},\agamma) &=& (\av_1,\av_2)^{\bot}
    % &\text{where}& \av_i = \expr(T_i,\agamma)\\[1pt]
    \expr (\trfst{T},\agamma) &=& \av_1
    &\text{where} &(\av_1,\av_2)^t = \expr(T,\agamma)\\[1pt]
    \expr (\trsnd{T},\agamma) &=& \av_2
    &\text{where}& (\av_1,\av_2)^t = \expr(T,\agamma)\\[1pt]
    % \expr (\trinl{T},\agamma) &=& \vinl{\expr(T,\agamma)}^\bot\\[1pt]
    %  \expr (\trinr{T},\agamma) &=&  \vinr{\expr(T,\agamma)}^\bot\\ [1pt]
    \expr (\trcaseshortl{T}{x}{T_1},\agamma) &=&
    \expr(T_1,\agamma[x\mapsto \av])
    &\text{where}& \expr(T,\agamma) = \vinl{\av}^t\\[1pt]
     \expr (\trcaseshortr{T}{y}{T_2},\agamma) &=&
     \expr(T_2,\agamma[y\mapsto \av])
     &\text{where} & \expr(T,\agamma) = \vinr{\av}^t\\[1pt]
    \expr (\exfunk{\kappa},\agamma) &=& \vclos{\kappa}{\agamma}^{\bot}\\[1pt]
    \expr (\trappk{\kappa}{T_1}{T_2}{f}{x}{T},\agamma) &=&
    \expr(T,\agamma'[f\bindsto \vclos{\kappa}{\agamma'}^t, x \bindsto
    \av) 
&\text{where}& \vclos{\kappa}{\agamma'}^t
    =\expr(T_1,\agamma))\\
    &&&\text{and}& \av = \expr(T_2,\agamma)\\
    % \expr (\trroll{T},\agamma) &=& \vroll{\expr(T,\agamma)}^\bot\\[1pt]
    \expr (\trunroll{T},\agamma) &=& \av & \text{where}&
    \vroll{\av}^t = \expr(T,\agamma)
\end{array}
\]
%\hrule
   \caption{Expression provenance extraction (selected cases).}
\label{fig:expr-from-trace}
\end{figure}

%%% Local Variables: 
%%% mode: latex
%%% TeX-master: "main"
%%% End: 

% It would also be
% straightforward to define a translation from traces to provenance
% graphs (for example, Open Provenance Model graphs~\cite{opm}).
%  --- a
% workshop paper~\cite{acar10tapp} gives an OPM-like trace semantics for
% a database query language based on monadic comprehensions.

\begin{example}
 Continuing with the $map$ example from Section~\ref{sec:intro-example},
the result of the expressione-provenance extraction semantics applied to
this program is  $[2^{L_1+1},2^L,4^{L_3+1}]$, as shown graphically in
Figure~\ref{fig:prov-illustration}.  This shows that the second result element is
copied from $y$ and the other two arguments are computed by
incrementing the first and last elements of the input, respectively.
Observe that this is strictly more informative than the where-provenance.
\end{example}

The correctness property for expression provenance states that the
expression annotation correctly recomputes the value it annotates.  To
formalize this, we use the auxiliary definitions $\occ(),\occnb()$
introduced for where-provenance.  Let $h : \Loc \to \Val$ be a function from locations
to values, and let $h(t)$ be the value obtained by evaluating
annotation term $t$ with values from $h$ substituted for locations in
$t$.  We say that $h$ is \emph{consistent} with $\av$ if whenever $w^t
\in \occnb(\av)$, we have $h(t) = |w|$.  Similarly, $h$ is consistent
with $\agamma$ if whenever $w^t \in \occnb(\agamma)$, we have $h(t) =
|w|$.  We note that for any distinctly-annotated value, for example
$\path(\gamma)$, there is always a consistent mapping $h$, obtained by
mapping $\lbl$ to $|w|$ whenever $w^\lbl \in \occ(\agamma)$.

  \begin{example}
    Consider $\gamma = [x \mapsto (1,2),y\mapsto \vinl{3}]$ and
    $\agamma = \path(\gamma) = [x \mapsto (1^{x.1},2^{x.2})^x, y
    \mapsto \vinl{3^y.1}]$.  Then the consistent mapping $h$ is
    defined as follows:
    \[
    \begin{array}{rclcrclcrcl}
      h(x) &=& (1,2) &&
      h(x.1) &=& 1 &&
      h(x.2) &=& 2\\
      h(y) &=& \vinl{3} &&
      h(y.1) &=& 3  
    \end{array}
    \]
    Observe in particular that this illustrates that $|\agamma|$ is
    typically not a consistent mapping for $\agamma$ (since in this
    example, $|\agamma| = \gamma \neq h$).
  \end{example}

\begin{theorem}\label{thm:expr}
  Suppose $|\agamma|,e \red v',T$.  Then if $h$ is consistent with
  $\agamma$, then $h$ is also consistent with $\expr(T,\agamma)$.
\end{theorem}
\begin{proof}
  Similar to the proof of \thmref{where}.  See \appref{expr-proof}.
\end{proof} 
% It follows that if $\agamma = \path(\gamma)$ and $h(\pi) =
% \gamma[\pi]$ is the mapping from paths in $\gamma$ to values, then if an
% annotated value $w^t$ appears in $\expr(T,\agamma)$ with $t \neq
% \bot$, then $\agamma(t) = h(t) = |v|$.

\paragraph{Dependency provenance.}
To extract dependency provenance (adapting the definition
from~\cite{cheney11mscs}) we will use annotations $\phi$ that are sets
of source locations $\{\lbl_1,\ldots,\lbl_n\}$, and we take the
default annotation $\bot$ to be the empty set $\emptyset$.  Initial annotations
consist of disjoint singleton sets $\{\lbl\}$.  We define $(\av)^{+a}$
to mean adding annotations $a$ to the top-level of $\av = w^a$; that
is, $(w^a)^{+b} = w^{a\cup b}$.  We define
$\dep(T,\agamma)$ using the following propagation functions:
\begin{eqnarray*}
    \dep_c, \dep_\kappa &=& \emptyset\\
  \dep_1,\dep_2,\dep_L,\dep_R,\dep_{\kwapp},\dep_{\kwunroll}
  &=& \lambda (x,y). x \cup y\\
  \dep_\kwf&=& \lambda (a_1,\ldots,a_n).  a_1 \cup \cdots \cup a_n
\end{eqnarray*}
This semantics is based on
the dynamic provenance tracking semantics given by
Cheney~et~al.~\cite{cheney11mscs}, generalized to \pl.
 \figref{dep-from-trace} shows the generic semantics specialized to
  dependency-provenance.  
\begin{figure}[tb]
\[  \small\begin{array}{rclll}
   % \dep(\trvar{x},\agamma) &=& \agamma(x)
   %  \\[1pt]
   %  \dep (\trlet{T_1}{x}{T_2},\agamma) &=&
   %  \dep(T_2,\agamma[x:=\dep(T_1,\agamma)])\\[1pt]
   \dep (\trc,\agamma) &=& \vc^\emptyset
\\[1pt]
    \dep (\trf{T_1,\ldots,T_n},\agamma) &=&
    (\hat{\oplus}(c_1,\ldots,c_n))^{\bigcup_i \phi_i}
&\text{where}&c_i^{\phi_i} = \dep(T_i,\agamma)
\\[1pt]
%    \dep (\trpair{T_1}{T_2},\agamma) &=&
 %   (\dep(T_1,\agamma),\dep(T_2,\agamma))^\emptyset
%\\[1pt]
    \dep (\trfst{T},\agamma) &=& (\av_1)^{+\phi}
&\text{where}&(\av_1,\av_2)^\phi = \dep(T,\agamma)
\\[1pt]
  \dep (\trsnd{T},\agamma) &=& (\av_2)^{+ \phi}
   &\text{where}&(\av_1,\av_2)^\phi = \dep(T,\agamma) \\[1pt]
%   \dep (\trinl{T},\agamma) &=& \vinl{\dep(T,\agamma)}^\emptyset
%\\[1pt]
 % \dep (\trinr{T},\agamma) &=&
 %  \vinr{\dep(T,\agamma)}^\emptyset \\ [1pt]
\dep (\trcaseshortl{T}{x}{T_1},\agamma) &=&
\dep(T_1,\agamma[x\bindsto \av])^{+\phi}
&\text{where}& \vinl{\av}^\phi = \dep(T,\agamma)
\\[1pt]
 \dep (\trcaseshortr{T}{y}{T_2},\agamma) &=&
  \dep(T_2,\agamma[y\bindsto \av])^{+\phi} &\text{where}&
  \vinr{\av}^\phi = \dep(T,\agamma)
  \\[1pt]
   \dep (\trfunk{\kappa},\agamma) &=&
    \vclos{\kappa}{\agamma}^\emptyset
\\[1pt]
    \dep (\trappk{\kappa}{T_1}{T_2}{f}{x}{T},\agamma) &=&  
    \dep(T,\agamma'[f\bindsto\vclos{\kappa}{\agamma'}^\phi,  x
    \bindsto \av])^{+\phi} 
&\text{where}&\vclos{\kappa}{\agamma'}^\phi =\dep(T_1,\agamma)
\\
&&
&\text{and}& \av =  \dep(T_2,\agamma)
\\[1pt]
 % \dep (\trroll{T},\agamma) &=& \vroll{\dep(T,\agamma)}^\emptyset
 %  \\[1pt]
  \dep (\trunroll{T},\agamma) &=& (\av)^{+\phi} &\text{where}&
  \vroll{\av}^\phi = \dep(T,\agamma)
 \end{array}\]
%\hrule
       \caption{Dependency provenance extraction.}
\label{fig:dep-from-trace}
\end{figure}

%%% Local Variables: 
%%% mode: latex
%%% TeX-master: "main"
%%% End: 

\begin{example}
 Continuing with the $map$ example from Section~\ref{sec:intro-example},
the result of the dependency-provenance extraction semantics applied to
this program is  $[2^{L_1,L},2^{L_1,L},4^{L_3,L}]$, as shown graphically in
Figure~\ref{fig:prov-illustration}.  This shows that all three
arguments depend on both $L$ and on the respective element of the input list.
This information is not computable from the where-provenance or
expression-provenance, or vice versa.  
\end{example}

This definition satisfies the \emph{dependency-correctness} property
introduced in~\cite{cheney11mscs}.  
As explained in Cheney et al.~\cite{cheney11mscs},
  dependency-correctness is intuitively motivated by analogy to
  dependency-tracking and information flow analyses, following the
  dependency core calculus of Abadi et al.~\cite{abadi99popl}.
  Analogously to noninterference in information flow security, we
  define an auxiliary relation $\eqxat{\lbl}$, where $v \eqxat{\lbl}
  v'$ intuitively says that two annotated values are equal except
  (possibly) at parts labeled by $\lbl$, defined as shown in
  \figref{eq-except-at}.  The $\eqxat{l}$ relation is reflexive, symmetric and
  transitive, and in particular $c \eqxat{\lbl} c$ for any $\lbl$,
  since $c = c$, whereas $c^\lbl \eqxat{\lbl} d^\lbl$ holds even
  though $c \neq d$, because both are labeled by $\lbl$.

  As discussed in~\cite{cheney11mscs}, for distinctly-annotated
  values, $v\eqxat{\lbl}v'$ holds if and only if $v$ and $v'$ are of
  the form $C[v_0]$ and $C[v_0']$, where $C[]$ is a context capturing
  the common parts of $v$ and $v'$ (above $\lbl$), and $v_0$ and
  $v_0'$ are subvalues showing where $v$ and $v'$ differ (below
  $\lbl)$.  However, during propagation of dependency annotations,
  values do not remain distinctly-annotated, and the $\eqxat{\lbl}$
  relation is an appropriate generalization of this property.  

\begin{figure}[tb]
  \fbox{$\av \eqxat{\lbl} \av'$}
  \begin{smathpar}
    \inferrule{w \eqxat{\lbl} w'}{w^\phi \eqxat{\lbl} (w')^\phi}
    \and
    \inferrule{\lbl \in \phi \cap \phi'}{w^\phi \eqxat{\lbl} (w')^{\phi'}}
\and
\inferrule{\strut}{c \eqxat{\lbl} c}\\
\inferrule{\av_1 \eqxat{\lbl} \av_1'\\ \av_2 \eqxat{\lbl} \av_2'}
{(\av_1,\av_2) \eqxat{\lbl} (\av_1',\av_2')}
\and
\inferrule{\av \eqxat{\lbl} \av'}
{\vinl{\av} \eqxat{\lbl} \vinl{\av'}}
\and
\inferrule{\av \eqxat{\lbl} \av'}
{\vinr{\av} \eqxat{\lbl} \vinr{\av'}}
\and
\inferrule{\av \eqxat{\lbl} \av'}
{\vroll{\av} \eqxat{\lbl} \vroll{\av'}}
\and
\inferrule{\gamma \eqxat{\lbl} \gamma'}
{\vclos{\kappa}{\gamma}\eqxat{\lbl} \vclos{\kappa}{\gamma'}}
\\
\gamma \eqxat{\lbl} \gamma' \iff \forall x \in \dom(\gamma) \cup\dom(\gamma').~ \gamma(x) \eqxat{\lbl} \gamma'(x)
 \end{smathpar}
  \caption{Equal-except-at relation}
  \label{fig:eq-except-at}
\end{figure}
%%% Local Variables: 
%%% mode: latex
%%% TeX-master: "paper-tr"
%%% End: 

Then we can show:
\begin{theorem}\label{thm:dep}
  Suppose $|\agamma|,e \red v,T$ and $\agamma' \eqxat{\lbl} \agamma$
  and $|\agamma'|,e \eval v',T'$.  Then we have $\dep(T,\agamma)
  \eqxat{\lbl} \dep(T',\agamma')$.
\end{theorem}
 \begin{proof}
    See \appref{dep-proof}.
  \end{proof}
This says that the label of a value in the input propagates to all
parts of the output where changing the value can have an impact on the
result.  

  \begin{example}
    Revisiting the previous example, dependency-correctness has
    several implications for output $[2^{L_1,L},2^{L_2,L},
    4^{L_3,L}]$.  Taking $\lbl=L$, dependency-correctness tells us
    that if the value of $y$ were changed, all three parts of the
    output list might change.  If $\lbl=L_1$, then
    dependency-correctness implies that the output will be of the form
    $[v,2,4]$ for some value $v$ (which by type-safety must be an
    integer also).  Similarly, dependency-correctness implies that if
    $L_2$ or $L_3$ change then the output will be of the form $[2,v,4]$
   or $[2,2,v]$ respectively.  

   Dependency-correctness does not provide a guarantee concerning the
   effects of multiple, independent changes at different locations
   (although an approximation of this information is available by
   using a single location that includes both changes).  Also, in all cases,
   the structure of the output list cannot change, because we cannot
   change the length of the input list by changing $y$ or changing the
   values of elements of the list.
  \end{example}

  \begin{example}\label{ex:first-argument-not-always-bot}
    Consider a trace $T=\mathtt{fst}(x)$ evaluated in environment
    $\gamma = [x\mapsto (1^a,2^b)^c]$, whose result is
    $\mathsf{D}(T,\gamma) = 1^{D_{1}(\{a\},\{b\})} = 1^{a,b}$.  This
    can naturally happen if the pair is a part of the (annotated)
    input, rather than being constructed by the program.  This example
    illustrates that although the annotations of pairs, sum
    injections, and other constructors are hard-wired to be $\bot$,
    this does not mean that the binary functions
    $\extract_1,\extract_2$ etc. are always called with $\bot$ as the
    first argument.
\end{example}

\begin{remark}
  Cheney et al.~\cite{cheney11mscs} considered a query language with
  nested collection types (similar to that used by Buneman et
  al.~\cite{buneman08tods}).  That language included an equality
  operation at all types, and the dependency provenance semantics for
  equality made use of another operation $\|\av\|$ that collects all
  of the annotations in $\av$.  Here, we only consider equality at
  base types, and so it suffices to consider only local annotations
  during propagation.
\end{remark}

\section{Disclosure and obfuscation analysis}
\label{sec:analysis}

In the previous sections we have defined a trace model for $\plz$ and
defined certain classes of trace queries, provenance views, and
introduced technical machinery such as paths and partial values.  In
this section we put these components to work by investigating the
disclosure and obfuscation problems for $\plz$ traces and provenance
views.  We confine attention to queries that test properties of the
input or output.  Investigating queries that capture properties of the
trace is more difficult, since traces involve variable binding,
whereas for values we have restricted attention to queries formulated
in terms of partial values.

\subsection{Patterns, partial traces, and trace queries}

  In section ~\ref{sec:background}, we reviewed and refined a
  general provenance framework with definitions of disclosure and
  obfuscation, formulated in terms of abstract sets of traces.  We now
  introduce additional concepts needed to formulate the \pl model of
  provenance as an instance of the abstract provenance framework in
  section~\ref{sec:background}, so that we can analyze the security properties of
  \pl-traces.  Specifically, we will consider a consistent triple
  $(\gamma,T,v)$ as an abstract trace, we will define some provenance
  queries over such traces, and we will consider some approaches to
  defining provenance views of the traces.  The queries and views rely
  on notions of patterns and partial traces; specifically, we will
  consider queries based on testing whether a partial value is present
  in the input or output, and we will consider views based on deleting
  information from the trace, input or output.  

We introduce patterns for values, environments and traces.  The syntax
of patterns (pattern environments) is similar to that of values
(respectively environments), extended with special \emph{holes}:
\begin{eqnarray*}
  p &::=& \vc \mid \vpair{p_1}{p_2} \mid \vinl{p} \mid \vinr{p} \mid
  \vroll{p} \mid \vclos{\kappa}{\rho} \mid \vany \mid \vhole
\\
\rho &::=& [x_1 \mapsto p_1,\ldots,x_n \mapsto p_n]
\end{eqnarray*}
Patterns actually denote binary relations on values.  The hole symbol
$\vhole$ denotes the total relation, while the exact-match symbol
$\vany$ denotes the identity relation.  The $\vany$ pattern is a
technical device used later in this section in backward disclosure slicing; we
sometimes refer to \emph{$\vany$-free patterns} that do not contain
$\vany$.

We say that $v$ matches $v'$ modulo $p$ (written $v \eqat{p} v'$) if
$v$ and $v'$ match the structure of $p$, and are equal at
corresponding positions denoted by $\vany$.  Moreover, we write $p
\sqcup p'$ for the least upper bound (join) of two patterns and define
$p \sqleq p'$ to hold if $p' = p \sqcup p'$.  Rules defining
$\eqat{p}$ and $\sqcup$ are given in Figures~\ref{fig:eqat}
and~\ref{fig:lub}.  In the equations defining $p \sqcup
\vany$, we use notation $p[\vany/\vhole]$ to denote the result of
replacing all occurrences of $\vhole$ in $p$ with $\vany$.

When $p \sqsubseteq v$, we write $v|_p$ for the pattern obtained by
replacing all of the $\vany$-holes in $p$ with the corresponding
values in $v$, defined as follows:
\begin{eqnarray*}
  v|_\vhole &=& \vhole\\
v|_\vany &=& v\\
\vc|_\vc &=& \vc\\
(v_1,v_2)|_{(p_1,p_2)} &=& (v_1|_{p_1},v_2|_{p_2})\\
\vinl{v}|_{\vinl{p}} &=& \vinl{v|_p}\\
\vinr{v}|_{\vinr{p}} &=& \vinr{v|_p}\\
\vroll{v}|_{\vroll{p}} &=& \vroll{v|_p}\\
\vclos{\kappa}{\gamma}|_{\vclos{\kappa}{\rho}}&=&
\vclos{\kappa}{\gamma|_\rho}\\
{}[x_1 \mapsto v_1, \ldots,x_n \mapsto v_n]|_{[x_1 \mapsto p_1,
  \ldots,x_n \mapsto p_n]} &=& [x_1 := v_1|_{p_1},\ldots, x_n :=v_n|_{p_n}]
\end{eqnarray*}
For example, $(1,2) |_{(\vany,\vhole)} = (1,\vhole)$.

\begin{lemma}\label{lem:eqat-properties}
  For any $v$ and $p \sqleq v$, we have $v \eqat{p} v|_p$.
\end{lemma}
\begin{lemma}
      The set $\{p \mid p \sqleq v\}$ of partial values matching a given
    value $v$ is an upper semilattice with least element $\hole$,
    greatest element $v$ and the least upper bound operation $\sqcup$
    as defined in \figref{lub}.
  \end{lemma}
\begin{lemma}
  For each $p$, $ \eqat{p}$ is a partial equivalence relation.  Moreover,
  whenever $p \sqcup p'$ is defined, we have $(\eqat{p \sqcup p'}) =
  (\eqat{p}) \cap (\eqat{p'})$, and whenever $p \sqleq p'$ we have
  $(\eqat{p}) \supseteq (\eqat{p'})$.
 \end{lemma}
 \begin{proof}
   To show $\eqat{p}$ is a partial equivalence relation, we must show
   that it is symmetric and transitive (but not necessarily
   reflexive).  Symmetry follows by straightforward induction on
   derivations.  Transitivity is by induction on $p$.  The second part
   follows by induction on the (partial recursive) definition of
   $\sqcup$.  The third part follows from the second by the fact that
   $p \sqleq p'$ holds if and only if $p \sqcup p' = p'$.  Full
   details of the proof are in \appref{eqat-properties-proof}.
 \end{proof}
 \begin{lemma}\label{lem:eqat-and}
   If $v \eqat{p} v'$ then $p \sqleq v$ and $p
   \sqleq v'$.  Conversely, if $p \sqleq v$ then $v \eqat{p} v$.
 \end{lemma}
 \begin{proof}
   By induction on derivations.  By symmetry it suffices to show by induction that $v
   \eqat{p} v'$ implies $p \sqleq v$.   The second part is straightforward. 
 \end{proof}

% It is worth noting that $\eqpat{p}$ actually says something stronger
% than the conclusion of the above lemma: namely, it says that $v$ and
% $v'$ match or mismatch $p$ in ``the same way''.  For example, $(2,2)
% \eqpat{(3,3)} (4,4)$  holds, because both corresponding components of
% pairs differ from the pattern, but $(2,3) \eqpat{(3,3)} (3,4)$ does not
% hold, because different parts of the two values match or mismatch the
% pattern.  

\begin{figure}[tb]
\fbox{$v \eqat{p} v'$}
\vspace{-3mm}
\begin{smathpar}
\inferrule*{
\strut
}
{v \eqat{\vhole} v'}
\and
\inferrule*
{
\strut
}
{
v \eqat{\vany} v
}
\and
\inferrule*
{
\strut
}
{
\vc \eqat{\vc} \vc
}
\and
\inferrule*
{
  v_1 \eqat{p_1} v_1'
  \\
  v_2 \eqat{p_2} v_2'
}
{
  \vpair{v_1}{v_2} \eqat{(p_1,p_2)} \vpair{v_1'}{v_2'}
}
\and
\inferrule*{v \eqat{p} v'}
{\vinl{v} \eqat{\vinl{p}} \vinl{v'}}
\and
\inferrule*{v \eqat{p} v'}
{\vinr{v} \eqat{\vinr{p}} \vinr{v'}}
\and
\inferrule*{v \eqat{p} v'}
{\vroll{v} \eqat{\vroll{p}} \vroll{v'}}
\and
\inferrule*{
%\kappa \eqat{\kappa_0} \kappa'\\
\gamma \eqat{\rho}\gamma'
}
{
\vclos{\kappa}{\gamma} \eqat{\vclos{\kappa}{\rho}} \vclos{\kappa}{\gamma'}
}
\end{smathpar}
\begin{eqnarray*}
  \gamma \eqat{\rho} \gamma' &\iff &\forall x \in
  \dom(\rho). ~\gamma(x) \eqat{\rho(x)} \gamma'(x)
%\\
 % \gamma \eqat{\vany} \gamma' &\iff & \gamma = \gamma'\\
 % \gamma \eqat{\vhole} \gamma'& \iff &\top
\end{eqnarray*}
\caption{Equality modulo patterns.}
\label{fig:eqat}
\end{figure}

%%% Local Variables: 
%%% mode: latex
%%% TeX-master: "main"
%%% End: 

\begin{figure}[tb]
  \[\small \begin{array}{rcl}
    \vhole \sqcup p = p \sqcup \vhole &=& p 
\\[1mm] 
    \vany \sqcup p = p \sqcup \vany &=& p[\vany/\vhole]\\[1mm]
   \vpair{p_1}{p_2} \sqcup \vpair{p_1'}{p_2'} &=& \vpair{p_1\sqcup
      p_1'}{p_2\sqcup p_2'} \\[1mm]
 \vc \sqcup \vc &= &\vc \\[1mm]
    \vinl{p} \sqcup \vinl{p'} &=& \vinl{p\sqcup p'} \\[1mm]
   \vinr{p} \sqcup \vinr{p'} &=& \vinr{p\sqcup p'} \\[1mm]
    \vroll{p} \sqcup \vroll{p'} &=& \vroll{p\sqcup p'} \\[1mm]
\vclos{\kappa}{\rho} \sqcup \vclos{\kappa}{\rho'} &=& \vclos{\kappa}{\rho \sqcup \rho'}\\[1mm]
    (\rho \sqcup \rho')(x) &=& \left\{
      \begin{array}{ll}
        \rho(x) \sqcup \rho'(x) & x \in \dom(\rho) \cup \dom(\rho')\\
        \rho(x) & x \in \dom(\rho) \backslash \dom(\rho')\\
        \rho'(x) & x \in \dom(\rho') \backslash \dom(\rho)
     \end{array}\right.
 \end{array}\]
\caption{Least upper bounds of patterns and environments.}
\label{fig:lub}
\end{figure}

We also consider partial traces, usually written $S$, which are trace
expressions where some subexpressions have been replaced with $\hole$:
\begin{eqnarray*}
  S &::=& \cdots \mid \tremp
\end{eqnarray*}
As with patterns, we write $S \sqsubseteq T$ to indicate that $T$
matches $S$, that is, $S$ can be made equal to $T$ by filling in some
holes.

As mentioned at the beginning of this section, for the purpose of
disclosure and obfuscation analysis, we will consider the ``traces''
to be triples $(\gamma,T,v)$ where $T$ is consistent with $\gamma$ and
$v$, that is, $\gamma,T \trrun v$.  We refer to such a triple as a
\emph{consistent triple}.  
We consider trace or provenance queries built out of partial values
and partial traces.

\begin{definition}
  \begin{enumerate}
  \item Let $\phi(\gamma)$ be a predicate on input environments.  An
    \emph{input query} $\IN{\gamma.\phi(\gamma)}$ is defined as
    $\{(\gamma,T,v) \mid \gamma,T \trrun v \text{ and }
    \phi(\gamma)\}$.  (Here, $\IN$ binds
    $x$ in  $\phi(x)$.)  As a special case, we write $\IN_\rho$ for
    $\IN{\gamma.(\rho \sqsubseteq \gamma)}$.  
  \item Let $\phi(v)$ be a predicate on output values.  An
    \emph{output query} $\OUT{v.\phi(v)}$ is defined as $ \{(\gamma,T,v)
    \mid \gamma,T \trrun v \text{ and } \phi(v)\}$.  (Here, $\OUT$
    binds $x$ in $\phi(x)$.) As a special
    case, we write $\OUT_p$ for $ \OUT{v.(p \sqsubseteq v)}$.
  \end{enumerate}

      \begin{remark}
    The $\IN{\gamma.\phi(\gamma)}$ and $\OUT{v.\phi(v)}$ notations are
    chosen to resemble quantifiers; they should be read as ``In the
    input of the trace, $\phi$ holds'' or ``In the output of
    the trace, $\phi$ holds''.  One can also think of them as
    higher-order functions, for example $\mathsf{IN} : (\Env \to \Bool)
    \to \Traces \to \Bool$ and $\mathsf{OUT} : (\Val \to \Bool) \to
    \Traces \to \Bool$, and regard $\IN{\gamma.\phi(\gamma)}$ and
      $\OUT{v.\phi(v)}$ respectively as syntactic sugar for
      $\mathsf{IN}(\Lambda \gamma.\phi(\gamma)$ and
      $\mathsf{OUT}(\Lambda v. \phi(v))$.
    \end{remark}
  % \item A \emph{trace query} $\TRACE_S$ is defined as $\TRACE_S =
%   \{(\gamma,T,v) \mid S \subseteq v\}$.
% \item 
%   Suppose $\phi$ is a binary predicate relating annotated environments
%   and annotated values.  A \emph{provenance query}
%   $\PROV_{\extract,\phi}$ is defined as $\PROV_{\extract,\phi} =
%   \{(\gamma,T,v) \mid \phi(F(T,\hat{\gamma}),\hat(\gamma))\}$.
\end{definition}

  To analyze forms of provenance based on annotation (as considered in
  Section~\ref{sec:extraction}) we will also consider
  consistent annotated triples $(\agamma,T,\av)$ where
  $\av = \extract(T,\agamma)$. We will later also consider
  corresponding queries derived from different forms of provenance,
  based on annotated triples.

\subsection{Disclosure}

We first consider properties disclosed by various forms of provenance
considered above.  Both where-provenance and expression provenance 
disclose useful information about the input.  Dependency provenance
does not disclose input information in an easy-to-analyze way, but is useful
for obfuscation, as discussed in Section~\ref{sec:obfuscation}.

For where-provenance, we consider input queries
\[\IN^\where_{v_0,\pi} = \IN{\gamma.~ (\gamma[\pi] = v_0)})\] 
and output queries
\[\OUT^\where_{v_0,\pi} = \OUT{\hat{v}.~ (w^\pi \in
  \occ(\av) \wedge |w| = v_0)}\]
where $\pi$ is a path and $v_0$ is a value.  Such a query
tests whether $\gamma$ or $v$ contains a value $v_0$ with the
provided annotation.  

\begin{theorem}
  The where-provenance view $(\gamma,T,v) \mapsto \where(T,\path(\gamma))$
  positively discloses $\IN^\where_{v_0,\pi}$ via $\OUT^\where_{v_0,\pi}$.
\end{theorem}
\begin{proof}
  If $\OUT^\where_{v_0,\pi}$ holds of $\where(T,\path(\gamma))$ then by
  Theorem~\ref{thm:where} we know that $\gamma$ contains a copy of $v_0$
  annotated by $\pi$, hence $\IN^\where_{v_0,\pi}$ holds of $\path(\gamma)$.
\end{proof}

For expression-provenance, suppose $t$ is an expression
  over paths (that is, locations $\lbl$ in $t$ are paths $\pi$).  We
define $\gamma(t)$ to be the result of evaluating $t$ in $\gamma$ with
all paths $\pi$ replaced by their values $\gamma[\pi]$ in $\gamma$.
This is defined as follows:
\begin{eqnarray*}
  \gamma(x.\pi) &=& \gamma(x)[\pi]\\
  \gamma(\vc) &=& \vc\\
  \gamma(\oplus(t_1,\ldots,t_n)) &=& \hat{\oplus}(\gamma(t_1),\ldots,\gamma(t_n))\\
  \gamma(\bot) &=& \bot
\end{eqnarray*}

We consider queries
$\IN^\expr_{v_0,t} = \IN{\gamma.~(\gamma(t)=v_0)}$, where $t$ is an expression provenance
annotation and $v_0$ is a value.  Such a query tests whether
evaluating an expression $t$ over $\gamma$ yields the specified value.
For example, $\IN{\gamma.~x.1+y.2=4}$ holds for $\gamma=
[x=(1,2),y=(2,3)]$, because $\gamma(x.1) + \gamma(y.2) = 1 +3 = 4$.
We also consider output queries $\OUT^\expr_{v_0,t} =
\OUT{\av.~(w^t \in \occ(\av) \wedge |w| = v_0)}$, that simply test whether an annotated
copy of $v_0$ appears in the output with annotation $t$.
\begin{theorem}
  The expression-provenance view $(\gamma,T,v) \mapsto \expr(T,\path(\gamma))$
  positively discloses $\IN^\expr_{v_0,t}$ via $\OUT^\expr_{v_0,t}$.
\end{theorem}
\begin{proof}
  Similarly to where-provenance, using Theorem~\ref{thm:expr} we show
  that if $\OUT^\expr_{v_0,t}$ holds on $\expr(T,\gamma)$ then $\gamma(t) =
  v_0$, which implies $\IN^\expr_{v_0,t}$.
\end{proof}
  For example, if the annotated output is $42^{x + y}$, then we know
  that the annotated value is equal to the sum of $\gamma(x)$ and
  $\gamma(y)$, but we do not know anything more about the values of $x$
  and $y$ beyond the equation $x+y=42$.  However, if the output is
  $(42^{x+y}, 17^{x-y})$ then we know that $x$ and $y$ are the
  (unique) solution to the linear equations
  \begin{eqnarray*}
    x + y &=& 42\\
    x - y &=& 17\,,
  \end{eqnarray*}
  that is, $x = 29.5, y=12.5$.

Expression provenance and where-provenance are also related in the
following sense:
\begin{theorem}
  Where-provenance is computable from expression-provenance.
\end{theorem}
\begin{proof}
  Where-provenance annotations can be extracted from
  expression-provenance annotations by mapping locations $\lbl$ to
  themselves and all other expressions to $\bot$.  
\end{proof}
Hence, any query
disclosed by where-provenance is disclosed by expression-provenance,
and any query obfuscated by expression-provenance is also obfuscated
by where-provenance.

\begin{figure}[tb!]
\fbox{$p,T \bwdslice S,\rho$}
\vspace{-3mm}\\
\begin{smathpar}
      \inferrule*
      {
        \strut
      }
      {
        \vhole, T \bwdslice \tremp,[]
      }
\and
\inferrule*
      {
\strut     }
      {
        p, \trvar{x} \bwdslice \trvar{x}, [x \mapsto p]
      }
\and      \inferrule*{
        \strut
      }
      {
        \vc,\trc \bwdslice \trc , []
      }
\and
    \inferrule*{\strut}
      {
        \vclos{\kappa}{\rho},\trfunk{\kappa} \bwdslice \trfunk{\kappa},\rho
      }
% \and
%     \inferrule*{\kappa' \neq \kappa}
%       {
%         \vclos{\kappa'}{\rho},\trfunk{\kappa} \bwdslice \trfunk{\kappa},\tremp
%       }
%
\and
      \inferrule*
      {
        p_2, T_2 \bwdslice S_2,\rho_2[x\mapsto p_1]\\
        p_1,T_1 \bwdslice S_1,\rho_1
      }
      {
        p_2, \trlet{T_1}{x}{T_2} 
        \bwdslice 
        \trlet{S_1}{x}{S_2}, \rho_1 \sqcup \rho_2
      }
\and
      \inferrule*{
       \vany,T_1 \bwdslice S_1,\rho_1
        \quad \cdots \quad
        \vany,T_n \bwdslice S_n,\rho_n
      }
      {
        p,\trf{T_1,\ldots,T_n} \bwdslice \trf{S_1,\ldots,S_n},\rho_1 \sqcup
        \cdots \sqcup \rho_n
      }
      \and
      \inferrule*
      {
        p_1, T_1 \bwdslice S_1 , \rho_1
        \\
        p_2, T_2 \bwdslice S_2 , \rho_2
      }
      {
        (p_1,p_2),\trpair{T_1}{T_2} 
        \bwdslice 
        \trpair{S_1}{S_2}, \rho_1 \sqcup \rho_2
      }
\and
      \inferrule*
      { \vpair{p}{\hole}, T \bwdslice S , \rho
      }
      {
        p,\trfst{T} \bwdslice \trfst{S}, \rho
      }
\and
      \inferrule*
      {
               \vpair{\hole}{p}, T \bwdslice S, \rho
      }
      {
        p,\trsnd{T} \bwdslice \trsnd{S}, \rho
      }
\and
     \inferrule*{p,T \bwdslice S, \rho}
      {
        \vinl{p},\trinl{T} \bwdslice \trinl{S},{\rho}
      }
%    \and
%       \inferrule*{p,T \bwdslice S, \rho}
%       {
%         \vinl{p},\trinr{T} \bwdslice \trinr{\tremp},{[]}
%       }
\and
      \inferrule*{p,T \bwdslice S, \rho}
      {
        \vinr{p},\trinr{T} \bwdslice \trinr{S},{\rho}
      }
% \and
%       \inferrule*{p,T \bwdslice S, \rho}
%       {
%         \vinr{p},\trinl{T} \bwdslice \trinl{\tremp},{[]}
%       }
 \and
      \inferrule*{p,T \bwdslice S, \rho}
      {
        \vroll{p},\trroll{T} \bwdslice \trroll{S},{\rho}
      }
  \and
      \inferrule*{\trroll{p},T \bwdslice S, \rho}
      {
        p,\trunroll{T} \bwdslice \trunroll{S},\rho
      }
\and
    \inferrule*
      {
       p_1, T_1 \bwdslice S_1, \rho_1[x_1\mapsto p]
       \\
        \vinl{p}, T \bwdslice S, \rho
      }
      {
       p_1, \trcaseml{m}{T}{x_1}{T_1}
        \bwdslice 
         \trcaseml{m}{S}{x_1}{S_1},\rho \sqcup \rho_1
      }
    \and
      \inferrule*
      {
       p_2, T_2 \bwdslice S_2, \rho_2[x_2\mapsto p]
        \\
        \vinr{p}, T \bwdslice S,\rho
      }
      {
        p_2, \trcasemr{m}{T}{x_2}{T_2}
        \bwdslice 
       \trcasemr{m}{S}{x_2}{S_2}, \rho \sqcup \rho_2
      }
\and
      \inferrule*
      {
       p, T \bwdslice S, \rho[f \mapsto p_1, x \mapsto p_2]
        \\
        p_1 \sqcup \vclos{\kappa}{\rho}, T_1 \bwdslice S_1,\rho_1
        \\
        p_2,T_2 \bwdslice S_2,\rho_2     }
      {
       p, \trappk{\kappa}{T_1}{T_2}{f}{x}{T} 
        \bwdslice 
        \trappk{\kappa}{S_1}{S_2}{f}{x}{S} ,\rho_1 \sqcup \rho_2
      }
\and
     \inferrule*{fvs(\kappa) = \{x_1,\ldots,x_n\}}
      {
        \vany,\trfunk{\kappa} \bwdslice \trfunk{\kappa},[x_1 \mapsto
        \vany,\ldots,x_n \mapsto \vany]
      }
     \and
      \inferrule*
      {
        \vany, T_1 \bwdslice S_1 , \rho_1
        \\
        \vany, T_2 \bwdslice S_2 , \rho_2
      }
      {
        \vany,\trpair{T_1}{T_2} 
        \bwdslice 
        \trpair{S_1}{S_2}, \rho_1 \sqcup \rho_2
      }
   \and
     \inferrule*{\vany,T \bwdslice S, \rho}
      {
        \vany,\trinl{T} \bwdslice \trinl{S},{\rho}
      }
    \and
      \inferrule*{\vany,T \bwdslice S, \rho}
      {
        \vany,\trinr{T} \bwdslice \trinr{S},{\rho}
      }
   \and       \inferrule*{\vany,T \bwdslice S, \rho}
      {
        \vany,\trroll{T} \bwdslice \trroll{S},{\rho}
      }
\end{smathpar}
\caption{Disclosure slicing.}
\label{fig:disclosure-slice}
\end{figure}

%%% Local Variables: 
%%% mode: latex
%%% TeX-master: "main"
%%% End: 

We now consider a form of \emph{trace slicing} that takes a partial
output value and removes information from the input and trace that is
not needed to disclose part of the output.  We show that such \emph{disclosure
  slices} also disclose generic provenance views
(Theorem~\ref{thm:extraction-from-slices}).  Thus, disclosure slices form a
quite general form of provenance in their own right.

\begin{definition}
  Let $\gamma,T \trrun v$, and suppose $S \sqsubseteq T$ and $\rho
  \sqsubseteq \gamma$.  We say $(\rho,S)$ is a \emph{disclosure slice}
  with respect to partial value $p$ if for all $\gamma' \sqsupseteq
  \rho$ and $T' \sqsupseteq S$ such that if $\gamma',T' \trrun v'$, we
  have $p \sqsubseteq v$ iff $p \sqsubseteq v'$.
\end{definition}
The intuition is that a disclosure slice should contain enough
information that any replay of a (completed) trace 
on a (completed) input environment (both extending the respective
components of the slice) yields a result that matches $p$; in
other words, the slice is a (possibly smaller) ``witness'' to the
construction of $p$ from the input. 
Note that by this definition, minimal disclosure slices exist (since
there are finitely many slices) but need not be unique.  For example,
both $\hole \vee \kwtrue$ and $\kwtrue \vee \hole$ are disclosure slices
showing that $\kwtrue \vee \kwtrue$ evaluates to $\kwtrue$, but $\hole
\vee \hole$ is not a disclosure slice.

Figure~\ref{fig:disclosure-slice} shows rules defining a disclosure
slicing judgment $p,T \bwdslice S,\rho$.  Basically, the idea is to
push a partial value backwards through a trace to obtain a partial
input environment and trace slice.  The partial input environment is
needed to handle local variables in traces.  In the rule for $\kwlet$,
we first slice through the body of the let, then identify the partial
value showing the needed parts of the let-bound value, and use that to
slice backwards through the first subtrace.  

\begin{example}
      To illustrate the behavior of $\kwlet$ and bound variables,
    consider:
  \[
  \inferrule*{ \inferrule*{ \inferrule*{ (\vhole,1),x \bwdslice x, [x
        \mapsto \vpair{\vhole}{1}] }{ 1,\trsnd{x} \bwdslice  \trsnd{x},
        [x\mapsto \vpair{\vhole}{1}]
      }\\
      \vhole,\trfst{x} \bwdslice \vhole,[]} {\vpair{1}{\vhole},
      \trpair{\trsnd{x}}{\trfst{x}}\bwdslice \trpair{\trsnd{x}}{\vhole},[x\mapsto
      \vpair{\vhole}{1}]
    }\\
    \inferrule*{
      \vhole,y \bwdslice \vhole,[]\\
      1,z \bwdslice z,[z\mapsto 1] }{ \vpair{\vhole}{1},(y,z)
      \bwdslice (\vhole,z),[z \mapsto 1] } }{ \vpair{1}{\vhole} ,
    \trlet{(y,z)}{x}{\trpair{\trsnd{x}}{\trfst{x}}} \bwdslice
    \trlet{(\vhole,z)}{x}{\trpair{\trsnd{x}}{\vhole}},[z\mapsto 1] }
  \]
\end{example}

  Slicing for conditionals (case expressions) follows a similar
  pattern to let-binding.  For example, consider a trace
  $\trcaseml{m}{T}{x_1}{T_1} $ indicating that case expression $m$ was
  executed with argument computed by $T$, evaluating to some value
  $\vinl{v'}$ and the trace of the body of the $\kwinl$-branch was
  $T_1$.  If we wish to slice this with respect to $p$, then we first
  slice the trace of the taken branch $T_1$ with respect to $T$,
  yielding trace slice $S_1$ and pattern environment
  $\rho_1[x_1\mapsto p]$.  Since we know that the result of $T$ must
  been of the form $\vinl{v'}$, we slice $T$ with respect to
  $\vinl{p}$, yielding slice $S$ and pattern environment $\rho$.  The
  final result is slice $\trcaseml{m}{S}{x_1}{S_1}$ with pattern
  environment $\rho \sqcup \rho_1$. 

  Slicing for application traces is similar to slicing for let and
  case constructs, but more complex due to
  the need to propagate partial values backwards through closure
  environments.  Specifically, a trace of the form
  $\trappk{\kappa}{T_1}{T_2}{f}{x}{T}$ is sliced with respect to
  output pattern $p$ as follows.  First, $T$ is sliced with respect to
  $p$, yielding slice $S$ and partial environment $\rho[f \mapsto
  p_1,x\mapsto v_2]$.  Here, $\rho$ is the part of the closure needed
  to rerun the body of the function call.  We then slice $T_1$, the
  subtrace that computed the called function, with respect to $p_1
  \sqcup \vclos{\kappa}{\rho}$, since we know from the trace that
  $\kappa$ was the called function and we know that the parts of the
  environment denoted $\rho$ were needed in the call, and we also know
  that $p_1$ was needed to apply recursive calls of $f$.  (In
  particular, if $f$ was not called recursively in $T$, then $p_1 =
  \vhole$.)  This yields a slice $S_1$ and pattern environment
  $\rho_1$.  We also slice $T_2$ with respect to $p_2$, obtaining a
  slice $S_2$ and pattern environment $\rho_2$ that show what part of
  the trace and input environment were needed to compute the function
  argument.  The final result is slice
  $\trappk{\kappa}{S_1}{S_2}{f}{x}{S}$ with pattern environment
  $\rho_1 \sqcup \rho_2$; any dependence on the environment in which
  the function closure was constructed is propagated to $\rho_1$ via
  the slicing subderivation for $T_1$.

Note also that the special $\vany$
patterns are used to slice backwards through primitive operations even
when we do not know the values of the inputs or results.
  This necessitates additional rules that deal with the cases where
  $p$ is $\vany$.
Another
possibility is to annotate the traces of primitive operations with
these values, an approach taken on related work on using traces for
program slicing~\cite{perera12icfp};
  however, this approach does not work as well in this setting since
  our disclosure slicing criterion involves replaying the trace on
  changed inputs.

\begin{lemma}
  If $\gamma,T \trrun v$ then for any $p \sqleq v$ there exists
  $S \sqsubseteq T$ and $\rho \sqsubseteq \gamma$ such that $p,T
  \bwdslice S,\rho$.  
\end{lemma}
  \begin{proof}
    The first part follows by induction on the structure of the
    derivation of $\gamma,T \trrun v$.  If $p = \vhole$ then the
    conclusion is immediate in any case. For each constructor case
    (pairs, $\kwinl$, $\kwinr$, $\kwfun$, $\kwroll$), if $p$ is not
    $\vhole$ then its toplevel constructor must match, so we can
    proceed by induction.  The other cases, for primitive operations,
    pair projection, cases, function application, and unroll, are
    straightforward because there is no restriction on $p$ (though we
    need to check that the invariant $p \sqleq v$ holds for the
    induction hypotheses).  
  \end{proof}

We define a function $\Disc_p(\gamma,T,v)$ on consistent triples
$(\gamma,T,v)$ as follows.
\[\Disc_p(\gamma,T,v) = \left\{ \begin{array}{ll}
(\gamma|_\rho,S)& \text{ if $p \sqleq v$ and $p,T \bwdslice
S,\rho$}\\
(\gamma|_\rho,S) &\text{ if $p \not\sqleq v$ and $\witness(p,v),T
  \bwdslice S,\rho$}
\end{array}\right.
\]
The idea is that when $p \sqleq v$, we slice using the
rules in Figure~\ref{fig:disclosure-slice} and then transform $\rho$
by filling in all $\vany$-holes with the corresponding values in
$\gamma$.  However, when $p \not\sqleq v$, we do not use $p$ to slice,
but instead use $\witness(p,v)$, a pattern that contains enough of $v$
to show how $v$ fails to match $p$.  The $\witness$ function is
defined in \figref{witness}.

\begin{figure}[tb]
  \begin{eqnarray*}
\witness(p,v) &=& \hole \quad \text{ if $p \sqleq v$}\\
    \witness(c',c) &=& c\\
    \witness((p_1,p_2),(v_1,v_2)) &=& \left\{
      \begin{array}{ll}
        (\witness(p_1,v_1),\vhole) & \text{if $p_1 \not\sqleq v_1$}\\
        (\vhole,\witness(p_2,v_2)) & \text{if $p_1 \sqleq v_1$ and $p_2 \not\sqleq v_2$}
      \end{array}\right.\\
\witness(p,(v_1,v_2)) &=& (\vhole,\vhole) \quad \text{otherwise}\\
    \witness(\vinl{p},\vinl{v}) &=& \vinl{\witness(p,v)}\\
    \witness(p,\vinl{v}) &=& \vinl{\vhole} \quad \text{otherwise}\\
    \witness(\vinr{p},\vinr{v}) &=& \vinr{\witness(p,v)}\\
    \witness(p,\vinr{v}) &=& \vinr{\vhole} \quad \text{otherwise}\\
    \witness(\vroll{p},\vroll{v}) &=& \vroll{\witness(p,v)}\\
    \witness(p,\vroll{v}) &=& \vroll{\vhole} \quad \text{otherwise}\\
    \witness(\vclos{\kappa}{\rho},\vclos{\kappa}{\gamma}) &=&
    \vclos{\kappa}{[x_1 \mapsto
      \witness(\rho(x_1),\gamma(x_1)),\ldots,x_n \mapsto
      \witness(\rho(x_n),\gamma(x_n))]}\\
&& \quad \text{ where $\dom(\rho)
      = \dom(\gamma) =\{x_1,\ldots,x_n\}$}\\
    \witness(\vhole,\vclos{\kappa}{\gamma}) &=&
    \vclos{\kappa}{[]} \quad \text{otherwise}
  \end{eqnarray*}
\caption{Witness function}\label{fig:witness}
\end{figure}

  \begin{example}
Recall the running example $map~f~xs$ where 
\[\gamma = [f \mapsto \exfun{f}{x}{\exif{x=y}{y}{x+1}},xs =
  [1,2,3], y=2]\] and yielding result $v = [2,2,4]$.  We write $a::l$
  for the list construction operator, that is, $[2,2,4] =
  2::2::4::[]$. Let $T$ be the trace obtained by running this example,
  i.e. $\gamma,map~f~xs\eval [2,2,4],T$.
\begin{itemize}
\item If $p = [\vhole,\vhole,\vhole]$ then $\Disc_p(\gamma,T) =
  ([f\mapsto\vhole,xs\mapsto [\vhole,\vhole,\vhole], y\mapsto \vhole]  ,S)$ where $S$ shows three
  recursive calls to $map$, each with a partial trace of $f$.
\item If $p = []$ then $\Disc_p(\gamma,T) = ([xs \mapsto \vhole ::
  \vhole],S)$ where $S$ shows one recursive call to $map~f~xs$ in
  which $xs$ is inspected and found to be of the form $v::vs$, the
  corresponding branch is taken and a nonempty list is constructed.
\item If $p = [2,\vhole,\vhole]$ then $\Disc_p(\gamma,T) = ([f\mapsto
  \exfun{f}{x}{\exif{x=y}{y}{x+1}},y \mapsto 2],S)$ where $S$ lists
    all three calls to $map~f~xs$, two partial calls to $f$ and one complete call to $f$ on $1$.
  \item  If $p = [\vhole,3,\vhole]$ then $\Disc_p(\gamma,T) =
    ([f\mapsto \exfun{f}{x}{\exif{x=y}{y}{x+1}},xs mapsto
      \vhole::2::\vhole,y \mapsto 2],S)$ where $S$ lists two
      calls to $map$, one partial call to $f$ and one  complete call to $f$ on $2$.
\end{itemize}
  \end{example}

\paragraph{Correctness of disclosure slicing.}
  We now establish the key properties of disclosure slicing,
  culminating in the main result that $\Disc_p$ discloses the output
  query $\OUT_p$ (Theorem~\ref{thm:disclosure-slicing-correct}).
\begin{lemma}
  \label{lem:witness-correct}
  If $p \not\sqleq v$ then $p \not\sqleq \witness(p,v)$; moreover, for
  any $v' \sqgeq \witness(p,v)$ we have $p
  \not\sqleq v'$.
\end{lemma}
  \begin{proof}
  The first part follows by induction on the structure of $v$,
  with secondary case analysis on the possible forms of $p$.  The second part is immediate.
  \end{proof}

The witness function can be replaced by any other function that has
this property (for example, we could alter $\witness$ to find a
minimum-size pattern witnessing $p \not\sqleq v$.)

Recall the definition of $v
\eqat{p} v'$ as shown in Figure~\ref{fig:eqat}.  Using this relation,
we can prove the correctness of the slicing relation as follows:
% \begin{lemma}\label{lem:slicing-correct}
%   Assume $\gamma,T \trrun v$ and $p,T \bwdslice S,\rho$.
% %   \begin{enumerate}
% %   \item
%     Then if $p \sqleq v$ then for all $\gamma' \eqpat{\rho} \gamma$ 
%     and $T' \sqsupseteq S$, if $\gamma',T' \trrun v'$ then $v' \eqpat{p}
%     v$.
% %   \item If $p \not\sqsubseteq v$ then for all $\gamma' \eqat{\rho}
% %     \gamma$ and $T' \sqsupseteq S$, if $\gamma',T' \trrun v'$ then $p
% %     \not\sqsubseteq v'$.
% %\end{enumerate}
% \end{lemma}
% \begin{proof}
% See \appref{slicing-correct-proof}.
% \end{proof}
\begin{lemma}\label{lem:slicing-correct}
  Assume $\gamma,T \trrun v$ and $p,T \bwdslice S,\rho$ where $p
  \sqleq v$.
    Then for all $\gamma' \eqat{\rho} \gamma$ 
    and $T' \sqsupseteq S$, if $\gamma',T' \trrun v'$ then $v' \eqat{p}
    v$.
\end{lemma}
\begin{proof}
See \appref{slicing-correct-proof}.
\end{proof}

Correctness follows as a consequence of the above property.  To
simplify the argument, we prove positive and negative disclosure
simultaneously using an auxiliary query on the provenance view.
Specifically, we define a function on sliced traces and environments
called $\REPLAY$ that, intuitively, computes a plausible output for the slice.
  Given slice $(\rho,S)$, we define the auxiliary function
  $\REPLAY$ as follows:
  \[
  \REPLAY(\rho,S) = choose(\{v' \mid \exists \gamma' \sqgeq \rho,T' \sqgeq
  S. ~\gamma',T' \trrun v'\})
  \]
  In other words, $\REPLAY(\rho,S)$ chooses one of the possible values
  obtainable by replaying a complete trace extending $S$ on a complete
  environment extending $\rho$, if such a value exists; otherwise, the
  result is arbitrary.  Here, $choose: \mathcal{P}(\Val) \to \Val$ is
  a choice function such that if $X \subseteq \Val$ and $X \neq
  \emptyset$ then $choose(X) \in X$.  If $X = \emptyset$, then
  $choose(X)$ is an arbitrary value, say $42$.
  \begin{remark}
    Observe that this is not a constructive definition.  We can use
    the Axiom of Choice to define $choose$, or define a linear
    ordering on values to avoid appealing to the Axiom of Choice;
    however, it is not obvious whether $\REPLAY$ itself is
    computable.  In any case, $\REPLAY$ is only needed as a technical
    device to help define the intermediate provenance query used to
    prove positive and negative disclosure; we never need to try to
    compute it directly.
  \end{remark}
\begin{lemma}
  $\Disc_p$ negatively discloses $\OUT_p$ via $\OUT_p \circ \REPLAY$.
\end{lemma}
\begin{proof}
We prove the contrapositive.    Suppose $\OUT_p (\gamma,T,v)$ holds, that is, $p \sqleq v$.  Then
  let $(\rho,S) = \Disc_p(\gamma,T,v)$ be the computed slice, where
  $p,T \bwdslice S,\rho$, and suppose $v' = \REPLAY(\rho,S)$, where
  $T' \sqgeq S$ and $\gamma' \sqgeq \rho$ are the complete trace and
  environment used by $\REPLAY$ to compute $\gamma',T' \trrun v'$.
  Thus, we have $\rho \sqsubseteq \gamma$ and $\rho \sqsubseteq
  \gamma'$, which together with the fact that $\rho$ is $\vany$-free
  (by definition of $\Disc$) implies $\gamma \eqat{\rho} \gamma'$.  By
  \lemref{slicing-correct} and \lemref{eqat-properties} this implies
  $v \eqat{p} v'$ so $p \sqleq v'$.
\end{proof}
% % To prove full disclosure, we need a stronger property, based on an
% % equivalence relation that captures a richer invariant about the
% % slicing algorithm, namely that slices retain enough information to
% % certify that $p$ matched the result or that $p$ failed to match the result.

% % This requires some additional rules in the slicing algorithm, and an
% % extended pattern-equivalence relation $\eqpat{p}$ which intuitively
% % holds of $v$ and $v'$ if they match or mismatch $p$ at the same
% % points.  (For example, $(2,2) \eqpat{(3,3)} (4,4)$ holds but not
% % $(2,2) \eqpat{(2,4)} (4,4)$.) Once a mismatch occurs, we ignore the
% % rest of the respective values.
% % \begin{lemma}
% %   If $v \eqpat{p} v'$ then $p \sqleq v$ if and only if $p \sqleq v'$.
% % \end{lemma}

\begin{lemma}
    $\Disc_p$ positively discloses $\OUT_p$ via $\OUT_p \circ \REPLAY$.
\end{lemma}
\begin{proof}
  We prove the contrapositive.  Suppose that $\OUT_p(\gamma,T,v)$
  fails, that is, $p \not\sqleq v$.  We need to show that $\OUT_p
  (\REPLAY(\Disc_p(\gamma,T,v)))$ also fails.  Since $p \not\sqleq v$,
  we know that $\Disc_p(\gamma,T,v) = (\rho,S)$ where $\witness(p,v),T
  \bwdslice S,\rho$.  Thus, $\REPLAY(\rho,S) = v'$ for some $v'$
  obtained by replaying $T' \sqgeq S$ and $\gamma' \sqgeq \rho$, that
  is, $\gamma',T' \trrun v'$.  Since $\rho \sqleq \gamma$ and $\rho$
  is $\vany$-free we have that $\gamma \eqat{\rho} \gamma'$.  So, by
  \lemref{slicing-correct}, we know that $v \eqat{\witness(p,v)} v'$
  holds, which implies $\witness(p,v) \sqleq v'$,
  and by \lemref{witness-correct} this implies $p \not\sqleq v'$.
  This is what we need to show to conclude $\OUT_p
  (\REPLAY(\Disc_p(\gamma,T,v))) = 0$.
\end{proof}

Then by Proposition~\ref{prop:pos-neg} and the previous two lemmas we have:
\begin{theorem}\label{thm:disclosure-slicing-correct}
      $\Disc_p$ discloses $\OUT_p$.
\end{theorem}

  This is the main result about disclosure; we previously established
  some disclosure results for more restricted computational
  models~\cite{cheney11csf}, but this is the first such result for a
  general-purpose language.  It means that the disclosure slicing
  algorithm can be used to identify a subset of the trace that is
  large enough to recompute a part of the output, provided the parts
  of the input specified by $\rho$ remain fixed.  As noted elsewhere,
  this may not be a minimal slice, but it can be much smaller than the
  original trace.  In particular, let $(x,e)$ be a program where $e$
  is an arbitrarily complex expression, evaluated in a context with
  $x$ bound to 42. If all we care about is the first component of the
  result then the slice with respect to $(\vany,\vhole)$ is $([x \mapsto 42],(x,\vhole))$, which can be
  arbitrarily smaller than the full trace.

  This does not mean that there is no room for improvement in the
  disclosure slicing algorithm, for example by taking advantage of
  program analyses that can identify dead code or subprograms whose
  values are constant: this information can be used to further shrink
  the trace.  Further investigation is needed to experiment with the
  syntactic disclosure slicing algorithm on realistic examples and
  identify areas for improvement.

\paragraph{Disclosure from slices.} 
Finally, we link disclosure for value patterns to disclosure for
generic provenance views.  Essentially, we show that for any
$\extract$, the disclosure slice for $p$ positively discloses the
$\extract$-provenance annotations of values matching $p$. Informally,
this means that disclosure slices provide a highly general form of
provenance specialized to a part of the output: one can compute and
reveal the disclosure slice and others can then compute any generic
provenance view from the slice, without rerunning the original
computation or consulting input data or subtraces that are dropped in
the slice.

To state the desired property, we need to lift $\eqat{-}$
to apply to annotated values.  The definition is similar to that for
unannotated values, with additional rules:
\[
\inferrule*{
v \eqat{p} w
}{\
v^a \eqat{p} w^a
}\quad
\inferrule*{
\strut
}{
v^a \eqat{\hole} w^b
}\quad
\inferrule*{
\strut
}{
v^a \eqat{\vany} v^a
}\]

\begin{theorem}\label{thm:extraction-from-slices}
  Assume $|\agamma|,T \trrun v$ and $p \sqleq v$.  Suppose $p,T
  \bwdslice S,\rho$.  Suppose that $\extract$ is a generic extraction
  function. Then the annotations associated with $p$ in
  $\extract(T,\agamma)$ can be correctly extracted from $S$ using only
  input parts needed by $\rho$.  That is, suppose we have $\agamma
  \eqat{\rho} \agamma'$ and $T' \sqgeq S$, where $|\agamma'|,T' \trrun v'$.
  Then we have $\extract(T,\agamma) \eqat{p} \extract(T',\agamma')$.
\end{theorem}
 \begin{proof} Straightforward induction on the structure of
    derivations of $p,T \bwdslice S,\rho$.  See \appref{extraction-from-slices-proof}.
  \end{proof}

  Observe that some minimal slices discard information needed for
  provenance extraction.  For example, given expression
  $\vee(x,\kwtrue)$, the minimal slice with respect to $\kwtrue$ is
  $(\tremp,\kwtrue)$. However, this slice makes it impossible for
  dependency or expression provenance extraction to produce the right
  answer, since in both cases the annotation on $x$ is needed.
  Moreover, if we ignore code and match pointers, our slicing
  algorithm appears to be minimal with respect to provenance
  extraction (that is, removing any more from a trace would produce
  slices that do not satisfy \thmref{extraction-from-slices}).  This
  supports our view that the trace slicing algorithm is a natural one
  for the purpose of generating provenance or explanations, despite
  its non-minimality with respect to the semantic replayability
  criterion.  

  An alternative approach to slicing based on a criterion for which
  minimal slices exist is explored in another recent
  paper~\cite{perera12icfp}. 
Intuitively, the difference
    arises because disclosure slices are defined in terms of a fixed
    semantics, whereas Perera et al.~\cite{perera12icfp} define a
    correct backward slice as one that contains enough information to
    recompute a given part of the output using an ad hoc replay
    semantics defined over expressions with holes.  This makes a nice
    theory but means that we are required to include information in
    the slice that is not required in a disclosure slice.

    Note that it is typical for a notion oftypical for a notion of
    witness to lack unique minimal solutions (e.g. why-provenance in
    databases is defined as the set of minimal witnesses to a query
    result~\cite{buneman01icdt}) and for minimal slices to be
    non-computable.  For example, in the original work on
    slicing by Weiser~\cite{weiser81icse} minimal slices are shown to
    exist but are not computable.  Similarly, since \pl is
    Turing-complete, it is easy to show that it is undecidable to
    determine whether a given partial trace is a (minimal) disclosure
    slice.

\subsection{Obfuscation}\label{sec:obfuscation}

We now consider obfuscating properties of the input.  We first consider
what can be obfuscated by the standard provenance views.
Where-provenance, essentially, obfuscates anything that can never be
copied to the output or affect the control flow of something that is
copied to the output.  Similarly, expression provenance obfuscates any
part of the input that never participates in or influences expression annotations.
In both cases, we can potentially learn about parts of the input that
affected control flow, however.  For example, $\exif{x=1}{1}{y}$
does not obfuscate the value of $x$ in either model, provided $y$
comes from the input, since we can inspect the annotation of the result
to determine that $x=1$ or $x \neq 1$.  

This illustrates a possibly counterintuitive fact: obfuscation of the
query that tests whether $x=1$ fails
if we can ever learn anything about the result of the query, even if
we cannot determine the exact value of $x$.  Thus, where-provenance
and expression-provenance do not provide particularly strong
obfuscation properties, since they do not take control-flow into account.
Given that we want to ensure obfuscation, we consider conservative
techniques that accept (or construct) only provenance views that
successfully obfuscate, but may reject some views or construct views
that are unnecessarily opaque.

There are several ways to erase information from traces (or other
provenance views) to ensure obfuscation of input properties.  One way
is to re-use the static analysis of dependency provenance
(in~\cite{cheney11mscs}, for example) to identify and
  elide parts of the output
that suffice to make it impossible to guess sensitive parts of the
input.  Alternatively we can use dynamic dependency provenance to
increase precision, by propagating dependency tracking information
from the input to the output.

This is similar to using static analysis or dynamic labels for
information flow security; the difference is one of emphasis.  In
information flow security, we usually identify high- or low-security
locations and try to certify that high-security data does not
affect the computation of low-security data; here, instead, we
identify a high-security property of the trace (e.g. that the
input satisfies a certain formula) and try to determine what parts of
the output do not depend on sensitive inputs, and hence can be safely
included in the provenance view.  However, these techniques do not
provide guidance about what parts of the trace can be safely
included in the provenance view.

Here, we develop an alternative approach based on directly analyzing
and slicing traces.
  Consider a pattern $\rho \sqsubseteq \gamma$, in which the parts of
  $\gamma$ that are considered confidential have been replaced by
  $\vhole$.
We construct an
\emph{obfuscation slice} by re-evaluating $T$ on $\rho$ as much as
possible, to compute a sliced trace $S$ and partial value $p$.  We
erase parts of $T$ and of the original output value that depend on the
erased parts of $\rho$.  Thus, any part of the trace or output value
that remains in the obfuscation slice is irrelevant to the sensitive
part of the input, and cannot be used to guess it.

Figure~\ref{fig:obfuscation-slice} shows a syntactic algorithm for
computing obfuscation slices as described above, defined via a
judgment $\rho,T \fwdslice p,S$, which takes a partial input
environment $\rho$ and trace $T$ as input and computes a partial
output $p$ and sliced trace $S$.  Many rules are
essentially generalizations of the rules for evaluation to allow for
partial inputs, outputs and traces.  
The rules of interest, near the
bottom of the figure, show how to handle attempts to compute that
encounter holes in places where a value constructor is expected.  When
this happens, we essentially propagate the hole result and return a
hole trace.  This may be unnecessarily draconian for some cases, but
is necessary for the case and application traces where the trace form
gives clues about the control flow.

\begin{figure}[tb!]
\fbox{$\rho,T \fwdslice p,S$}
\vspace{-3mm}\\
\begin{smathpar}
\inferrule*
      {
\strut     }
      {
        \rho, \trvar{x} \fwdslice  \rho(x), \trvar{x}
      }
\and      \inferrule*{
        \strut
      }
      {
        \rho,\trc \fwdslice \vc, \trc 
      }
\and
    \inferrule*{\strut}
      {
       \rho,\trfunk{\kappa} \fwdslice \vclos{\kappa}{\rho}, \trfunk{\kappa}
      }
\and
      \inferrule*
      {
        \rho,T_1 \fwdslice p_1,S_1\\
        \rho[x\mapsto p_1], T_2 \fwdslice p_2,S_2
     }
      {
        \rho, \trlet{T_1}{x}{T_2} 
        \fwdslice 
       p_2,\trlet{S_1}{x}{S_2}
      }
\and
      \inferrule*{
       \rho,T_1 \fwdslice v_1,S_1
        \quad \cdots \quad
        \rho,T_n \fwdslice v_n,S_n
      }
      {
        \rho,\trf{T_1,\ldots,T_n} \fwdslice \kwf(v_1,\ldots,v_n), \trf{S_1,\ldots,S_n}
      }
     \and
      \inferrule*
      {
        \rho, T_1 \fwdslice p_1,S_1
        \\
        \rho, T_2 \fwdslice p_2,S_2 
      }
      {
        \rho,\trpair{T_1}{T_2} 
        \fwdslice 
        \vpair{p_1}{p_2},
        \trpair{S_1}{S_2}
      }
\and
      \inferrule*
      { \rho, T \fwdslice (p_1,p_2),S
      }
      {
        \rho,\trfst{T} \fwdslice p_1,\trfst{S}
      }
\and
      \inferrule*
      { \rho, T \fwdslice (p_1,p_2),S
      }
      {
        \rho,\trsnd{T} \fwdslice p_2,\trsnd{S}
      }
\and
     \inferrule*{\rho,T \fwdslice p,S}
      {
        \rho,\trinl{T} \fwdslice \vinl{p},\trinl{S}
      }
  \and
      \inferrule*{\rho,T \fwdslice p,S}
      {
        \rho,\trinr{T} \fwdslice \vinr{p},\trinr{S}
      }
\and
    \inferrule*
      {
        \rho, T \fwdslice \vinl{p},S
        \\
        \rho[x_1\mapsto p], T_1 \fwdslice p_1,S_1
        }
      {
       \rho,\trcaseml{m}{T}{x_1}{T_1}
        \fwdslice 
         p_1,\trcaseml{m}{S}{x_1}{S_1}
      }
\and
    \inferrule*
      {
        \rho, T \fwdslice \vinr{p},S
        \\
        \rho[x_2\mapsto p], T_2 \fwdslice p_2,S_2
        }
      {
       \rho,\trcasemr{m}{T}{x_2}{T_2}
        \fwdslice 
         p_2,\trcasemr{m}{S}{x_2}{S_2}
      }
\and
      \inferrule*{\rho,T \fwdslice p,S}
      {
        \rho,\trroll{T} \fwdslice \trroll{p},\trroll{S}
      }
\and
      \inferrule*{\rho,T \fwdslice \trroll{p},S}
      {
        \rho,\trunroll{T} \fwdslice p,\trunroll{S}
      }
\and
      \inferrule*
      {
        \rho, T_1 \fwdslice \vclos{\kappa}{\rho_0},S_1
        \\
        \rho,T_2 \fwdslice p_2,S_2
        \\
        \rho_0[f \mapsto \vclos{\kappa}{\rho_0}, x \mapsto p_2], T
        \fwdslice p,S
      }
    {
      \rho,\trappk{\kappa}{T_1}{T_2}{f}{x}{T} 
      \fwdslice 
      p,\trappk{\kappa}{S_1}{S_2}{f}{x}{S}
      }
\and
      \inferrule*{
       \rho,T_i \fwdslice \vhole,S_i \\ 
(\text{for some $i \in 1,\ldots,n$})
     }
      {
        \rho,\trf{T_1,\ldots,T_n} \fwdslice \vhole,\tremp
      }
\and
      \inferrule*
      { \rho, T \fwdslice \vhole,S
      }
      {
        \rho,\trfst{T} \fwdslice \vhole,\tremp
      }
\and
      \inferrule*
      { \rho, T \fwdslice \vhole,S
      }
      {
        \rho,\trsnd{T} \fwdslice \vhole,\tremp
      }
\and
      \inferrule*{\rho,T \fwdslice \vhole,S}
      {
        \rho,\trunroll{T} \fwdslice \vhole,\tremp
      }
    \and
    \inferrule*
      {
        \rho, T \fwdslice \vhole,S
       }
      {
       \rho,\trcaseml{m}{T}{x_1}{T_1}
        \fwdslice 
         \vhole,\tremp
      }
    \and
    \inferrule*
      {
        \rho, T \fwdslice \vhole,S
       }
      {
       \rho,\trcasemr{m}{T}{x_1}{T_1}
        \fwdslice 
         \vhole,\tremp
      }
\and
      \inferrule*
      {
        \rho, T_1 \fwdslice \vhole,S_1
      }
   {
      \rho,\trappk{\kappa}{T_1}{T_2}{f}{x}{T} 
      \fwdslice 
     \vhole,\tremp
      }
\end{smathpar}
\caption{Obfuscation slicing.}
\label{fig:obfuscation-slice}
\end{figure}

%%% Local Variables: 
%%% mode: latex
%%% TeX-master: "main"
%%% End: 

  \begin{example}\label{ex:disc-slicing}
    To illustrate the behavior of $\fwdslice$, consider again a simple
    program that swaps the elements of a pair:
  \[\small
\inferrule*{
 \inferrule*{
      [z\mapsto 1],y \fwdslice \vhole,\vhole\\
      [z\mapsto 1],z \fwdslice 1,z }{ [z \mapsto 1],(y,z)
      \fwdslice \vpair{\vhole}{1}, (\vhole,z) }
\\
  \inferrule*{ \inferrule*{\gamma,x \fwdslice(\vhole,1),x }{ \gamma,\trsnd{x} \fwdslice
       1, \trsnd{x}
      }\\
      \inferrule*{\gamma,x \fwdslice (\vhole,1),x}{
\gamma,\trfst{x} \fwdslice \vhole,\trfst{x}}} {\gamma,
      \trpair{\trsnd{x}}{\trfst{x}}\fwdslice \vpair{1}{\vhole}, \trpair{\trsnd{x}}{\trfst{x}}
    }
}{ 
 [z\mapsto 1] ,
    \trlet{(y,z)}{x}{\trpair{\trsnd{x}}{\trfst{x}}} \fwdslice
  \vpair{1}{\vhole}, \trlet{(\vhole,z)}{x}{\trpair{\trsnd{x}}{\trfst{x}}} }
  \]
where $\gamma = [z\mapsto 1, x \mapsto (\vhole,1)]$.  Notice that it
is impossible to guess the value of $y$ used in the original run from
the slice $\trlet{(\vhole,z)}{x}{\trpair{\trsnd{x}}{\trfst{x}}}$ or
  partial result $(1,\vhole)$.
\end{example}
  
      \paragraph{Correctness of obfuscation slicing.}
    We now show the correctness of obfuscation slicing in the
    sense that the slicing algorithm supports positive obfuscation.
  \begin{lemma}\label{lem:obfuscation-slicing-functional}
  If $\gamma,T \trrun v$ and $\rho \sqsubseteq \gamma$ is $\vany$-free
  then there exist unique $p \sqsubseteq v$ and $S \sqsubseteq T$ such
  that $\rho,T \fwdslice p,S$.
\end{lemma}
\begin{proof}
  First, we show that if $\gamma,T \trrun v$ and $\rho \sqleq \gamma$ is
  $\vany$-free then there exists $p
  \sqsubseteq v$ and $S \sqsubseteq T$ such that $\rho,T \fwdslice
  p,S$.  Uniqueness is straightforward by induction over derivations
  of $\rho,T \fwdslice p,S$.
\end{proof}

Accordingly, we define a partial function $\Obf_\rho(\gamma,T,v)$ as
$(p,S)$ where $\rho,T \fwdslice p,S$ and $p \sqsubseteq v$. We can
show that this function is total for well-formed, partial traces
and $\vany$-free  input environments.

  \begin{example}
    Consider again the running  $map~f~xs$ example, with $\gamma,T,v$ as in Example~\ref{ex:disc-slicing}.  
    \begin{itemize}
    \item If $\rho =  [f \mapsto \exfun{f}{x}{\exif{x=y}{y}{x+1}},xs =
  [1,2,3], y=\vhole]$ then $\Obf_\rho(\gamma,T,v) =
  ([\vhole,\vhole,\vhole],S)$ where $S$ shows three recursive calls to
  $map~f$ and three partial calls to $f$ where the parts of the trace showing
  the results of the conditional
  tests $x = y$ in $f$ are deleted.

\item If $\rho =  [f \mapsto \vhole,xs =
  [1,2,3], y=2]$ then $\Obf_\rho(\gamma,T,v) =
  ([\vhole,\vhole,\vhole],S)$ where $S$ shows three recursive calls to
  $map~f$ where the traces showing the execution of $f$ are deleted.
\item If $\rho =  [f \mapsto \exfun{f}{x}{\exif{x=y}{y}{x+1}},xs =
  \vhole::\vhole], y=2]$ then $\Obf_\rho(\gamma,T,v) =
  (\vhole::\vhole,S)$ where $S$ shows two recursive calls to
  $map~f$ and one call to $f$, where information about the control
  flow branch taken after testing $x=y$ is omitted.
    \end{itemize}
  \end{example}

\begin{lemma}\label{lem:obfuscation-slicing-correctness}
  If $\gamma,e \eval v,T$ and $\rho \sqsubseteq \gamma$ and $\rho,T
  \fwdslice p,S$ then for all $\gamma'
  \sqsupseteq \rho$, if $\gamma',e \eval v',T'$ then $\rho,T'
  \fwdslice p,S$ and $p \sqleq v'$.
\end{lemma}
\begin{proof}
  See \appref{obfuscation-slicing-correctness-proof}.
\end{proof}

Finally, before considering the main correctness result for
$\Obf_\rho$, we note a technical issue: In our language, every base type
happens to have at least two values, so we can always instantiate a
hole at base type in at least two ways.  Similarly, pairs, functions
and so on involving base types can always be instantiated in several
ways.  However, in general we cannot assume that every type has more
than one ground value.  We say that a type is \emph{nonsingular} if it
has at least two different values, and in the following result we
restrict attention to patterns containing holes of nonsingular types:
\begin{theorem}
  For traces generated by terminating expressions, and $\rho$ with
 holes of nonsingular types, and $\rho' \sqsupset \rho$, we have
  $\Obf_\rho$ positively obfuscates $\IN_{\rho'}$.
\end{theorem}

\begin{proof}
  Suppose $\IN_{\rho'}$ holds of $(\gamma,T,v)$ where $\rho' \sqsupset
  \rho$.  Then $\rho \sqsubset \rho' \sqsubseteq \gamma$.  Moreover,
  since the inclusion is strict, and since $\rho $ contains holes of
  nonsingular type, $\rho$ must contain holes that can be replaced
  with different values, so there exists another $\gamma' \sqsupseteq
  \rho$ that differs from $\rho'$.  Since $T$ was generated by a
  terminating expression, we know that $\gamma',e \eval v',T'$ can be
  derived for some $v',T'$.  By
  Lemma~\ref{lem:obfuscation-slicing-correctness} we know that
  $\rho,T' \fwdslice p,S$, hence $\Obf_\rho(\gamma',T',v') = (p,S) =
  \Obf_\rho(\gamma,T,v)$, as required.
\end{proof}

  This is the main result about obfuscation.  As with disclosure,
  previously some properties of obfuscation were established for
  limited computational models~\cite{cheney11csf}, but this is the
  first such result to be established for a general-purpose
  programming language.  This result shows that the obfuscation
  slicing algorithm (a syntactic traversal of the trace that
  propagates ``holes'' forward) provides a safe approach to positive
  obfuscation.  This means that given a pattern $\rho'$ identifying a
  sensitive part of the input, for any triple $(\gamma,T,v)$, the
  syntactic algorithm yields a subtrace $(p,S)$ such that there
  exists $(\gamma',T',v')$ whose obfuscation slice is also
  $(p,S)$ but such that $\gamma'$ does not contain $\rho$.  Thus,
  we cannot deduce that $\rho'$ is present in $\gamma'$ from
  $(p,S)$.

% Need to statically determine whether we can learn enough about the trace 

% Goal: Given a trace query $Q$ and a view $P$, determine whether $P$ discloses $Q$.

% Simple case: Both $Q$ and $P$ use the same underlying extraction function.

Negative obfuscation may also hold for the obfuscation slicing
algorithm, but if so it appears more difficult to prove: we would have
to show that if $\rho$ does not match the input $\gamma$, then there
is another $\gamma'$ that does match $\rho$ but produces the same
obfuscation slice as $\rho$.  Calculating such a $\gamma'$ is not
straightforward if the expression $e$ can diverge, because even finding a
different input on which $e$ terminates is generally an undecidable
problem.  However, even under an assumption of termination, it is not
clear how to compute obfuscation slices to ensure that all traces on
inputs that avoid a certain pattern are indistinguishable from traces
on inputs that do contain the pattern.

\subsection{Discussion}

%\odo{Tie the results of the paper together.}

The analysis in section 5.1 gives novel characterizations of what
information is disclosed by where-provenance and expression
provenance.  Essentially, where-provenance discloses information about
what parts of the input are copied to the output, while expression
provenance additionally discloses information about how parts of the
input can be combined to compute parts of the output.  Both forms
ignore the control flow of the program.  The analysis in section 5.1
also shows (in a formal sense) that where-provenance and expression
provenance are closely related: one can obtain where-provenance from
expressions simply by erasure.  Moreover, we can obtain a number of
other intermediate provenance models, by extracting  information
compositionally from expression-provenance annotations.

The disclosure slicing algorithm is based on an interesting insight
(which we are exploring in concurrent work on
slicing~\cite{perera12icfp}): at a technical level, the information we
need for program comprehension via slicing (to understand how a
program has evaluated its inputs to produce outputs) is quite similar
to what we need for provenance.  Our past work on dependency
provenance identified connections between provenance and
slicing~\cite{cheney11mscs} which we have explored in more recent
work~\cite{perera12icfp} that employs slicing techniques similar to
disclosure slicing.

Obfuscation seems to be fundamentally more difficult to obtain than
disclosure.  From an intuitive point of view, this is not surprising;
however, it is interesting to see where the complications arise at a
technical level, and how these interact with conventional forms of provenance.  For example, both where- and expression provenance
effectively disclose certain information about the output given the
input (or vice versa), while dependency provenance does not appear to
disclose information in a particularly direct way.  On the other hand,
since it was inspired in part by information flow security techniques,
dependency provenance does seem to obfuscate information about the
input, but cannot directly tell us how much of the trace it is safe to
provide while still obfuscating a part of the input.  

Obfuscation slicing, which is based on a similar idea to dependency
provenance, does allow us to provide part of the trace in the
provenance view while obfuscating sensitive input.  However, we were
only able to obtain a positive obfuscation result for slicing.
  We do not currently have either a proof of negative obfuscation or a
  counterexample to it for the obfuscation slicing algorithm.  This
  means that whenever the query actually holds, we cannot be certain
  of this from the provenance view; however, when the input query
  fails it may be possible to tell this from the view.  Negative
  obfuscation seems more difficult to prove than positive obfuscation,
  at least for the input queries we considered.
This is
unsurprising, since as also found in~\cite{cheney11csf}, the
definition of obfuscation is more complex.  

  It is interesting to consider whether alternative definitions of
  disclosure or obfuscation could lead to more satisfying results.  As
  explained at the end of Section~\ref{sec:obfuscation}, the root of
  the difficulty with negative obfuscation seems to be the difficulty
  of analyzing the program to find alternative inputs that enable the
  program to complete and lead to the same obfuscation slice.

  One alternative could be to model the knowledge of the attacker
  about the possible traces more explicitly (e.g. assume the attacker
  knows the original program).  This seems orthogonal to the problem
  of proving negative obfuscation: it should complicate both positive
  and negative problems.  Another alternative could be to adopt a
  probabilistic or information-theoretic definition of obfuscation
  that makes it easier to provide both positive and negative
  obfuscation.  These are possibilities for future work.
%%% Local Variables: 
%%% mode: latex
%%% TeX-master: "main"
%%% End: 

% LocalWords:  Obf subexpression computable iff subtraces minimality subtrace
% LocalWords:  replayability determinizing determinacy transductions eqat versa
% LocalWords:  contrapositive unannotated counterintuitive nonsingular
% LocalWords:  compositionally

\section{Related Work} \label{sec:related}

There is a huge, and growing, literature on
provenance~\cite{bose05cs,cheney09ftdb,DBLP:journals/sigmod/SimmhanPG05,moreau10ftws},
but there is little work on formal models of provenance and no
previous work on provenance in a general-purpose higher-order
language.  Since we already covered prior work on provenance security
in the introduction, we confine our comparison to closely related work
on formal techniques for provenance, and on related ideas in
programming languages and language-based security.

\paragraph{Provenance.}
This work differs from previous work on provenance in databases in
several important ways.  First, we consider a general purpose,
higher-order language, whereas previous work considers database query
languages of limited expressiveness (e.g., monotone query languages),
which include unordered collection types with monadic iteration
operations but not sum types, recursive types or first-class
functions.  Second, we aim to record traces adequate to answer a wide
range of provenance queries in this general setting, whereas previous
work has focused on particular kinds of queries (e.g.,
where-provenance~\cite{buneman01icdt,buneman08tods},
why-provenance~\cite{buneman01icdt},
how-provenance~\cite{green07pods,foster08pods}).  % None of these
% techniques records enough information for our fidelity property,
% though how-provenance has been used for incremental
% recomputation~\cite{taylor06sigmod}.
% For example, where-provenance cannot update the output
% after modifications to source.  Dependency provenance can provide an
% over-approximation of how the output may be affected but cannot
% provide precise updates.
% Finally, our work aims for maximum generality; some work on provenance in databases has considered
% extracting certain forms from others.  For example, the \emph{semiring
%   provenance} model of Green et al.~\cite{green07pods} can express
% some other models such as why-provenance and lineage, but cannot
% express where-provenance and vice
% versa~\cite{buneman08pods,cheney09ftdb}.  There is currently no common
% formalism analogous to traces that generalizes these techniques.  We
% are working on generalizing our ideas to database query languages,
% which pose several additional challenges.

Provenance has also been studied extensively for scientific workflow
systems~\cite{bose05cs,DBLP:journals/sigmod/SimmhanPG05,davidson08sigmod}.
Many workflow provenance systems record additional information to
support replaying the computation (analogous to our fidelity property)
or provenance queries focusing on explaining parts of the result
(analogous to our extraction and slicing techniques).  Most work in
this area describes the provenance tracking behavior of a system
through examples and does not give a formal semantics that could be
used to prove correctness properties; furthermore, there has been
little work (and there is currently no consensus) on what the
appropriate correctness properties are.
An exception is Hidders et al.~\cite{hidders07dils}, which is the closest
workflow provenance work to ours.  They model
workflows using a core database query language extended with
nondeterministic, external function calls, and partially formalize a
semantics of \emph{runs}, or sets of triples $(\gamma,e,v)$ labeling an
operational derivation tree.  They
also discuss extracting \emph{subruns} which seem similar to slices,
and extracting provenance information from runs.
However, their definitions of
subrun and provenance extraction are complex, incomplete, and not
accompanied by precise statements or proofs of correctness properties.  
 Further progress on formalizing their approach has been made
  recently in a workshop paper by Acar et al.~\cite{acar10tapp};
  however, provenance extraction and trace slicing are not addressed
  in~\cite{acar10tapp}.

  \paragraph{Other related topics.}  Our trace model is partly
  inspired by previous work on self-adjusting
  computation~\cite{acar06toplas}, where execution traces are used to
  efficiently recompute functional programs under arbitrary
  modifications to their inputs.  Previous work on self-adjusting
  computation has not investigated trace slicing techniques or a
  relationship between traces and provenance.  Unlike self-adjusting
  traces, our traces are intended as data that can be manipulated and
  queried by users, with recomputation just one of many competing
  requirements.  Provenance-like ideas have also appeared in the
  context of \emph{alignment} in bidirectional
  computation~\cite{bohannon08popl} and language-based techniques for
  \emph{audit}~\cite{jia08icfp,vaughan08csf} More recently, Dimoulas
  et al. identified an intriguing connection between provenance and
  notions of correctness for blame assignment in contracts.  They
  introduce semantic properties that, they suggest, may be related to
  provenance~\cite{dimoulas11popl}.  However, to our knowledge no
  formal relationships between provenance and self-adjusting
  computation, bidirectional computation, or blame have been
  developed.

    Finally, our model of execution traces for \pl is closely related
    to that used in a recent publication~\cite{perera12icfp}; however,
    the technical contributions, slicing algorithms and the
    correctness criteria are different.  In this paper, we consider
    trace slicing algorithms that provide disclosure or obfuscation
    properties, while in~\cite{perera12icfp} we consider trace and
    program slicing techniques that satisfy a different
    consistency property, aimed at comparing different runs of
    a program for debugging or program understanding.  At a semantic
    level, the most important difference is that in this article our
    definitions are in terms of a standard operational semantics over
    standard values, which partial and annotated values need to
    respect; in the work on program slicing we consider a variant
    operational semantics over partial values (which is similar in
    some respects to the obfuscation slicing algorithm).  Perera et
    al.~\cite{perera12icfp} make several additional contributions,
    including algorithms for extracting program slices from trace
    slices and for constructing \emph{differential slices} that can
    highlight the exact location of a bug in the source program.
    Investigating the applicability of these ideas to provenance or
    provenance security is future work.

%%% Local Variables: 
%%% mode: latex
%%% TeX-master: "main"
%%% End: 

% LocalWords:  workflows replayability recomputation workflow Hidders et al
% LocalWords:  nondeterministic subruns subrun Redex updatable Biswas Swamy
% LocalWords:  unevaluating semirings Acar postprocessing Perera gprof OCaml
% LocalWords:  Kishon Hudak Chong Dimoulas unordered monadic

\section{Conclusions}\label{sec:concl}

While the importance of understanding provenance and its security
characteristics has been widely documented, to date there has been
little work on formal modeling of either provenance or its
security.  In this article, we elaborate upon the ideas introduced in
previous work~\cite{cheney11csf}, by instantiating the formal
framework proposed there with a general-purpose functional programming
language and a natural notion of execution traces.  We showed how more
conventional forms of provenance can be extracted from such traces via
a generic provenance extraction mechanism.  Furthermore, we studied
the key notions of disclosure and obfuscation in this context.  In the
process we identified weaker \emph{positive} and \emph{negative}
variants of disclosure and obfuscation, based on the observation that
the original definitions seem too strong to be satisfied often in
practice.  Our main results include algorithms for \emph{disclosure
  slicing}, which traverses a trace backwards to retain information
needed to certify how an output was produced, and \emph{obfuscation
  slicing}, which reruns a trace on partial input (excluding sensitive
parts of the input), yielding a partial trace and partial output that
excludes all information that could help a principal learn sensitive
data.

To summarize, our main contribution is the development of a general
model of provenance in the form of a core calculus that instruments
runs of programs with detailed execution traces.  We validated the
design of this calculus by showing that traces generalize other known
forms of provenance and by studying their disclosure and obfuscation
properties. There are many possible avenues for future work,
including:
\begin{itemize}
\item identifying richer languages for defining trace queries or
  provenance views
\item developing and implementing practical algorithms for trace
  slicing, and relating these to program slicing~\cite{perera12icfp}
\item developing a more uniform approach to the different forms of replay,
  extraction, and slicing
\item extending trace and provenance models to handle references,
  exceptions, input/output, concurrency, nondeterminism,
  communication, etc.
\end{itemize}

%%% Local Variables: 
%%% mode: latex
%%% TeX-master: "main"
%%% End: 

% LocalWords:  unevaluation recomputation nondeterminism

\paragraph{Acknowledgments}
Effort sponsored by the Air Force Office of Scientific Research, Air
Force Material Command, USAF, under grant number FA8655-13-1-3006. The
U.S. Government and University of Edinburgh are authorized to
reproduce and distribute reprints for their purposes notwithstanding
any copyright notation thereon.  Cheney is supported by a Royal
Society University Research Fellowship, by the EU FP7 DIACHRON
project, and EPSRC grant EP/K020218/1.  Parts of this research were
done while Acar and Perera were at Max-Planck Institute for Software
Systems, Kaiserslautern, Germany, and while Perera was a PhD student
at the University of Birmingham.  Acar is partially supported by an EU
ERC grant (2012-StG 308246---DeepSea) and an NSF grant (CCF-1320563).

\bibliographystyle{abbrv}
\bibliography{paper}

\newpage
\appendix

\section{Proofs}\label{app:proofs}

\subsection{Proof of  \thmref{where}}
\label{app:where-proof}

\begin{proof}[Proof of \thmref{where}]
We prove by induction on the structure of derivations that if
$|\agamma|,e\red v,T$, then $\occnb(\where(T,\agamma)) \subseteq
\occnb(\agamma)$.

\begin{itemize}
\item Base cases involving constants and variables are
  trivial.
\item Cases involving constructors (pairs, $\kwinl$, $\kwinr$,
  $\kwroll$, closures) and primitive operations ($\oplus$) are
  straightforward since the newly-constructed value is annotated with
  $\bot$.

\item Cases involving pair projections ($\kwfst$, $\kwsnd$) and
  $\kwunroll$ are straightforward, since the returned value is a
  subvalue of the value returned by a subexpression.
\item For a derivation of the form:
   \begin{displaymath}
      \inferrule { \kwinl(x_1).e_1 \in m \\
        |\agamma|,e_1 \red \vinl{v}, T
        \\
        |\agamma|[x_1 \mapsto v],e_1 \red v_1, T_1} { \strut
        |\agamma|,\excasem{e}{m} \red v_1,
        \trcaseml{m}{T}{x_1}{T_1} }
    \end{displaymath}
    By induction, $\occnb(\where(T,\agamma)) \subseteq
    \occnb(\agamma)$.  Moreover, $\where(T,\agamma) = \vinl{\av}^a$
    for some $\av$ with $|\av| = v$.  So, $\occnb(\av) \subseteq
    \where(T,\agamma) \subseteq \occnb(\agamma)$. Also by induction,
    $\occnb(\where(T_1,\agamma[x_1\mapsto \av]) \subseteq
    \occnb(\agamma[x_1 \mapsto \av])$.  Hence,
\[\occnb(\where(\trcaseml{m}{T}{x_1}{T_1},\agamma[x_1\mapsto
\av]) = \occnb(\where(T_1,\agamma[x_1 \mapsto \av])  \subseteq
\occnb(\agamma[x_1 \mapsto \av]) = \occnb(\agamma) \cup \occnb(\av)
\subseteq \occnb(\agamma)\]
\item The case for $\kw{case}$ where the right branch is taken is symmetric.
\item For function application, if the derivation is of the form:
    \begin{displaymath}
      \inferrule { |\agamma|,e_1 \red \vclos{\fn{f}{x}{e}}{\gamma_0} ,
        T_1
        \\
        |\agamma|,e_2 \red v_2,T_2
        \\
        \gamma_0[f\mapsto \vclos{\fn{f}{x}{e}}{\gamma_0} ,x\mapsto
        v_2], e \red v, T } 
      { 
        |\agamma|,\exapp{e_1}{e_2} \red v,\trapp{f}{x}{e}{T_1}{T_2}{T} 
      }
    \end{displaymath}
    then by validity we know that $\where(T_1,\agamma) = \vclos{\fn{f}{x}{e}}{\agamma_0}^a$
    with $|\agamma_0| = \agamma$ and $\where(T_2,\agamma) = \av_2$
    with $|\av_2| = v_2$.  By induction, we also know:
    \[
   \occnb( \vclos{\fn{f}{x}{e}}{\agamma_0}^a) \subseteq \occnb(\agamma)
    \qquad
    \occnb(\av_2) \subseteq \occnb(\agamma)
    \]
   Hence, 
    \[\gamma_0[f\mapsto \vclos{\fn{f}{x}{e}}{\gamma_0} ,x\mapsto
        v_2] = |\agamma_0[f\mapsto\vclos{\fn{f}{x}{e}}{\agamma_0}^a,x
        \mapsto \av_2] |
        \]
So the induction hypothesis applies to the third subderivation,
yielding:
 \[
    \where(T,\agamma_0[f\mapsto\vclos{\fn{f}{x}{e}}{\agamma_0}^a,x
        \mapsto \av_2]) \subseteq \occnb(\agamma_0[f\mapsto\vclos{\fn{f}{x}{e}}{\agamma_0}^a,x
        \mapsto \av_2])
\]
Thus, by the definition of $\where$ for application-traces, we have:
\begin{eqnarray*}
  \where(\trapp{f}{x}{e}{T_1}{T_2}{T},\agamma) &=& 
  \where(T,\agamma_0[f\mapsto\vclos{\fn{f}{x}{e}}{\agamma_0}^a,x
        \mapsto \av_2]) \\
        &\subseteq & \occnb(\agamma_0[f\mapsto\vclos{\fn{f}{x}{e}}{\agamma_0}^a,x
        \mapsto \av_2])\\
        &=& \occnb(\agamma_0) \cup
        \occnb(\vclos{\fn{f}{x}{e}}{\agamma_0}^a) \cup \occnb(\av_2)\\
&\subseteq &      \occnb(\agamma)
\end{eqnarray*}
\end{itemize}
\end{proof}

\subsection{Proof of  \thmref{expr}}
\label{app:expr-proof}

\begin{proof}[Proof of \thmref{expr}]

We prove by induction on the structure of derivations that if
$|\agamma|,e\red v,T$, then for any $h$ consistent with $\agamma$, we
have that $h$ is also consistent with $\expr(T,\agamma)$.
\begin{itemize}
\item Base cases involving constants and variables are
  trivial.
\item Cases involving constructors (pairs, $\kwinl$, $\kwinr$,
  $\kwroll$, closures) are
  straightforward since the newly-constructed value is annotated with
  $\bot$.
\item For primitive operations, consider a primitive function evaluation:
\[
   \inferrule*
    {
     |\agamma|,  e_1 \red v_1,T_1\\
\cdots\\
     |\agamma|, e_n \red v_n,T_n
   }
    {
      |\agamma|, \kwf(e_1,\ldots,e_n) \red \hat{\oplus}(v_1,\ldots,v_n), \trf{T_1,\ldots,T_n}
    }
\]
Suppose $h$ is consistent with $\agamma$.  By induction, $h$ is consistent with $\expr(T_i,\agamma)$ for each
$i$.  This means that each of the results $v_i^{t_i}$ satisfies
$h(t_i) = |\expr(T_i,\agamma)| = v_i$.  Thus, $\hat{\oplus}(v_1,\ldots,v_n) =
\hat{\oplus}(h(t_1),\ldots,h(t_n)) = h(\exf{t_1,\ldots,t_n})$, which
implies that $h$ is consistent with $\expr(T,\agamma) =
(\hat{\oplus}(v_1,\ldots,v_n))^{\exf{t_1,\ldots,t_n}}$, as desired.
\item The remaining cases follow similar reasoning to that for
  where-provenance.
\item For a derivation of the form:
   \begin{displaymath}
      \inferrule { \kwinl(x_1).e_1 \in m \\
        |\agamma|,e_1 \red \vinl{v}, T
        \\
        |\agamma|[x_1 \mapsto v],e_1 \red v_1, T_1} { \strut
        |\agamma|,\excasem{e}{m} \red v_1,
        \trcaseml{m}{T}{x_1}{T_1} }
    \end{displaymath}
    Suppose $h$ is consistent with $\agamma$.  By induction, $h$ is
    consistent with $\expr(T,\agamma)$.  Moreover, $\expr(T,\agamma) =
    \vinl{\av}^a$ for some $\av$ with $|\av| = v$.  Thus, $h$ is
    consistent with $\av$ and $\agamma[x_1\mapsto \av]$, so by
    induction, $h$ is consistent with $\expr(T_1,\agamma[x_1\mapsto
    \av])$.  Since $\expr(\trcaseml{m}{T}{x_1}{T_1},\agamma)) =
    \expr(T_1,\agamma[x_1 \mapsto \av])$, it follows that $h$ is
    consistent with $\expr(\trcaseml{m}{T}{x_1}{T_1},\agamma)) $.
\item The case for $\kw{case}$ where the right branch is taken is symmetric.
\item For function application, if the derivation is of the form:
    \begin{displaymath}
      \inferrule { |\agamma|,e_1 \red \vclos{\fn{f}{x}{e}}{\gamma_0} ,
        T_1
        \\
        |\agamma|,e_2 \red v_2,T_2
        \\
        \gamma_0[f\mapsto \vclos{\fn{f}{x}{e}}{\gamma_0} ,x\mapsto
        v_2], e \red v, T } 
      { 
        |\agamma|,\exapp{e_1}{e_2} \red v,\trapp{f}{x}{e}{T_1}{T_2}{T} 
      }
    \end{displaymath}
    then by validity we know that $\expr(T_1,\agamma) =
    \vclos{\fn{f}{x}{e}}{\agamma_0}^a$ with $|\agamma_0| = \agamma$
    and $\expr(T_2,\agamma) = \av_2$ with $|\av_2| = v_2$.  By
    induction, we also know $h$ is consistent with
    $\vclos{\fn{f}{x}{e}}{\agamma_0}^a$ and $\av_2$, so $h$ is
    consistent with
    $\agamma_0[f\mapsto\vclos{\fn{f}{x}{e}}{\agamma_0}^a,x \mapsto
    \av_2]$.  By induction on the third subderivation, we have that
    $h$ is consistent with
    $\expr(T,\agamma_0[f\mapsto\vclos{\fn{f}{x}{e}}{\agamma_0}^a,x
    \mapsto \av_2]) $.  To conclude, since
\[\expr(\trapp{f}{x}{e}{T_1}{T_2}{T},\agamma)  = \expr(T,\agamma_0[f\mapsto\vclos{\fn{f}{x}{e}}{\agamma_0}^a,x
    \mapsto \av_2])\;,\]
we know that $h$ is consistent with $\expr(\trapp{f}{x}{e}{T_1}{T_2}{T},\agamma) $.
\end{itemize}
\end{proof}

\subsection{Proof of \thmref{dep}}
\label{app:dep-proof}

\begin{lemma}
~
\begin{enumerate}
\item   If $\av_1 \eqxat{\lbl} \av_2$ then $\av_1^{+a} \eqxat{\lbl} \av_2^{+a}$.
\item If $\lbl \in a \cap b$ then $\av_1^{+a} \eqxat{\lbl} \av_2^{+b}$.
\end{enumerate}
\end{lemma}
\begin{proof}
  Similar to a property proved in~\cite{cheney11mscs}; the only new
  cases are for closures, and are straightforward.
\end{proof}

\begin{proof}[Proof of \thmref{dep}]
  Proof proceeds by induction on the structure of the derivation of
  $|\agamma|,e \red v,T$.  There are many straightforward cases,
  similar to those proved in ~\cite{cheney11mscs}.  We show the proof
  cases for case and application traces; the other cases use similar
  techniques.
  \begin{itemize}
  \item  If the derivation is of the form:
\begin{displaymath}
    \inferrule
    {
      (\exinl{x_1}.e_1 \in m)\\
      |\agamma|,e \red \vinl{v}, T
      \\
      |\agamma|[x_1\mapsto v], e_1 \red v_1, T_1
    }
    {
      \strut
      |\agamma|,\excasem{e}{m} \red v_1, \trcaseml{m}{T}{x_1}{T_1}
    }
\end{displaymath}
   then by inversion of $|\agamma'|, \trcaseml{m}{T}{x_1}{T_1} \trrun
   v'$  the derivation must be of the form:
\begin{displaymath}
\inferrule
{
|\agamma'|, T \trrun \vinl{v}
\\
|\agamma'|[x_1 \mapsto v], T_1 \trrun v_1
}
{
|\agamma'|, \trcaseml{m}{T}{x_1}{T_1}
\trrun
v_1
}
\end{displaymath}
Then by validity of $\dep$ and induction we know that $\vinl{\av}^a =
\dep(T,\agamma) \eqxat{\lbl} \dep(T',\agamma) = \vinl{\av'}^{b}$, where
$|\av| = v$ and $|\av'| = v'$.  Then $|\agamma[x \mapsto \av]| =
|\agamma|[x\mapsto v]$ and $|\agamma'[x\mapsto \av']| =
|\agamma'|[x\mapsto v']$.  Now, if $\lbl \in a \cap b$ then we are
done: it follows immediately that
\[\dep(\trcaseml{m}{T}{x_1}{T_1},\agamma) = (\av_1)^{+a} \eqxat{\lbl}
(\av_1')^{+b}
= \dep(\trcaseml{m}{T'}{x_1}{T_1'},\agamma')
\] 
Otherwise, we must have $\av \eqxat{\lbl} \av'$ and $a = b$; hence,
$\agamma[x\mapsto \av] \eqxat{\lbl} \agamma'[x\mapsto \av']$.  So, by
induction, $\av_1 = \dep(T_1,\agamma[x \mapsto \av]) \eqxat{\lbl}
\dep(T_1',\agamma'[x\mapsto \av']) = \av_1'$.  Then it follows immediately that
\[\dep(\trcaseml{m}{T}{x_1}{T_1},\agamma) = (\av_1)^{+a} \eqxat{\lbl}
(\av_1')^{+a}  = \dep(\trcaseml{m}{T'}{x_1}{T_1'},\agamma')
\] 

\item Application:  Suppose the derivation is of the form:
  \begin{displaymath}
    \inferrule
    {
      |\agamma|,e_1 \red \vclos{\kappa}{\gamma_0},T_1
      \\
      (\kappa = \fn{f}{x}{e})
      \\
      |\agamma|,e_2 \red v_2,T_2
      \\
      \gamma_0[f \mapsto \vclos{\kappa}{\gamma_0},x \mapsto v_2], e \red v, T
   }
    {
      |\agamma|,\exapp{e_1}{e_2} \red v, \trappk{\kappa}{T_1}{T_2}{f}{x}{T}
    }
  \end{displaymath}
  By validity we know that $ \dep(T_1,\agamma) =
  \vclos{\kappa}{\agamma_0}^a$ for some $a,\agamma_0$
  with $|\agamma_0| = \gamma_0$.
  Similarly, $\dep(T_2,\agamma) = \av_2$ for some $\av_2$
  with $|\av_2| = v_2$.  Finally, 
  $\dep(T,\agamma_0[f\mapsto\vclos{\kappa}{\agamma_0}^a,x\mapsto
  \av_2]) = \av$ for some $\av$ with $|\av| = v$.

Then  the replay derivation is of the form:
  \[
    \inferrule
{
|\agamma'|,T_1 \trrun \vclos{\kappa}{\gamma_0'}
\\
|\agamma'|,T_2 \trrun v_2'
\\
\gamma_0'[f\mapsto\vclos{\kappa}{\gamma_0'}, x\mapsto v_2'],T \trrun v'
}
{
|\agamma'|,\trappk{\kappa}{T_1}{T_2}{f}{x}{T} \trrun v'
}
  \]
  First, note that by fidelity and validity $\dep(T_1',\agamma') =
  \vclos{\kappa}{\agamma_0'}^b$ for some $b,\agamma_0'$ with
  $|\agamma_0'| = \gamma_0'$.  Similarly, $\dep(T_2',\agamma') =
  \av_2'$ for some $\av_2'$ with $|\av_2'| = v_2'$.  Finally,
  $\dep(T',\agamma_0'[f\mapsto\vclos{\kappa}{\agamma_0'}^b,x\mapsto
  \av_2']) = \av'$ for some $\av'$ with $|\av'| = v'$.

  Then, by induction, we know that
  \[\vclos{\kappa}{\agamma_0}^a = \dep(T_1,\agamma) \eqxat{\lbl}
  \dep(T_1',\agamma') = \vclos{\kappa}{\agamma_0'}^b \qquad
\av_2 = \dep(T_2,\agamma) \eqxat{\lbl} \dep(T_2',\agamma') = \av_2'\]
  Now, there are two cases.  If $\lbl \in a
  \cap b$, then we are done since we can derive:
  \[\dep(\trappk{\kappa}{T_1}{T_2}{f}{x}{T},\agamma) = \av^a \eqxat{\lbl}
  (\av')^b = \dep(\trappk{\kappa}{T_1'}{T_2'}{f}{x}{T'},\agamma')
  \]
  Otherwise, we know that $a = b$ and $\agamma_0
  \eqxat{\lbl}\agamma_0'$, hence also:
  \[
  \inferrule{\agamma_0 \eqxat{\lbl} \agamma_0' \\
    \vclos{\kappa}{\agamma_0}^a \eqxat{\lbl}
    \vclos{\kappa}{\agamma_0'}^b\\
    \av_2 \eqxat{\lbl} \av_2'
    }{
      \agamma_0 [f\mapsto \vclos{\kappa}{\agamma_0}^a,x\mapsto \av_2]
      \eqxat{\lbl} \agamma_0' [f\mapsto \vclos{\kappa}{\agamma_0'}^a,x\mapsto \av_2']
    }
    \]
    Then by induction on the remaining subderivations, we have $\av
    \eqxat{\lbl} \av'$, from which we can infer $\av^{+a} \eqxat{\lbl}
    (\av')^{+a}$.

\end{itemize}
\end{proof}

\subsection{Proof of \lemref{eqat-properties}}

\label{app:eqat-properties-proof}

\begin{lemma}\label{lem:eqat-vany}
  For any $p$, we have $(\eqat{p[\vany/\vhole]}) = (\eqat{\vany}) \cap (\eqat{p})$.
\end{lemma}
\begin{proof}
Induction on $p$.
  \begin{itemize}
  \item If $p = \vhole$, then the result is immediate since
    $(\eqat{\vany}) \cap (\eqat{\vhole }) = (\eqat{\vany})$.
\item If $p = \vany$, then the result is immediate.
\item If $p = (p_1,p_2)$, then $p[\vany/\vhole] =
  (p_1[\vany/\vhole],p_2[\vany/\vhole])$, so we reason as follows:
  \begin{eqnarray*}
    (v_1,v_2)\eqat{ (p_1[\vany/\vhole],p_2[\vany/\vhole])} (v_1',v_2')
    &\iff& 
v_1 \eqat{p_1[\vany/\vhole]} v_1' \text{ and } v_2
\eqat{p_2[\vany/\vhole]} v_2'\\
    &\iff& 
v_1 \eqat{\vany} v_1' \text{ and } v_1 \eqat{p_1} v_1' 
\text{ and }
v_2 \eqat{\vany} v_2'
\text{ and } v_2
\eqat{p_2} v_2'\\
    &\iff& 
(v_1,v_2) \eqat{\vpair{\vany}{\vany}} (v_1',v_2') \text{ and } (v_1,v_2) \eqat{(p_1,p_2)} (v_1',v_2') \\
    &\iff& 
(v_1,v_2) \eqat{\vany} (v_1',v_2') \text{ and } (v_1,v_2) \eqat{(p_1,p_2)} (v_1',v_2') 
\end{eqnarray*}
\item The cases for $p = C(p')$ or $p = \vclos{\kappa}{\rho}$ are
  similar to the case for pairing.
  \end{itemize}
\end{proof}

\begin{proof}[Proof of \lemref{eqat-properties}]
Symmetry of $(\eqat{p})$ follows by straightforward induction on derivations.  

We show transitivity by induction on $p$.
   \begin{itemize}
   \item If $p = \vhole$, then transitivity is obvious as $(\eqat{p})$ is the total relation.
\item If $p = \vany$, then transitivity is obvious as $(\eqat{\vany})$
  is the identity relation.
\item If $p = \vc$ then suppose $v_1 \eqat{\vc} v_2$ and $v_2
  \eqat{\vc} v_3$.  Then $v_1,v_2,v_3$ are all constants. By
  inversion, we must have $v_1 = v_2 = v_3 = \vc$ so we have $v_1 = \vc
  \eqat{\vc} \vc =  v_3$.  
\item If $p = C(p')$, then suppose $v_1 \eqat{C(p')}
  v_2$ and $v_2 \eqat{C(p')} v_3$.  By inversion, we must have
\[
\inferrule{v_1' \eqat{p'} v_2'}
{C(v_1') \eqat{ C(p')} C(v_2')}
\quad
\inferrule{v_2' \eqat{p'} v_3'}
{C(v_2') \eqat{ C(p')} C(v_3')}
\]
where $v_i = C(v_i')$ for each $i$.  In this case, by induction we have $v_1'
\eqat{p'} v_3'$ so we can conclude $C(v_1') \eqat{C(p')} C(v_3')$ as desired.
\item If $p = \vpair{p_1}{p_2}$ then transitivity follows immediately
  by induction.
\item If $p = \vclos{\kappa}{\rho}$, then the reasoning is similar to
  that for $p = C(p')$. 
  \end{itemize}

We show by induction on  pairs $(p,p')$ such that $p \sqcup p'$
exists, that $(\eqat{p \sqcup p'}) =
(\eqat{p}) \cap (\eqat{p'})$.
   \begin{itemize}
   \item If one of the patterns (say, $p$) is $\vhole$ then $\vhole \sqcup p' = p'$, so $(\eqat{\vhole \sqcup p'}) =
(\eqat{p'}) = (\eqat{\vhole}) \cap (\eqat{p'}) $ since
$\eqat{\vhole}$ is total.
\item If one of the patterns (say, $p$) is $\vany$ then $\vany \sqcup p' = p'[\vany/\vhole]$, so $(\eqat{\vany \sqcup p'}) =
(\eqat{p'[\vany/\vhole]}) = (\eqat{\vany}) \cap (\eqat{p'}) $, where
the second equation is \lemref{eqat-vany}.

\item If $\vpair{p_1}{p_2} \sqcup \vpair{p_1'}{p_2'} = \vpair{p_1
    \sqcup p_1'}{p_2 \sqcup p_2'} $ then 
  \begin{eqnarray*}
(v_1,v_2)
  \eqat{\vpair{p_1
    \sqcup p_1'}{p_2 \sqcup p_2'} } (v_1',v_2')
&\iff& v_1 \eqat{p_1 \sqcup p_1'} v_1' 
\text{ and } v_2 \eqat{p_2 \sqcup p_2'} v_2'
\\
&\iff& v_1 \eqat{p_1}  v_1' \text{ and } v_1 \eqat{p_1'} v_1'
\text{ and } v_2 \eqat{p_2} v_2'
\text{ and } v_2 \eqat{p_2'} v_2'
\\
&\iff& (v_1,v_2) \eqat{\vpair{p_1}{p_2}}  (v_1',v_2') 
\text{ and }  (v_1,v_2) \eqat{\vpair{p_1'}{p_2'}}  (v_1',v_2') 
  \end{eqnarray*}
\item For the cases $C(p) \sqcup C(p')$ and $ \vclos{\kappa}{\rho}
  \sqcup \vclos{\kappa}{\rho'}$ the reasoning
  is similar to the case for pairing.
   \end{itemize}

\end{proof}

\subsection{Proof of
  \lemref{slicing-correct}}\label{app:slicing-correct-proof}

\begin{proof}[Proof of \lemref{slicing-correct}]
  The proof is by induction on the derivation of $p, T \bwdslice
  S,\rho$ and inversion on $\gamma,T \trrun v$.  We use
  \lemref{eqat-properties} freely without comment to infer, for example,
  $\gamma \eqat{\rho_i} \gamma'$ from $\gamma\eqat{\rho_1 \sqcup
    \rho_2} \gamma'$.

\paragraph{Empty pattern.}  If the derivation is of the form:
\begin{mathpar}
  \inferrule{\strut}
  {
    \vhole,T \bwdslice \tremp,\vhole
  }
\end{mathpar}
then clearly, for any $\gamma' \eqat{\vhole} \gamma$ and $T' \sqgeq
\tremp$, if $\agamma',T' \trrun v'$ then $v' \eqat{\vhole} v$.

\paragraph{Variable.}
If the derivations are of the form:
\begin{mathpar}
\inferrule
      {
        \strut
      }
      {
        p, \trvar{x}\bwdslice \trvar{x},[x\mapsto p]
      }
~~\mbox{and}~~
\inferrule
{
  \strut
}
{
  \gamma,\trvar{x} \trrun \gamma(x)
}
\end{mathpar}
Then $T' \sqgeq \trvar{x}$ implies $T' = \trvar{x}$, so 
\begin{mathpar}
\inferrule
{
  \strut
}
{
  \gamma',\trvar{x} \eval \gamma'(x)
}
\end{mathpar}
Since by assumption $\gamma' \eqat{[x\mapsto p]}\gamma$, we conclude
that $\gamma'(x) \eqat{p} \gamma(x)$, as desired.

\paragraph{Let.}  If the derivations are of the form:
\begin{mathpar}
\inferrule
{p_2,T_2 \bwdslice S_2,\rho_2[x\mapsto p_1]\\
p_1,T_1 \bwdslice S_1,\rho_1
}
{
p_2,\trlet{T_1}{x}{T_2} \bwdslice \trlet{S_1}{x}{S_2},\rho_1 \sqcup
\rho_2
}
  ~~\mbox{and}~~
\inferrule{
  \gamma,T_1\trrun v_1\\
\gamma[x\mapsto v_1] , T_2 \trrun v_2
}{
  \gamma,\trlet{T_1}{x}{T_2} \trrun v_2
}
\end{mathpar}
Then by inversion we must have $T' = \trlet{T_1'}{x}{T_2'}$ where
$T_i' \sqgeq S_i$, and:
\begin{mathpar}
\inferrule{
  \gamma',T_1'\trrun v_1'\\
\gamma'[x\mapsto v_1'] , T_2 '\trrun v_2'
}{
  \gamma',\trlet{T_1'}{x}{T_2'} \trrun v_2'
}
\end{mathpar}
We also know that $\gamma \eqat{\rho_1} \gamma'$ and $\gamma
\eqat{\rho_2} \gamma'$.  Then, by induction, we have $v_1 \eqat{p_1}
v_1'$, hence we know that $\gamma[x\mapsto v_1] \eqat{\rho_1[x\mapsto
  p_1]} \gamma'[x\mapsto v_1']$.  So, the induction hypothesis applies
to $p_2,T_2 \bwdslice S_2, \rho_2[x\mapsto p_1]$, and we can conclude
that $v_2 \eqat{p_2}{ v_2'}$.

\paragraph{Constant trace.}
If the trace has the form $\trc$, then we have:

\begin{mathpar}
      \inferrule{
        \strut
      }
      {
        p,\trc \bwdslice \trc,\vhole
      }
~~\mbox{and}~~
\inferrule
{\strut}
{\gamma,\trc \trrun \trc}
\end{mathpar}
Then clearly $\gamma',\trc \trrun c$ and $c
\eqat{p} c$.

\paragraph{Primitives.}
If the trace has the form $\trf{\vec{T}} = \trf{T_1, \ldots, T_n}$,
then we have:
\begin{mathpar}
\inferrule
{
  \vany,T_1 \bwdslice S_1 ,\rho_1
  \\  
  \ldots 
  \\
  \vany,T_n \bwdslice S_n,\rho_n
}
{
  p,\trf{T_1, \ldots, T_n} \bwdslice \trf{S_1, \ldots, S_n},\rho_1
  \sqcup \cdots \sqcup \rho_n
}
~~\mbox{and}~~
\inferrule
{
  \gamma,T_1 \trrun v_1 \\ \ldots  \\ \gamma,T_n \trrun v_n
}
{
\gamma,  \trf{T_1, \ldots, T_n}
  \trrun 
  \hat{\oplus}(v_1, \ldots, v_n)
}\;.
\end{mathpar}
Suppose $T' \sqgeq \trf{S_1,\ldots,S_n}$.  This implies $T' =
\trf{T_1',\ldots,T_n'}$ where $T_i' \sqgeq S_i$ for each $i$.
Moreover, by inversion we must have:
\begin{mathpar}
\inferrule
{
  \gamma,T_1' \trrun v_1' \\ \ldots  \\ \gamma,T_n' \trrun v_n'
}
{
\gamma,  \trf{T_1', \ldots, T_n'}
  \trrun
  \hat{\oplus}(v_1', \ldots, v_n').
}
\end{mathpar}

By induction, we have for all $1 \le i \le n$, that $v_i' \eqat{\vany}
v_i$, that is, $v_i' = v_i$.  Hence, we can conclude that
$\hat{\oplus}(v_1',\ldots,v_n') \eqat{p} \hat{\oplus}(v_1,\ldots,v_n)$ since
both sides are equal.

\paragraph{Pairs/pair patterns.}
If the derivation is of the form:
\begin{mathpar}
\inferrule
{
  p_1, T_1 \bwdslice  S_1,\rho_1\\
  p_2, T_2 \bwdslice  S_2,\rho_2
}
{
  \vpair{p_1}{p_2},\trpair{T_1}{T_2} 
  \bwdslice 
  \trpair{S_1}{S_2},\rho_1 \sqcup \rho_2
}
~~\mbox{and}~~
\inferrule
{
\gamma,T_1 \trrun v_1
\\
\gamma,T_2 \trrun v_2
}
{
  \gamma,\trpair{T_1}{T_2}
  \eval  \vpair{v_{1}}{v_{2}}
}.
\end{mathpar}
Then, as in the previous case we know $T' = (T_1',T_2')$ where $T_i'
\sqgeq S_i$ and
 \begin{mathpar}
\inferrule
{
\gamma',T_1' \trrun v_1'
\\
\gamma',T_2' \trrun v_2'
}
{
  \gamma',\trpair{T_1'}{T_2'}
  \trrun  \vpair{v_{1}'}{v_{2}'}
}.
\end{mathpar} 

By induction, we have $v_1 \eqat{p_1} v_1'$ and $v_2 \eqat{p_2} v_2'$,
from which it follows that $(v_1,v_2) \eqat{(p_1,p_2)} (v_1',v_2')$.

\paragraph{First.}
If the derivation is of the form:

\begin{mathpar}
      \inferrule
      {
        \vpair{p}{\vhole}, T \bwdslice S,\rho
      }
      {
        p,\trfst{T} \bwdslice \trfst{S},\rho
      }
~~\mbox{and}~~
\inferrule
{
\gamma,T \trrun \vpair{v_{1}}{v_{2}}
}
{
\gamma,\trfst{T}
\trrun
v_{1}
}
\end{mathpar}
then, by inversion, we know that $T' = \trfst{T_0'}$ for some $T_0'
\sqgeq S$, such that
\begin{mathpar}
\inferrule
{
\gamma',T_0' \eval \vpair{v_{1}'}{v_{2}'},T''
}
{
\gamma',\trfst{T_0'}
\eval
v_{1}',\trfst{T''}
}
\end{mathpar}
By induction, we have $\vpair{v_1'}{v_2'} \eqat{\vpair{p}{\vhole} }
\vpair{v_1}{v_2}$, which implies that $v_1' \eqat{p} v_1$.

\paragraph{Second.}
Symmetric to $\kwfst$ case.

\paragraph{Inl.}
Suppose the derivations are of the form:
\begin{mathpar}
  \inferrule{p,T \bwdslice S,\rho}
  {\vinl{p},\trinl{T} \bwdslice \trinl{S},\rho}
~~\mbox{and}~~
\inferrule{\gamma,T \trrun v}
{\gamma,\trinl{T}\trrun \vinl{v}}
\end{mathpar}
Then we must have $T' = \trinl{T_0'}$ where $T_0' \sqgeq S$, and:
\begin{mathpar}
\inferrule{\gamma',T_0' \trrun v'}
{\gamma',\trinl{T_0'}\trrun \vinl{v'}}
\end{mathpar}
So, by induction, we know that $v \eqat{p} v'$ and so $\vinl{v}
\eqat{\vinl{p}} \vinl{v'}$.

\paragraph{Inr.}
 Symmetric to $\kwinl$ case.

\paragraph{Case/L.}
If the derivations are of the form:
\begin{mathpar}
\inferrule 
{ 
p_1, T_1 \bwdslice S_1,\rho_1[x_1\mapsto p]  \\
\vinl{p},T \bwdslice S,\rho
}
{ 
  p_1, \trcaseml{m}{T}{x_1}{T_1} \bwdslice
  \trcaseml{m}{S}{x_1}{S_1},\rho \sqcup \rho_1
}
\\
 \mbox{and}
\\
\inferrule
{
\gamma,T \trrun \vinl{v}
\\
\gamma[x_1\mapsto v],T_1 \trrun v_1
}
{
\gamma,\trcaseml{m}{T}{x_1}{T_1} 
\trrun
v_1
}
\end{mathpar}
Then $T' = \trcaseml{m}{T_0'}{x_1}{T_1'}$ where $T_0' \sqgeq S$ and $T_1'
\sqgeq S_1$.  The only way for the replay judgment
$\gamma,T' \trrun v'$ to be derived is:
\[\inferrule
{
\gamma',T_0' \trrun \vinl{v_0'}
\\
\gamma[x_1\mapsto v_0'],T_1' \trrun v'
}
{
\gamma,\trcaseml{m}{T_0'}{x_1}{T_1'} 
\trrun
v'
}\]
so by induction we can conclude $\vinl{v} \eqat{\vinl{p}}
\vinl{v_0'}$, which in turn implies $v \eqat{p}
v_0'$.  Thus,
$\gamma[x\mapsto v] \eqat{\rho_1[x\mapsto p]} \gamma'[x\mapsto v_0']$,
  from which it follows by induction that $v_1 \eqat{p_1} v'$.

\paragraph{Case/R.} Symmetric to the previous case.

\paragraph{Function abstraction.}
If the derivations have the form:
\begin{mathpar}
\inferrule
{
\strut
}
{
\vclos{\kappa}{\rho}, \trfunk{\kappa} \bwdslice \trfunk{\kappa},\rho
}
\quad\mbox{and}\quad
\inferrule
{
\strut
}
{
\gamma,\trfunk{\kappa} \trrun \vclos{\kappa}{\gamma}
}
\end{mathpar}
then $T'$ must be of the form
$\trfunk{\kappa}$, with derivation:
\begin{mathpar}
\inferrule
{
\strut
}
{
\gamma',\trfunk{\kappa} \trrun \vclos{\kappa}{\gamma'}
}
\end{mathpar}
Hence, we can conclude $\vclos{\kappa}{\gamma}
\eqat{\vclos{\kappa}{\rho}} \vclos{\kappa}{\gamma'}$ immediately
from $\gamma \eqat{\rho} \gamma'$.

\paragraph{Application.}
If the derivations are of the form:
\begin{mathpar}
 \inferrule
      {
       p, T \bwdslice S, \rho[f \mapsto p_1, x \mapsto p_2]
        \\
        p_1 \sqcup \vclos{\kappa}{\rho}, T_1 \bwdslice S_1,\rho_1
        \\
        p_2,T_2 \bwdslice S_2,\rho_2     }
      {
       p, \trappk{\kappa}{T_1}{T_2}{f}{x}{T} 
        \bwdslice 
        \trappk{\kappa}{S_1}{S_2}{f}{x}{S} ,\rho_1 \sqcup \rho_2
      }
\\
\mbox{and}
\\
\inferrule
{
\gamma,T_1 \trrun \vclos{\kappa}{\gamma_0}
\\
\gamma,T_2 \trrun v_2
\\
\gamma_0[f\mapsto\vclos{\kappa}{\gamma_0}, x\mapsto v_2],T \trrun v
}
{
\gamma,\trappk{\kappa}{T_1}{T_2}{f}{x}{T} \trrun v
}
\end{mathpar}
Then we know that $T' = \trappk{\kappa}{T_1'}{T_2'}{f}{x}{T_0'}$ where
$T_1' \sqgeq S_1$ and $T_2' \sqgeq S_2$ and $T_0' \sqgeq S$.  The
replay derivation of $\gamma',T' \trrun v'$ must be of the form:
\begin{mathpar}
\inferrule
{
\gamma',T_1' \trrun \vclos{\kappa}{\gamma_0'}
\\
\gamma',T_2' \trrun v_2'
\\
\gamma_0'[f\mapsto\vclos{\kappa}{\gamma_0'}, x\mapsto v_2'],T_0' \trrun v'
}
{
\gamma',\trappk{\kappa}{T_1'}{T_2'}{f}{x}{T_0'} \trrun v'
}
\end{mathpar}
First, by induction on the first subderivation we know that
$\vclos{\kappa}{\gamma_0} \eqat{p_1 \sqcup
  \vclos{\kappa}{\rho}}\vclos{\kappa}{\gamma_0'}$.  Here, recall that
$p_1$ is a value pattern for the function argument obtained from
slicing the body.  By inversion, we have
\[\vclos{\kappa}{\gamma_0} \eqat{p_1}\vclos{\kappa}{\gamma_0'} \qquad \gamma_0 \eqat{\rho}\gamma_0'\]
By induction, we also have that $v_2 \eqat{p_2} v_2'$.  Hence, putting
the above observations together, we have:
\[
\gamma_0[f\mapsto \vclos{\kappa}{\gamma_0},x\mapsto v_2] \eqat{\rho[f
  \mapsto p_1,x \mapsto p_2]} \gamma_0'[f \mapsto
\vclos{\kappa}{\gamma_0'}, x \mapsto v_2']
\]
Thus, the induction hypothesis applies again and we can conclude that
$v \eqat{p} v'$.

\paragraph{Roll and unroll.}  These cases are straightforward, similar
to those for pairs and projection.

\paragraph{Pairs/wildcard.}
If the trace has the form $\trpair{T_1}{T_2}$, then we have:

\begin{mathpar}
\inferrule
{
  \vany, T_1 \bwdslice S_1,\rho_1
  \\
  \vany, T_2 \bwdslice S_2,\rho_2
}
{
  \vany,\trpair{T_1}{T_2} 
  \bwdslice 
  \trpair{S_1}{S_2},\rho_1 \sqcup \rho_2
}
~~\mbox{and}~~
\inferrule
{
\gamma,T_1 \trrun v_1
\\
\gamma,T_2 \trrun v_2
}
{
  \gamma,\trpair{T_1}{T_2}
  \trrun
  \vpair{v_{1}}{v_{2}}
}.
\end{mathpar}
Then $T' \sqgeq (S_1,S_2)$ so $T'$ must be of the form $(T_1',T_2')$
with $T_1 \sqgeq S_1$ and $T_2 \sqgeq S_2$, and we must have
\begin{mathpar}
\inferrule
{
\gamma',T_1 \trrun v_1'
\\
\gamma',T_2 \trrun v_2'
}
{
  \gamma',\trpair{T_1'}{T_2'}
  \trrun  \vpair{v_{1}'}{v_{2}'}
}.
\end{mathpar}
By induction, we have $v_1 \eqat{\vany} v_1'$ and $v_2 \eqat{\vany}
v_2'$, which implies $(v_1,v_2) \eqat{\vany} (v_1',v_2')$, as required.
\paragraph{Other wildcard cases}
Other cases involving wildcards are similar to the above.
% \paragraph{Cases where the pattern and trace do not match} These cases
% are vacuous because we assume that $p \sqleq v$, which means that
% there is no way we could have derived $\gamma,T \trrun v$ in the first
% place.
\end{proof}

\subsection{Proof of \thmref{extraction-from-slices}}
\label{app:extraction-from-slices-proof}

\begin{proof}[Proof of \thmref{extraction-from-slices}]
  The proof is by induction on the structure of the derivation of $p,T
  \bwdslice S,\rho$.  We show that if $|\agamma|, T \trrun
  |\extract(T,\agamma)|$ then for any $T',\agamma'$, if $\agamma
  \eqat{\rho} \agamma'$ and $S \sqleq T'$ and $|\agamma'|,T' \trrun
  |\extract(T',\agamma')|$ then $\extract(T,\agamma) \eqat{p} \extract(T',\agamma')$.

 \begin{itemize}
  \item If the slicing derivation is of the form:
      \[
\inferrule
      {
        \strut
      }
      {
        \vhole, T\bwdslice \tremp,\vhole
      }
\]
then we are done: the conclusion is trivial since $\extract(T,\agamma)
\eqat{\vhole} \extract(T',\agamma')$.

\item If the slicing derivation is of the form:
  \[
  \inferrule
  {
    \strut
  }
  {
    p, \trvar{x} \bwdslice \trvar{x}, [x\mapsto p]
  }
  \]
  then we reason as follows:
  \[
  \extract(\trvar{x},\agamma) = \agamma(x) \eqat{p} \agamma'(x) =
  \extract(\trvar{x},\agamma')
  \]
  where $\agamma(x) \eqat{p} \agamma'(x)$ follows from the assumption
  that $\agamma \eqat{[x\mapsto p]} \agamma'$.

\item If the slicing derivation is of the form:
\[
      \inferrule*
      {
        p_2, T_2 \bwdslice S_2,\rho_2[x\mapsto p_1]\\
        p_1,T_1 \bwdslice S_1,\rho_1
      }
      {
        p_2, \trlet{T_1}{x}{T_2} 
        \bwdslice 
        \trlet{S_1}{x}{S_2}, \rho_1 \sqcup \rho_2
      }
\]
then let $T_1' \sqgeq S_1, T_2' \sqgeq S_2$ and $\agamma' \eqat{\rho}
\agamma$ be given.  By induction and validity we have 
\[\av_1 = \extract(T_1,\agamma) \eqat{p_1} \extract(T_1',\agamma') = \av_1'\]
thus, we also know that $\agamma[x\mapsto \av_1]
\eqat{\rho_1[x\mapsto p_1]} \agamma'[x\mapsto \av_1']$.  So, by
induction, we also have:
\[\av_2 = \extract(T_2,\agamma[x\mapsto \av_1]) \eqat{p_2} \extract(T_2',\agamma'[x\mapsto \av_1']) = \av_2'\]
Thus, 
\[\extract( \trlet{T_1}{x}{T_2} ,\agamma) = \av_2 \eqat{p_2} \av_2' = \extract(\trlet{T_1'}{x}{T_2'},\agamma')\]

\item If the slicing derivation is of the form:
  \[
  \inferrule{
    \strut
  }
  {
    p,\trc \bwdslice \trc,\vhole
  }
  \]
then again we are done as $\extract(\trc,\agamma) = c^{\extract_c} =
\extract(\trc,\agamma')$.

\item If the slicing derivation is of the form:
      \[
      \inferrule{
        \vany,\vec{T} \bwdslice \vec{S},\rho
      }
      {
        p,\trf{\vec{T}} \bwdslice \trf{\vec{S}},\bigsqcup \vec{\rho}
      }
      \]
      then let $\agamma',\vec{T'}$ be given with $\agamma'
      \eqat{\bigsqcup \vec{\rho}} \agamma$ and $\vec{T'} \sqgeq
      \vec{S}$.  By induction, we know that $\extract(T_i,\agamma)
      \eqat{\vany} \extract(T_i',\agamma')$ (that is, $\extract(T_i,\agamma) =
      \extract(T_i',\agamma')$) holds for each $i$.  Let $\av_i =
      \extract(T_i,\agamma) = \extract(T_i',\agamma')$ for each $i$.  Then we
      can reason as follows:
      \[\extract(\trf{\vec{T}},\agamma)) = 
      \hat{\oplus}(|\av_1|,\ldots,|\av_n|)^{\extract_\kwf(a_1,\ldots,a_n)} \eqat{p} 
      \hat{\oplus}(|\av_1|,\ldots,|\av_n|)^{\extract_\kwf(a_1,\ldots,a_n) } = 
      \extract(\trf{\vec{T'}},\agamma')
      \]
\item If the slicing derivation is of the form:
      \[
      \inferrule
      {
        p_1, T_1 \bwdslice  S_1,\rho_1
\\
        p_2, T_2 \bwdslice  S_2,\rho_2
     }
      {
        (p_1,p_2),\trpair{T_1}{T_2} 
        \bwdslice 
        \trpair{S_1}{S_2},\rho_1 \sqcup \rho_2
      }
      \]
      then let $\gamma', T_1',T_2'$ be given with $T_1' \sqgeq S_1,
      T_2' \sqgeq S_2$ and $\agamma' \eqat{\rho_1 \sqcup \rho_2}
      \agamma$.  By induction we know that $\extract(T_1 ,\agamma) \eqat{p_1}
      \extract(T_1' ,\agamma')$  and $\extract(T_2 ,\agamma) \eqat{p_2}
      \extract(T_2' ,\agamma')$  so we can conclude:
 \begin{eqnarray*}
        \extract(\trpair{T_1}{T_2} ,\agamma)
        &=& (\extract(T_1 ,\agamma ),\extract(T_2,\agamma ))^\bot \\
        &\eqat{(p_1,p_2)}& (\extract(T_1' ,\agamma'),\extract(T_2',\agamma'))^\bot\\
        &=& \extract(\trpair{T_1'}{T_2'} ,\agamma')
      \end{eqnarray*}

\item If the slicing derivation is of the form:
      \[
      \inferrule
      {
        (p,\vhole), T \bwdslice S,\rho
      }
      {
        p,\trfst{T} \bwdslice \trfst{S},\rho
      }
      \]
      then fix $T' \sqgeq S$ and $\agamma' \eqat{\rho} \agamma$.  If
      $p = \vhole$, then the conclusion is immediate.  By
      induction we know that $\extract(T ,\rho) \eqat{(p,\vhole)}
      \extract(T',\rho)$.  Then by inversion on the replay derivation and
      validity we know
      that 
      \[
      \extract(T,\rho) = (\av_1,\av_2)^a \eqat{(p,\vhole)}  
      (\av_1',\av_2')^{a'} = \extract(T',\agamma')
      \]
      Since $p \neq \vhole$, this implies $a = a'$ and $v_1^b = \av_1 \eqat{p} \av_1' =
      (v_1')^{b'}$, hence, $v_1 \eqat{p} v_1'$ and $b = b'$.  We can
      conclude by reasoning as follows:
      \[
      \extract(\trfst{T} ,\agamma) 
      = v_1^{\extract_1(a,b)} \eqat{p} (v_1')^{\extract_1(a',b')}
      = \extract(\trfst{T'} ,\rho)
      \]

\item If the slicing derivation is of the form:
      \[
      \inferrule
      {
        (\vhole,p), T \bwdslice S,\rho
      }
      {
        p,\trsnd{T} \bwdslice \trsnd{S},\rho
      }
      \]
then the reasoning is symmetric to the previous case.

\item If the slicing derivation is of the form:
\[
\inferrule
{
  p,T \bwdslice S,\rho
}
{
  \vinl{p},\trinl{T} \bwdslice \trinl{S},\rho
}
\]
then fix $T' \sqgeq S$ and $\agamma' \eqat{\rho} \agamma$.  By
induction 
we know that $\extract(T,\agamma) \eqat{p} \extract(T',\agamma')$, so it
follows directly that 
\[\extract(\trinl{T},\agamma)  =
\vinl{\extract(T,\agamma)}^\bot \eqat{\vinl{p}}
\vinl{\extract(T',\agamma')}^\bot = \extract(\trinl{T'},\agamma')\]

\item  If the slicing derivation is of the form:
\[
\inferrule
{
  p,T \bwdslice S,\rho
}
{
  \vinr{p},\trinr{T} \bwdslice \trinr{S},\rho
}
\]
then the reasoning is symmetric to the previous case.
\item  If the slicing derivation is of the form:
\[
      \inferrule*
      {
       p_1, T_1 \bwdslice S_1, \rho_1[x_1\mapsto p]
       \\
        \vinl{p}, T \bwdslice S, \rho
      }
      {
       p_1, \trcaseml{m}{T}{x_1}{T_1}
        \bwdslice 
         \trcaseml{m}{S}{x_1}{S_1},\rho \sqcup \rho_1
      }
\]
then fix $T_0' \sqgeq S$, $T_1' \sqgeq S_1$ and $\agamma'\eqat{\rho\sqcup \rho_1}
\agamma$.  If $p = \vhole$ then the conclusion is immediate.  By induction we know that $\extract(T,\agamma) \eqat{\vinl{p}}
\extract(T_0',\agamma)$, and by validity this means that $\extract(T,\agamma) =
(\vinl{\av})^a$ and $\extract(T_0',\agamma') = (\vinl{\av'})^a$ where $\av
\eqat{p} \av'$.  Thus, $\agamma[x_1\mapsto \av] \eqat{\rho_1[x_1\mapsto p]}
  \agamma'[x_1\mapsto \av']$, so by induction we also have
  \[v_1^b = \extract(T_1,\agamma[x_1\mapsto \av]) \eqat{p_1} \extract(T_1',\agamma'[x_1\mapsto
  \av']) = (v_1')^{b'}\;.\]
  Moreover, since $p \neq \vhole$ we know $b = b'$ and $v_1
  \eqat{p_1} v_1'$, so:
\[\extract( \trcaseml{m}{T}{x_1}{T_1},\agamma) = v_1^{\extract_L(a,b)}
\eqat{p_1}  (v_1')^{\extract_L(a,b')}= \extract( \trcaseml{m}{T_0'}{x_1}{T_1'},\agamma')\]

\item If the slicing derivation is of the form:
\[\inferrule*
      {
       p_2, T_2 \bwdslice S_2, \rho_2[x_2\mapsto p]
        \\
        \vinr{p}, T \bwdslice S,\rho
      }
      {
        p_2, \trcasemr{m}{T}{x_2}{T_2}
        \bwdslice 
       \trcasemr{m}{S}{x_2}{S_2}, \rho \sqcup \rho_2
      }\]
then the reasoning is symmetric to the previous case.

\item If the slicing derivation is of the form:
      \[
      \inferrule{\strut}
      {
        \vclos{\kappa}{\rho},\trfunk{\kappa} \bwdslice \trfunk{\kappa},\rho
      }
      \]
      then let $T' \sqgeq \trfunk{\kappa}$ and $\agamma' \eqat{\rho}
      \agamma$ be given; note that $T' = \trfunk{\kappa}$.  We can
      conclude immediately that
\[
\extract(\trfunk{\kappa},\agamma) = \vclos{\kappa}{\agamma} ^\bot\eqat{\rho}
\vclos{\kappa}{\agamma'} ^\bot= \extract(\trfunk{\kappa},\agamma')
\]

\item If the slicing derivation is of the form:
      \[
      \inferrule
      {
       p, T \bwdslice S,\rho[f\mapsto p_1,x\mapsto p_2]
       \\
      p_1 \sqcup \vclos{\kappa}{\rho}, T_1 \bwdslice S_1,\rho_1
        \\
        p_2, T_2 \bwdslice S_2,\rho_2
     }
      {
        p, \trappk{\kappa}{T_1}{T_2}{f}{x}{T} 
        \bwdslice 
        \trappk{\kappa}{S_1}{S_2}{f}{x}{S} ,\rho_1 \sqcup \rho_2
      }
\]
Let $T_1',T_2',T_0'$ and $\agamma'$ be given with $T_1' \sqgeq S_1$,
$T_2' \sqgeq S_2$ and $T_0' \sqgeq S$, and $\agamma' \eqat{\rho_1
  \sqcup \rho_2} \agamma$.  If $p = \vhole$, then the conclusion is
immediate.  Otherwise, by induction, validity, and inversion of
$\eqat{-}$ derivations, we know that:
\begin{eqnarray*}
  \vclos{\kappa}{\agamma_0}^a =
\extract(T_1 ,\agamma) &\eqat{p_1 \sqcup
  \vclos{\kappa}{\rho}}& \extract(T_1 ',\agamma')  = \vclos{\kappa}{\agamma_0'}^a \\
  \av_2 = \extract(T_2 ,\agamma) &\eqat{p_2} & \extract(T_2' ,\agamma') = \av_2'
\end{eqnarray*}
Thus, we also have $\agamma_0 \eqat{\rho} \agamma_0'$, so we can
obtain:
\[\agamma_0[f\mapsto \vclos{\kappa}{\agamma_0}^a,x\mapsto \av_2]
\eqat{\rho[f\mapsto p_1,x\mapsto p_2]} \agamma_0'[f\mapsto
\vclos{\kappa}{\agamma_0'}^a,x\mapsto \av_2']\]
By induction, it follows that 
\[v^b = \extract(T,\agamma_0[f\mapsto \vclos{\kappa}{\agamma_0},x\mapsto
\av_2]) \eqat{p} \extract(T_0',\agamma_0'[f\mapsto
\vclos{\kappa}{\agamma_0'},x\mapsto \av_2'] = {v'}^{b'}\]
This, together with the fact that $p \neq \vhole$, implies that $v
\eqat{p} v'$ and $b = b'$, so:
\[
\extract(\trappk{\kappa}{T_1}{T_2}{f}{x}{T} ,\agamma) =
v^{\extract_{\kwapp}(a,b)} \eqat{p}
(v')^{\extract_\kwapp(a,b')} = \extract(\trappk{\kappa}{T_1'}{T_2'}{f}{x}{T_0'},\agamma')
\]
\item The cases for $\kwroll$ and $\kwunroll$ are analogous
  to the cases for pairing and projection.
\item If the slicing derivation is of the form:
      \[
      \inferrule
      {
        \vany, T_1 \bwdslice S_1,\rho_1
        \\
        \vany, T_2 \bwdslice S_2,\rho_2
      }
      {
        \vany,\trpair{T_1}{T_2} 
        \bwdslice 
        \trpair{S_1}{S_2},\rho_1 \sqcup \rho_2
      }
      \]
      then fix $\agamma',T_1',T_2'$ with $T_i' \sqgeq S_i$ and
      $\agamma' \eqat{\rho_1\sqcup \rho_2} \agamma$.  By induction we
      know that $\extract(T_i,\agamma) \eqat{\vany} \extract(T_i',\agamma')$
      (that is, $\extract(T_i, \agamma) = \extract(T_i',\agamma')$) holds for
      each $i \in \{1,2\}$.  We reason as follows:
      \begin{eqnarray*}
       \extract(\trpair{T_1}{T_2} ,\agamma)
        &=& (\extract(T_1 ,\agamma),\extract(T_2,\agamma))^\bot \\
        &=& (\extract(T_1' ,\agamma'),\extract(T_2',\agamma'))^\bot\\
       &=& \extract(\trpair{T_1'}{T_2'} ,\agamma)
      \end{eqnarray*}

\item The other cases in which $p = \vany$ are straightforward, following
  similar reasoning to the above cases where $p$ starts with a value
  constructor.

% \item The remaining rules deal with cases where $p \not\sqleq v$,
%   so they are vacuous since we assumed that $p \sqleq v$ holds.
 \end{itemize}
\end{proof}

\subsection{Proof of \lemref{obfuscation-slicing-correctness}}
\label{app:obfuscation-slicing-correctness-proof}

\begin{proof}[Proof of  \lemref{obfuscation-slicing-correctness}]
The proof is by induction on the structure of derivations of $\rho,T
\fwdslice p,S$, and inversion on derivations of $\gamma,e \eval
v,T$.  
\begin{itemize}
\item If the derivations are of the form:
\[\inferrule*
      {
\strut     }
      {
        \rho, \trvar{x} \fwdslice  \rho(x), \trvar{x}
      }
\quad 
\inferrule*
    {
      \strut
    }
    {
      \gamma, x \red \gamma(x), \trvar{x}
    }
\]
then suppose $\gamma' \sqgeq \rho$ and $\gamma',x \eval v',T'$.  Then
$v' = \gamma'(x)$ and $T' = x$, so it suffices to observe that
\[\inferrule*
      {
\strut     }
      {
        \rho, \trvar{x} \fwdslice  \rho(x), \trvar{x}
      }
\quad 
\rho(x) \sqleq \gamma'(x)
\]
hold.

\item If the derivations are of the form:
\[
\inferrule*{
        \strut
      }
      {
        \rho,\trc \fwdslice \rho(x), \trc 
      }
\quad
 \inferrule*
    {\strut}{\gamma, \exc \red \exc,\trc}
\]
then suppose $\gamma' \sqgeq \rho$ where $\gamma',\exc \eval v',T'$.
By inversion the only way the latter can be derived is if $v' = \vc$
and $T' = \trc$.  So we can conclude by observing:
\[\inferrule*
      {
\strut     }
      {
        \rho, \exc \fwdslice  \vc, \trc
      }
\quad 
\vc \sqleq \vc
\]
\item If the derivations are of the form:
\[
  \inferrule*{\strut}
      {
       \rho,\trfunk{\kappa} \fwdslice \vclos{\kappa}{\rho}, \trfunk{\kappa}
      }
\quad
\inferrule*
    {
      \strut      
    }
    {
     \gamma,  \exfunk{\kappa} \red \vclos{\kappa}{\gamma},\trfunk{\kappa}
    }
\]
then suppose $\gamma' \sqgeq \rho$ is given where
$\gamma',\trfunk{\kappa} \eval v',T'$.  By inversion we must have $ v'
= 
\vclos{\kappa}{\gamma'}$ and $T' = \trfunk{\kappa}$.  Thus, we can
conclude by observing:
\[
  \inferrule*{\strut}
      {
       \rho,\trfunk{\kappa} \fwdslice \vclos{\kappa}{\rho}, \trfunk{\kappa}
      }
\quad
\vclos{\kappa}{\rho} \sqleq \vclos{\kappa}{\gamma'}
\]
\item
If the derivations are of the form:
\[
   \inferrule*
      {
        \rho,T_1 \fwdslice p_1,S_1\\
        \rho[x\mapsto p_1], T_2 \fwdslice p_2,S_2
     }
      {
        \rho, \trlet{T_1}{x}{T_2} 
        \fwdslice 
       p_2,\trlet{S_1}{x}{S_2}
      }
\quad
    \inferrule*
    {
      \gamma, e_1 \red v_1, T_1
      \\
      \gamma[x\mapsto v_1], e_2 \red v_2, T_2
   }
    {
      \gamma,\exletin{x  = e_1}{e_2} \red v_2, \trlet{T_1}{x}{T_2}
    }
\]
then suppose $\gamma' \sqleq \rho$ and $\gamma',\exletin{x=e_1}{e_2}
\eval v',T'$.   By inversion, this derivation is of the form:
\[
\inferrule*
    {
      \gamma', e_1 \red v_1', T_1'
      \\
      \gamma'[x\mapsto v_1'], e_2 \red v_2', T_2'
   }
    {
      \gamma',\exletin{x  = e_1}{e_2} \red v_2', \trlet{T_1'}{x}{T_2'}
    }
\]
and $v' = v_2'$ and $T' = \trlet{T_1'}{x}{T_2'}$.  By induction, we
have $\rho,T_1' \fwdslice p_1,S_1$ and $p_1 \sqleq v_1'$.  Thus,
$\gamma'[x\mapsto v_1'] \sqgeq \rho[x\mapsto p_1]$, so by induction,
we have $\rho[x\mapsto p_1],T_2' \fwdslice p_2,S_2$ where $p_2 \sqleq
v_2'$.  To conclude, we have:
\[
 \inferrule*
      {
        \rho,T_1' \fwdslice p_1,S_1\\
        \rho[x\mapsto p_1], T_2' \fwdslice p_2,S_2
     }
      {
        \rho, \trlet{T_1'}{x}{T_2'} 
        \fwdslice 
       p_2,\trlet{S_1}{x}{S_2}
      }
\quad p_2 \sqleq v_2'
\]
\item 
If the derivations are of the form:
\[
      \inferrule*{
       \rho,T_1 \fwdslice v_1,S_1
        \quad \cdots \quad
        \rho,T_n \fwdslice v_n,S_n
      }
      {
        \rho,\trf{T_1,\ldots,T_n} \fwdslice \kwf(v_1,\ldots,v_n), \trf{S_1,\ldots,S_n}
      }
\quad
    \inferrule*
    {
     \gamma,  e_1 \red v_1,T_1\\
     \cdots\\
     \gamma,e_n \red v_n,T_n
   }
    {
      \gamma, \kwf(e_1,\ldots,e_n) \red \hat{\oplus}(v_1,\ldots,v_n), \trf{T_1,\ldots,T_n}
    }
\]
then suppose $\gamma' \sqgeq \rho$ and $\gamma',\kwf(e_1,\ldots,e_n) \red
v',T'$.  By inversion the
derivation must be of the form:
\[
    \inferrule*
    {
     \gamma',  e_1 \red v'_1,T'_1\\
     \cdots\\
     \gamma',  e_n \red v'_n,T'_n
}
  {
      \gamma', \kwf(e_1,\ldots,e_n) \red \hat{\oplus}(v'_1,\ldots,v'_n), \trf{T'_1,\ldots,T'_n}
    }
\]
where $v' = \hat{\oplus}(v_1',\ldots,v_n')$ and $T' = \trf{T'_1,\ldots,T'_n}$.
By induction, we know that for each $i$, $\rho,T'_i \fwdslice v_i,S_i$
and $v_i \sqleq v_i'$.  The latter implies $v_i =v_i'$ since $v_i$ is
a constant value.  Thus, we can conclude:
\[
\inferrule*{
       \rho,T_1' \fwdslice v_1,S_1
        \quad \cdots \quad
        \rho,T_n' \fwdslice v_n,S_n
      }
      {
        \rho,\trf{T_1',\ldots,T_n'} \fwdslice \kwf(v_1,\ldots,v_n), \trf{S_1,\ldots,S_n}
      }
\quad
\hat{\oplus}(v_1,\ldots,v_n) = \hat{\oplus}(v_1',\ldots,v_n') = v'
\]
\item 
If the derivations are of the form:
\[
          \inferrule*{
       \rho,T_i \fwdslice \vhole,S_i \\ 
(\text{for some $i \in 1,\ldots,n$})
     }
      {
        \rho,\trf{T_1,\ldots,T_n} \fwdslice \vhole,\tremp
      }
\quad
    \inferrule*
    {
     \gamma,  e_1 \red v_1,T_1\\
\cdots\\
     \gamma,  e_n \red v_n,T_n
   }
    {
      \gamma, \kwf(e_1,\ldots,e_n) \red \hat{\oplus}(v_1,\ldots,v_n), \trf{T_1,\ldots,T_n}
    }
\]
then suppose $\gamma' \sqgeq \rho$ and $\gamma',\kwf(e_1,\ldots,e_n) \red
v',T'$.  By inversion the
derivation must be of the form:
\[
    \inferrule*
    {
     \gamma',  e_1 \red v'_1,T'_1\\
\cdots\\
     \gamma',  e_n \red v'_n,T'_n
    }
    {
      \gamma', \kwf(e_1,\ldots,e_n) \red \hat{\oplus}(v'_1,\ldots,v'_n), \trf{T'_1,\ldots,T'_n}
    }
\]
where $v' = \hat{\oplus}(v'_1,\ldots,v'_n)$ and $T' = \trf{T'_1,\ldots,T'_n}$.
By induction, we know that $\rho,T'_i \fwdslice \vhole,S_i$
and $\vhole \sqleq v_i'$.  Thus, we can conclude:
\[
\inferrule*{
             \rho,T'_i \fwdslice \vhole,S_i \\ 
(\text{for some $i \in 1,\ldots,n$})
      }
      {
        \rho,\trf{T_1',\ldots,T_n'} \fwdslice \vhole, \hole
      }
\quad
\vhole \sqleq \hat{\oplus}(v_1',\ldots,v_n') = v'
\]
\item 
Suppose the derivations are of the form:
\[
      \inferrule*
      {
        \rho, T_1 \fwdslice p_1,S_1
        \\
        \rho, T_2 \fwdslice p_2,S_2 
      }
      {
        \rho,\trpair{T_1}{T_2} 
        \fwdslice 
        \vpair{p_1}{p_2},
        \trpair{S_1}{S_2}
      }
\quad
    \inferrule*
    {
      \gamma,  e_1 \red v_1, T_1
      \\
      \gamma,  e_2 \red v_2, T_2
    }
    {
     \gamma,  \expair{e_1}{e_2} \red \expair{v_1}{v_2}, \trpair{T_1}{T_2}
    }
\]
and suppose $\gamma' \sqgeq \rho$ is given, where $\gamma', \expair{e_1}{e_2} \eval
v',T'$.  By inversion, the derivation must have the form:
\[
   \inferrule*
    {
      \gamma',  e_1 \red v_1', T_1'
      \\
      \gamma',  e_2 \red v_2', T_2'
    }
    {
     \gamma',  \expair{e_1}{e_2} \red \expair{v_1'}{v_2'}, \trpair{T_1'}{T_2'}
    }
\]
so $v' = \vpair{v_1'}{v_2'}$ and $T' = \trpair{T_1'}{T_2'}$.  By
  induction we have $\rho,T_1' \fwdslice p_1,S_1$ and $p_1 \sqleq
  v_1'$ and $\rho,T_2' \fwdslice p_2,S_2$ and $p_2 \sqleq v_2'$.
So we can conclude:
\[
  \inferrule*
      {
        \rho, T_1' \fwdslice p_1,S_1
        \\
        \rho, T_2' \fwdslice p_2,S_2 
      }
      {
        \rho,\trpair{T_1'}{T_2'} 
        \fwdslice 
        \vpair{p_1}{p_2},
        \trpair{S_1}{S_2}
      }
\quad 
\vpair{p_1}{p_2} \sqleq \vpair{v_1'}{v_2'}
\]
\item 
Suppose the derivations are of the form:
\[
 \inferrule*
      { \rho, T \fwdslice (p_1,p_2),S
      }
      {
        \rho,\trfst{T} \fwdslice p_1,\trfst{S}
      }
\quad
    \inferrule*
    {
\gamma,  e \red \expair{v_1}{v_2}, T
}
    {
\gamma, \exfst{e} \red v_1, \trfst{T}
}
\]
and suppose $\gamma' \sqgeq \rho$ is given, where $\gamma', \exfst{e} \eval
v',T'$.  By inversion, the derivation must have the form:
\[
 \inferrule*
    {
\gamma',  e \red \expair{v_1'}{v_2'}, T'
}
    {
\gamma', \exfst{e} \red v_1', \trfst{T'}
}
\]
so $v' = v_1'$ and $T' = \trfst{T'}$.  By induction, we know that
$\rho, T' \fwdslice \vpair{p_1}{p_2},S$ where $\vpair{p_1}{p_2} \sqleq
\vpair{v_1'}{v_2'}$. 
So we can conclude:
\[
 \inferrule*
      { \rho, T' \fwdslice (p_1,p_2),S
      }
      {
        \rho,\trfst{T'} \fwdslice p_1,\trfst{S}
      }
\quad 
p_1 \sqleq v_1'
\]
\item 
Suppose the derivations are of the form:
\[
   \inferrule*
      { \rho, T \fwdslice \vhole,S
      }
      {
        \rho,\trfst{T} \fwdslice \vhole,\tremp
      }
\quad
    \inferrule*
    {
\gamma,  e \red \expair{v_1}{v_2}, T
}
    {
\gamma, \exfst{e} \red v_1, \trfst{T}
}
\]
and suppose $\gamma' \sqgeq \rho$ is given, where $\gamma', \exfst{e} \eval
v',T'$.  By inversion, the derivation must have the form:
\[
 \inferrule*
    {
\gamma',  e \red \expair{v_1'}{v_2'}, T'
}
    {
\gamma', \exfst{e} \red v_1', \trfst{T'}
}
\]
so $v' = v_1'$ and $T' = \trfst{T'}$.  By induction, we know that
$\rho, T' \fwdslice \vpair{p_1}{p_2},S$ where $\vhole \sqleq
\vpair{v_1'}{v_2'}$. 
So we can conclude:
\[
 \inferrule*
      { \rho, T' \fwdslice \vhole,S
      }
      {
        \rho,\trfst{T'} \fwdslice \vhole,\vhole
      }
\quad 
\vhole \sqleq v_1'
\]
\item The cases for $\exsnd{e}$ are symmetric.
\item 
Suppose the derivations are of the form:
\[
     \inferrule*{\rho,T \fwdslice p,S}
      {
        \rho,\trinl{T} \fwdslice \vinl{p},\trinl{S}
      }
\quad
    \inferrule*
    {
      \gamma , e \red v, T
    }
    {
      \gamma , \exinl{e} \red \vinl{v}, \trinl{T} 
    }
\]
and suppose $\gamma' \sqgeq \rho$ is given, where $\gamma', \exinl{e}\eval
v',T'$.  By inversion, the derivation must have the form:
\[
    \inferrule*
    {
      \gamma' , e \red v_0', T_0'
    }
    {
      \gamma' , \exinl{e} \red \vinl{v'}, \trinl{T_0'} 
    }
\]
so $v' = \vinl{v_0'}$ and $T' = \trinl{T_0'}$.  By induction we have
$\rho,T_0' \fwdslice p,S$ and $p \sqleq v_0'$.
So we can conclude:
\[
     \inferrule*{\rho,T' \fwdslice p,S}
      {
        \rho,\trinl{T'} \fwdslice \vinl{p},\trinl{S}
      }
\quad
\vinl{p} \sqleq \vinl{v_0'}
\]
%
%
%%%%%%%%%%%%%%%%%%%%%%%%%%%%%%%%%%%%%%%%%%%%%%%%%%%%%%%%%%%%%%%%%%%%%%%%%%%%%%%
\item The case for $\vinl{e}$ is symmetric.
\item 
Suppose the derivations are of the form:
\[
    \inferrule*
      {
        \rho, T \fwdslice \vinl{p},S
        \\
        \rho[x_1\mapsto p], T_1 \fwdslice p_1,S_1
        }
      {
       \rho,\trcaseml{m}{T}{x_1}{T_1}
        \fwdslice 
         p_1,\trcaseml{m}{S}{x_1}{S_1}
      }
\]\[
\inferrule*
    {
     (\exinl{x_1}.e_1 \in m)\\
      \gamma,e \red \vinl{v}, T
      \\
      \gamma[x_1\mapsto v], e_1 \red v_1, T_1
   }
    {
      \strut
      \gamma,\excasem{e}{m} \red v_1, \trcaseml{m}{T}{x_1}{T_1}
    }
\]
and suppose $\gamma' \sqgeq \rho$ is given, where $\gamma',
\excasem{e}{m} \eval
v',T'$.  By inversion of this derivation, there are two cases.  If the derivation is of
the form:
\[
\inferrule*
    {
     (\exinl{x_1}.e_1 \in m)\\
      \gamma',e \red \vinl{v_0'},T_0'
      \\
      \gamma'[x_1\mapsto v_0'], e_1 \red v_1', T_1'
   }
    {
      \strut
      \gamma',\excasem{e}{m} \red v_1', \trcaseml{m}{T'}{x_1}{T_1'}
    }
\]
then $v' = v_1'$ and $T' = \trcaseml{m}{T'}{x_1}{T_1'}$.  By
induction, we have $\rho,T_0' \fwdslice \vinl{p},S$ and $\vinl{p}
\sqleq \vinl{v_0'}$, so $p \sqleq v_0'$ and $\gamma'[x\mapsto v_0']
\sqgeq \rho[x\mapsto p]$.  Again by induction, we have $\rho[x\mapsto
p],T_1' \fwdslice p_1,S_1$ where $p_1 \sqleq v_1'$.  
So we can conclude:
\[
 \inferrule*
      {
        \rho, T' \fwdslice \vinl{p},S
        \\
        \rho[x_1\mapsto p], T_1' \fwdslice p_1,S_1
        }
      {
       \rho,\trcaseml{m}{T'}{x_1}{T_1'}
        \fwdslice 
         p_1,\trcaseml{m}{S}{x_1}{S_1}
      }
\quad
p_1 \sqleq v_1'
\]

If the derivation of $\gamma',
\excasem{e}{m} \eval
v',T'$ is of the form 
\[
    \inferrule*
    {
      (\exinr{x_2}.e_2 \in m)\\
      \gamma,e \red \vinr{v}, T
      \\
      \gamma[x_2\mapsto v], e_2 \red v_2, T_2
    }
    {
      \strut
      \gamma,\excasem{e}{m} \red v_2, \trcasemr{m}{T}{x_2}{T_2}
    }
\]
then by induction we can derive $\rho,T' \fwdslice \vinl{p},S$
and $\vinl{p} \sqleq \vinr{v_0'}$, which is absurd, so this case is
vacuous.
\item 
Suppose the derivations are of the form:
\[
       \inferrule*
      {
        \rho, T \fwdslice \vhole,S
       }
      {
       \rho,\trcaseml{m}{T}{x_1}{T_1}
        \fwdslice 
         \vhole,\tremp
      }
\quad
\inferrule*
    {
     (\exinl{x_1}.e_1 \in m)\\
      \gamma,e \red \vinl{v}, T
      \\
      \gamma[x_1\mapsto v], e_1 \red v_1, T_1
   }
    {
      \strut
      \gamma,\excasem{e}{m} \red v_1, \trcaseml{m}{T}{x_1}{T_1}
    }
\]
and suppose $\gamma' \sqgeq \rho$ is given, where $\gamma',
\excasem{e}{m} \eval
v',T'$.  By inversion of this derivation, there are two cases.  If the derivation is of
the form:
\[
\inferrule*
    {
     (\exinl{x_1}.e_1 \in m)\\
      \gamma',e \red \vinl{v_0'},T_0'
      \\
      \gamma'[x_1\mapsto v_0'], e_1 \red v_1', T_1'
   }
    {
      \strut
      \gamma',\excasem{e}{m} \red v_1', \trcaseml{m}{T'}{x_1}{T_1'}
    }
\]
then $v' = v_1'$ and $T' = \trcaseml{m}{T'}{x_1}{T_1'}$.  By
induction, we know that $\rho,T_0' \fwdslice \vhole,S$ and $\vhole
\sqleq \vinl{v_0'}$, so we can conclude:
\[
   \inferrule*
      {
        \rho, T_0' \fwdslice \vhole,S
       }
      {
       \rho,\trcaseml{m}{T_0'}{x_1}{T_1'}
        \fwdslice 
         \vhole,\tremp
      }
\quad \vhole \sqleq v_1'
\]

If the derivation of $\gamma',
\excasem{e}{m} \eval
v',T'$ is of the form
\[
 \inferrule*
    {
      (\exinr{x_2}.e_2 \in m)\\
      \gamma,e \red \vinr{v_0'}, T_0'
      \\
      \gamma[x_2\mapsto v], e_2 \red v_2', T_2'
    }
    {
      \strut
      \gamma,\excasem{e}{m} \red v_2', \trcasemr{m}{T'}{x_2}{T_2'}
    }
\]
then the same reasoning applies: by induction on the first
subderivation we can obtain $\rho,T_0' \fwdslice \vhole,S$ and $\vhole
\sqleq \vinr{v_0'}$, and conclude:
\[
   \inferrule*
      {
        \rho, T_0' \fwdslice \vhole,S
       }
      {
       \rho,\trcasemr{m}{T_0'}{x_1}{T_1'}
        \fwdslice 
         \vhole,\tremp
      }
\quad \vhole \sqleq v_2'
\]
\item The cases for $\trcasemr{m}{T}{x_2}{T_2}$ are symmetric.
\item 
If the derivations are of the form:
\[
      \inferrule*
      {
        \rho, T_1 \fwdslice \vclos{\kappa}{\rho_0},S_1
        \\
        \rho,T_2 \fwdslice p_2,S_2
        \\
        \rho_0[f \mapsto \vclos{\kappa}{\rho_0}, x \mapsto p_2], T
        \fwdslice p,S
      }
    {
      \rho,\trappk{\kappa}{T_1}{T_2}{f}{x}{T} 
      \fwdslice 
      p,\trappk{\kappa}{S_1}{S_2}{f}{x}{S}
      }
\]
\[
\inferrule*
    {
      \gamma,e_1 \red \vclos{\kappa}{\gamma_0},T_1
      \\
      (\kappa = \fn{f}{x}{e})
      \\
      \gamma,e_2 \red v_2,T_2
      \\
      \gamma_0[f \mapsto \vclos{\kappa}{\gamma_0},x \mapsto v_2], e \red v, T
   }
    {
      \gamma,\exapp{e_1}{e_2} \red v, \trappk{\kappa}{T_1}{T_2}{f}{x}{T}
    }
\]
then suppose $\gamma' \sqgeq \rho$ and $\gamma',\exapp{e_1}{e_2} \eval
v',T'$.  Then by inversion the derivation must have the form:
\[
\inferrule*
    {
      \gamma',e_1 \red \vclos{\kappa'}{\gamma_0'},T'_1
      \\
      (\kappa' = \fn{f}{x}{e'})
      \\
      \gamma',e_2 \red v_2',T'_2
      \\
      \gamma_0'[f \mapsto \vclos{\kappa'}{\gamma_0'},x \mapsto v_2'], e' \red v', T_0'
   }
    {
      \gamma',\exapp{e_1}{e_2} \red v', \trappk{\kappa'}{T'_1}{T'_2}{f}{x}{T_0'}
    }
\]
so $T' =  \trappk{\kappa'}{T'_1}{T'_2}{f}{x}{T_0'}$.
By induction, we know that $\rho,T'_1 \fwdslice
\vclos{\kappa}{\rho_0},S_1$ and $\vclos{\kappa}{\rho_0} \sqleq
\vclos{\kappa'}{\gamma_0'}$, from which it follows that $\kappa =
\kappa'$ and $\gamma_0' \sqgeq \rho_0$.  Similarly, by induction we
know that $\rho,T'_2 \fwdslice p_2,S_2$.  Furthermore, note that
$\rho_0[f\mapsto \vclos{\kappa}{\rho_0} , x \mapsto p_2] \sqleq
  \gamma_0'[f\mapsto \vclos{\kappa}{\gamma_0'}, x\mapsto v_2']$, and
  since $\kappa = \kappa'$, we have $e = e'$ so by
  induction on the third subderivation we have 
\[
\rho_0[f\mapsto \vclos{\kappa}{\rho_0} , x \mapsto p_2], T_0'
  \fwdslice p,S
\]
and $p \sqleq v'$.  To conclude, we have:
\[
      \inferrule*
      {
        \rho, T_1' \fwdslice \vclos{\kappa}{\rho_0},S_1
        \\
        \rho,T_2' \fwdslice p_2,S_2
        \\
        \rho_0[f \mapsto \vclos{\kappa}{\rho_0}, x \mapsto p_2], T_0'
        \fwdslice p,S
      }
    {
      \rho,\trappk{\kappa}{T_1'}{T_2'}{f}{x}{T_0'} 
      \fwdslice 
      p,\trappk{\kappa}{S_1}{S_2}{f}{x}{S}
      }
\quad p \sqleq v'
\]
as desired.
\item 
If the derivations are of the form:
\[
       \inferrule*
      {
        \rho, T_1 \fwdslice \vhole,S_1
      }
   {
      \rho,\trappk{\kappa}{T_1}{T_2}{f}{x}{T} 
      \fwdslice 
     \vhole,\tremp
      }
\]
\[
\inferrule*
    {
      \gamma,e_1 \red \vclos{\kappa}{\gamma_0},T_1
      \\
      (\kappa = \fn{f}{x}{e})
      \\
      \gamma,e_2 \red v_2,T_2
      \\
      \gamma_0[f \mapsto \vclos{\kappa}{\gamma_0},x \mapsto v_2], e \red v, T
   }
    {
      \gamma,\exapp{e_1}{e_2} \red v, \trappk{\kappa}{T_1}{T_2}{f}{x}{T}
    }
\]
then suppose $\gamma' \sqgeq \rho$ and $\gamma',\exapp{e_1}{e_2} \eval
v',T'$.  Then by inversion the derivation must have the form:
\[
\inferrule*
    {
      \gamma',e_1 \red \vclos{\kappa'}{\gamma_0'},T'_1
      \\
      (\kappa' = \fn{f}{x}{e'})
      \\
      \gamma',e_2 \red v_2',T'_2
      \\
      \gamma_0'[f \mapsto \vclos{\kappa'}{\gamma_0'},x \mapsto v_2'], e' \red v', T_0'
   }
    {
      \gamma',\exapp{e_1}{e_2} \red v', \trappk{\kappa'}{T'_1}{T'_2}{f}{x}{T_0'}
    }
\]
so $T' =  \trappk{\kappa'}{T'_1}{T'_2}{f}{x}{T_0'}$.
By induction, we know that $\rho,T'_1 \fwdslice
\vhole,S_1$ and $\vhole \sqleq
\vclos{\kappa'}{\gamma_0'}$.  To conclude, we have:
\[
      \inferrule*
      {
        \rho, T'_1 \fwdslice \vhole,S_1
      }
   {
      \rho,\trappk{\kappa}{T'_1}{T'_2}{f}{x}{T_0'} 
      \fwdslice 
     \vhole,\tremp
      }
\quad \vhole \sqleq v'
\]
as desired. 
\item 
If the derivations are of the form:
\[
    \inferrule*{\rho,T \fwdslice p,S}
      {
        \rho,\trroll{T} \fwdslice \trroll{p},\trroll{S}
      }
\quad
\inferrule*
    {
      \gamma , e \red v, T
    }
    {
      \gamma , \exroll{e} \red \vroll{v}, \trroll{T} 
    }
\]
then let $\gamma' \sqgeq \rho$ be given, and assume
$\gamma',\exroll{e} \eval v',T'$.  By inversion the derivation must be
of the form:
\[
   \inferrule*
    {
      \gamma' , e \red v_0', T_0'
    }
    {
      \gamma' , \exroll{e} \red \vroll{v_0'}, \trroll{T_0'} 
    }
\]
so by induction we have $\rho,T_0' \fwdslice p,S$ and $p \sqleq v_0'$.  We
can conclude that:
\[
      \inferrule*{\rho,T_0' \fwdslice p,S}
      {
        \rho,\trroll{T_0'} \fwdslice \trroll{p},\trroll{S}
      }
\quad
 \vroll{p} \sqleq \vroll{v_0'}
\]
\item If the derivations are of the form:
\[
      \inferrule*{\rho,T \fwdslice \trroll{p},S}
      {
        \rho,\trunroll{T} \fwdslice p,\trunroll{S}
      }
\quad
    \inferrule*
    {
      \gamma , e \red \vroll{v}, T
    }
    {
      \gamma , \exunroll{e} \red v, \trunroll{T} 
    }
\]
then let $\gamma' \sqgeq \rho$ be given and assume
$\gamma',\exunroll{e} \eval v',T'$.  By inversion this derivation must
be of the form:
\[
    \inferrule*{
      \gamma' , e \red \vroll{v'}, T_0'
    }
    {
      \gamma' , \exunroll{e} \red v', \trunroll{T_0'} 
    }
\]
so $T' = \trunroll{T_0'}$.  By induction we have that $\rho,T_0'
\fwdslice \vroll{p},S$ where $\vroll{p} \sqleq \vroll{v'}$, so we can
    conclude:
\[
      \inferrule*{\rho,T_0' \fwdslice \trroll{p},S}
      {
        \rho,\trunroll{T_0'} \fwdslice p,\trunroll{S}
      }
\quad
p \sqleq v'
\]

\item If the derivations are of the form:
\[
     \inferrule*{\rho,T \fwdslice \vhole,S}
      {
        \rho,\trunroll{T} \fwdslice \vhole,\tremp
      }
\quad
    \inferrule*
    {
      \gamma , e \red \vroll{v}, T
    }
    {
      \gamma , \exunroll{e} \red v, \trunroll{T} 
    }
\]
then let $\gamma' \sqgeq \rho$ be given and assume
$\gamma',\exunroll{e} \eval v',T'$.  By inversion this derivation must
be of the form:
\[
    \inferrule*{
      \gamma' , e \red \vroll{v'}, T_0'
    }
    {
      \gamma' , \exunroll{e} \red v', \trunroll{T_0'} 
    }
\]
so $T' = \trunroll{T_0'}$.  By induction we have that $\rho,T_0'
\fwdslice \vhole,S$ where $\vhole \sqleq \vroll{v'}$, so we can
    conclude:
\[
      \inferrule*{\rho,T_0' \fwdslice \vhole,S}
      {
        \rho,\trunroll{T_0'} \fwdslice \vhole,\tremp
      }
\quad
\vhole \sqleq v'
\]
\end{itemize}
\end{proof}

%%% Local Variables: 
%%% mode: latex
%%% TeX-master: "main"
%%% End: 

% LocalWords:  iff subderivations subvalue subexpression subderivation expr
% LocalWords:  dep replayable antitone Inl Inr subcase wildcard wildcards eqat
% LocalWords:  vany

\end{document}